\documentclass[twocolumn]{aastex631}

\defcitealias{paper1_healy}{I}
\defcitealias{paper2_healy}{II}

\newcommand{\paperone}{{Paper \citetalias{paper1_healy}}}
\newcommand{\papertwo}{{Paper \citetalias{paper2_healy}}}

\newcommand{\msun}{$M_{\odot}$}
\newcommand{\rsun}{$R_{\odot}$}

\received{May 11, 2022}
\revised{November 23, 2022}
\accepted{December 19, 2022}

\shorttitle{Study of Stellar Spins}
\shortauthors{Healy et al.}

\graphicspath{{./}{figures/}}

\usepackage[figuresright]{rotating}

\begin{document}

\title{A Study of Stellar Spins in 15 Open Clusters}

\correspondingauthor{Brian F. Healy}
\email{healyb@umn.edu}

\author[0000-0002-7718-7884]{Brian F. Healy}
\affiliation{School of Physics and Astronomy\\
University of Minnesota \\
Minneapolis, MN 55455, USA}
\affiliation{William H. Miller III Department of Physics and Astronomy\\
Johns Hopkins University \\
3400 North Charles Street \\
Baltimore, MD 21218, USA}

\author[0000-0001-9165-9799]{P.~R. McCullough}
\affiliation{William H. Miller III Department of Physics and Astronomy\\
Johns Hopkins University \\
3400 North Charles Street \\
Baltimore, MD 21218, USA}
\affiliation{Space Telescope Science Institute \\
3700 San Martin Drive \\
Baltimore, MD 21218, USA}

\author[0000-0001-5761-6779]{Kevin C. Schlaufman}
\affiliation{William H. Miller III Department of Physics and Astronomy\\
Johns Hopkins University \\
3400 North Charles Street \\
Baltimore, MD 21218, USA}

\author[0000-0002-2365-2330]{Geza Kovacs}
\affiliation{Konkoly Observatory, Research Center for Astronomy and Earth Sciences, 
  E\"otv\"os Lor\'and Research Network \\ Budapest, Hungary}
 
\begin{abstract}
We analyze spectroscopic and photometric data to determine the projected inclinations of stars in 11 open clusters, placing constraints on the spin-axis distributions of six clusters. We combine these results with four additional clusters studied by \citet{paper1_healy} and \citet[][]{paper2_healy} to perform an ensemble analysis of their spins. We find that eight out of ten constrained clusters (80\%) have spin-axis orientations consistent with isotropy, and we establish a lower limit of four out of ten (40\%) isotropic clusters at 75\% confidence, assuming no correlation of spins between clusters. We also identify two clusters whose spin-axis distributions can be better described by a model consisting of an aligned fraction of stars combined with an isotropic distribution. However, the inclination values of these stars may be influenced by systematic error, and the small number of stars modeled as aligned in these two clusters precludes the interpretation that their stellar subsets are physically aligned. Overall, no cluster displays an unambiguous signature of spin alignment, and 97\% of the stars in our sample are consistent with isotropic orientations in their respective clusters. Our results offer support for the dominance of turbulence over ordered rotation in clumps and do not suggest alignment of rotation axes and magnetic fields in protostars.

\end{abstract}

\keywords{Open star clusters; Inclination; Stellar rotation; Star formation}

\section{Introduction} \label{sec:intro}
The process of star formation begins primarily within giant molecular clouds (GMCs). Within GMCs, ``clumps'' of higher density substructure form with comparable mass and length scales to those of open clusters \citep[][and references therein]{shu1987}. Large-scale supersonic turbulence is thought to help shape these clumps, which become gravitationally unstable and begin to collapse as they decouple from the overall turbulent flow in their GMC \citep[][and references therein]{lada2003}. Within clumps, material fragments into protostellar cores, which can accrete additional molecular gas before clearing the star-forming region and emerging as an open cluster. The spins of the resulting stars are drawn from the angular momentum of the material from which they formed \citep[][and references therein]{mckee2007}.

Within an open cluster, determining the orientation of stellar spin axes facilitates an inference of the balance of turbulent and rotational energy that existed during the formation of the cluster's stars. While the dominance of turbulent kinetic energy in a clump randomizes the resulting stellar spins, numerical simulations suggest a comparable contribution from ordered rotation could result in spin alignment for stars $> 0.7$ \msun\ via protostellar accretion of material sharing the clump's global angular momentum   \citep[][]{corsaro2017, reyraposo2018}. Previous determinations of stellar spin-axis orientations in open clusters have yielded mixed results, finding isotropic spins in the Alpha Per, Pleiades, and Praesepe clusters \citep[][]{jackson2010, jackson2018, jackson2019} and broadly aligned spins in NGC 6791, NGC 6811, \citep[][]{corsaro2017} and Praesepe \citep{kovacs2018}.

Our past contributions are also consistent with multiple explanations. \citet[][\paperone]{paper1_healy} determined the projected inclination ($\sin i$) values of NGC 2516 cluster members, finding that their distribution is consistent with either isotropy or moderate alignment. \citet[][\papertwo]{paper2_healy} performed a similar analysis of the Pleiades, Praesepe and M35 clusters. The results included the possibility of a subset of aligned stars in M35, but not at a statistically significant level. As seen with NGC 2516 in \paperone, \papertwo\ determined a similarly degenerate scenario for the Pleiades and Praesepe, with the data supporting either isotropy or moderate alignment.

In this paper, we broaden our study of stellar spins with an analysis of 11 open clusters, 10 of whose stars' spin-axis distributions have not been determined previously. Our analysis is again driven by the spectro-photometric method of determining projected inclinations, described by the following equation:
\begin{equation}
    v\sin i = \frac{2\pi R}{P}\sin i
    \label{eq:spectrophot}
\end{equation}
Above, $v\sin i$ is a star's projected rotation velocity, $R$ is the radius, and $P$ is the equatorial rotation period.

In expanding our sample, we not only aim to test the longstanding assumption of isotropic spins and learn the characteristics of each individual cluster, but we also seek to synthesize the results from the 11 clusters in this work and four from our previous papers to challenge the degenerate solutions determined in our previous works using insights from the overall ensemble of clusters.

With these goals in mind, the structure of this paper is as follows: Section \ref{sec:datasources} lists the sources of data we use to perform our analysis, which is detailed in Section \ref{sec:analysis}. Section \ref{sec:results} presents our results for this expanded study. In Section \ref{sec:discussion}, we interpret each cluster's results and discuss the broader inferences about star formation that we are afforded by our series of papers. Finally, we offer concluding thoughts and future outlooks in Section \ref{sec:conclusion}.

\section{Data Sources} \label{sec:datasources}
As in Papers \citetalias{paper1_healy} and \citetalias{paper2_healy}, we used the cluster membership tables of \citet[][CG18]{cantatgaudin2018} updated with Gaia EDR3 data \citep[][]{gaia_edr3_2020} to establish our initial lists of stars to study. Recent research has revealed large tidal tails extending from the central regions of open clusters \citep[][]{kounkel2019_gaia_cluster_halos, meingast2021_gaia_cluster_halos}. We continue to rely on the CG18 catalog since the cluster membership probability for individual stars in tidal tail regions is more uncertain than for stars in cluster centers, and past spectroscopic surveys did not targets regions outside the central regions to provide the necessary data for our study.

We constructed spectral energy distributions (SEDs) beginning with queries of the following catalogs also used in \papertwo: (CatWISE2020 \citep[][]{catwise_2020}, 2MASS \citep[][]{skrutskie2006}, Hipparcos/Tycho-2 \citep[][]{perryman1997, hoeg2000} and GALEX \citep[][]{galex}). Using Gaia match tables to find corresponding entries, we also added SkyMapper DR2 \citep[][]{skymapper_dr2} and SDSS DR13 \citep[][]{sdss_dr13} colors along with Pan-STARRS $g$-band photometry \citep[][]{panstarrs-1, panstarrs-1_database} to the SEDs when available. We also added Gaia $G$ magnitudes to each SED, with  mean uncertainties calculated using the flux values and errors and added in quadrature to the uncertainty in the $G$ magnitude zero-point \citep[][]{gaia2018_photometry}. The quality flags we applied to provide reliable magnitudes for fitting are listed in the Appendix of \citet[][]{hamer2022}.

All of the time-series photometry in our analysis is based on TESS images and products \citep[][]{ricker2015}. Table \ref{tab:vsini_sources} lists each cluster in this study along with its galactic latitude and longitude, distance via inversion of Gaia EDR3 parallaxes, age determined by this work, mass estimate when available, and the source of $v\sin i$ measurements. This table also provides the $v\sin i$ threshold we established for each dataset (see Section \ref{subsec:analysis_vsini}). For SDSS DR17 sources, we utilize data from the APOGEE-2 spectrograph \citep[][]{majewski2017}. Data for Pozzo 1 and NGC 2547 originally come from the Gaia-ESO survey \citep[][]{gilmore2012_gaiaeso}.

We also attempted to add another cluster, NGC 2264, to our sample using CoRoT-based rotation periods determined by \citet[][]{affer2013_corot_ngc2264}. We used these measurements in an attempt to overcome significant blending in TESS caused by the small angular extent of this cluster. We combined these periods with $v\sin i$ data from \citet[][]{jackson2016}, but we did not perform further analysis after determining that NGC 2264 would not yield enough $\sin i$ values ($\gtrsim$ 10-15, \citealp[][]{jackson2010, corsaro2017}) to constrain the inclination distribution.

\begin{table*}[]
    \begin{center}
    \caption{Clusters in this study (with alternate names), galactic longitude and latitude, Gaia EDR3 distance, mean ages and standard deviations from SED fitting posteriors, mass estimates, sources of $v\sin i$, and $v\sin i$ threshold established to select reliable measurements.}
    \label{tab:vsini_sources}
    \begin{tabular}{c|c|c|c|c|c|c|c}
        \hline
        \hline
         Cluster & $l$ [$^{\circ}$] & $b$ [$^{\circ}$] & d [pc] & Age [Myr] & M [\msun] & $v\sin i$ src. & Thresh. (km s$^{-1}$) \\
         \hline
         
         Collinder 69 ($\lambda$ Orionis cl.) & 195.118 & -12.070 & 392 $\pm$ 10 & 4.9 $\pm$ 0.9 & 150-250 [1,2] & [9] & 10 \\
         
         ASCC 16 & 201.204 & -18.375 & 347.4 $\pm$ 7.5 & 7.1 $\pm$ 2.2 & $\sim$ 250 [1] & [9] & 10 \\
         
         ASCC 19 & 204.884 & -19.371 & 355.5 $\pm$ 7.8 & 8.0 $\pm$ 5.4 & $\sim$ 400 [1] & [9] & 10 \\
         
         Gulliver 6 & 205.249 & -18.139 & 415 $\pm$ 12  & 8.6 $\pm$ 5.8 & --- & [9] & 10 \\
         
         Pozzo 1 ($\gamma$ Velorum cl.) & 262.727 & -7.713 & 346.6 $\pm$ 5.9 & 10.2 $\pm$ 3.7 & 400-2000 [3,1] & [10] & 10 \\
         
         BH 56 (vdB-Ha 56) & 264.457 & 1.582 & 910 $\pm$ 42 & 11.8 $\pm$ 3.8 & $\sim$ 150 [1] & [9] & 10 \\
         
         NGC 2547 & 264.426 & -8.605 & 386.8 $\pm$ 6.7 & 31.7 $\pm$ 7.5& 50-450 [1,4] & [11] & 5 \\
         
         Alpha Per cl. (Melotte 20) & 147.350 & -6.324 & 174.4 $\pm$ 2.1 & 58.3 $\pm$ 4.9 & 350-7000 [5,1] & [9] & 10 \\
         
         Blanco 1 & 14.906 & -79.304 & 236.3 $\pm$ 4.4 & 105 $\pm$ 15 & 300-3000 [6,1] & [12] & 5 \\
         
         NGC 2422 (M47) & 230.961 & 3.134 & 476 $\pm$ 11 & 155.9 $\pm$ 8.2 & 200-450 [1,7] & [13] & 5 \\
         
         NGC 2548 (M48) & 227.841 & 15.377 & 772 $\pm$ 31 & 455 $\pm$ 25 & 370-1100 [8,1] & [14] & 12 \\
         
         \hline
    \end{tabular}
    \end{center}
    \footnotesize{
    $^{1}$ \citealp{piskunov2008_cluster_masses}, 
    $^{2}$ \citealp{bayo2011_collinder_69},
    $^{3}$ \citealp{jeffries2009_gammavel},
    $^{4}$ \citealp{jeffries2004_ngc2547_mass}, 
    $^{5}$ \citealp{sheihki2016_alphaper_mass},
    $^{6}$ \citealp{zhang2020_blanco1_mass},
    $^{7}$ \citealp{prisinzano2003_ngc2422_mass},
    $^{8}$ \citealp{balaguernunez2006_ngc2548_mass},
    $^{9}$ SDSS DR17 \citep{sdss_dr17},
    $^{10}$ \citealp{jeffries2014_gammavel},
    $^{11}$ \citealp{jackson2016},
    $^{12}$ \citealp{mermilliod2009},
    $^{13}$ \citealp{bailey2018_ngc2422_vsini},
    $^{14}$ WIYN Open Cluster Study \citep{sun2020_m48_vsini}
    }
    
    \begin{center}
    \end{center}
\end{table*}

\section{Analysis} \label{sec:analysis}
To determine projected inclinations and analyze their distribution in each new cluster (along with a re-analysis of the previously studied clusters for uniformity), we followed the steps described in Papers \citetalias{paper1_healy} and \citetalias{paper2_healy} with minor changes. The following subsections list the techniques we used to determine each of the components of Equation \ref{eq:spectrophot}.

\subsection{Star Selection}
We began by identifying and removing CG18 members whose Gaia $G$ magnitude and $G_{\rm BP} - G_{\rm RP}$ color placed them on the equal-mass binary main sequence of their cluster's color-magnitude diagram. In addition, we removed binary and pulsating stars identified by SIMBAD \citep[][]{wenger2000}. Section 3.1 of \paperone\ describes the motivation for removing equal-mass binaries to avoid systematic errors in a cluster's inclination determinations. We avoid pulsating stars in order to minimize confusion between pulsation and rotation periods.

We also removed stars showing excessive astrometric noise with Gaia, as quantified by having a Reduced Unit Weight Error (RUWE) value greater than the 99th percentile of values estimated to represent single stars (e.g. Section 3.1 of \papertwo, \citealp[][]{belokurov2020}). In order to optimize the number of reliable resulting inclination determinations for each cluster, we established a CG18 membership probability threshold at $P_{\rm mem} > 0.5$ for the final part of the analysis. This threshold balanced our preferences for a sample of stars as large as possible while containing only probably cluster members.

\subsection{Radius Values}
We used the \texttt{isochrones} software \citep[][]{morton2015} to fit MIST stellar evolution models \citep[][]{choi2016} to each star's SED. We built SEDs from the catalogs listed in Section \ref{sec:datasources}. We applied an estimated correction to the stars' Gaia EDR3 parallaxes using \texttt{zero-point} \citep[][]{lindegren2020_gaia_edr3_parallax_bias}.

We referenced 3D Bayestar19 dust maps \citep[][]{green2019} to estimate reddening for clusters with declinations greater than $-30^{\circ}$. We supplemented these maps with a combination of 2D maps from \citet[][]{schlafly2011} and targeted studies. We list the full set of Gaussian priors we used for SED fitting along with references in Table \ref{tab:sed_priors}.

In some cases, we assigned the priors a greater uncertainty than quoted in individual studies in order to avoid over-constraining the resulting radius values. Due to the lack of a published age estimate for Gulliver 6, we set the same prior as ASCC 19 owing to the clusters' proximity, their potential to be part of an aggregate group of clusters \citep[e.g.][]{ascc_19_gulliver_6_piecka_2021, soubiran2018}, and their sharing of one member star according to CG18. We eliminated this star from further analysis due to its unknown constituency.

After completing SED fitting, we identified stars whose nested sampling global log-evidence was less than a threshold we determined empirically for each cluster (see Section 4.1 of \papertwo). This had the effect of removing 1-5\% of stars in each cluster with the poorest SED fits from further analysis.

\begin{table*}
\begin{center}
\caption{Gaussian prior values for isochrone fitting and their sources. The three priors are the cluster age in Myr, Fe/H metallicity in dex and $V$-band reddening in mag.}
\label{tab:sed_priors}
\begin{tabular}{cccc}
\hline
\hline
Cluster & Parameter & Value & Source \\
\hline
  \hline
  & age & $5.2 \pm 1.2$ & \citet[][]{bayo2011_collinder_69, BN2004_collinder_69} \\
 Collinder 69 & [Fe/H] & $-0.169 \pm 0.093$ & \citet[][]{sdss_dr17} \\
  & $A_{V}$ & $0.49 \pm 0.22$ & \citet[][]{green2019} \\
  \hline
   & age & $10 \pm 2.3$ & \citet[][]{bossini2019, kharchenko2013_ascc} \\
 ASCC 16 & [Fe/H] & $-0.092 \pm 0.087$ & \citet[][]{sdss_dr17} \\
  & $A_{V}$ & $0.188 \pm 0.065$ & \citet[][]{green2019} \\
  \hline
  & age & $20.0 \pm 9.5$ & \citet[][]{bossini2019, kharchenko2013_ascc} \\
 ASCC 19 & [Fe/H] & $-0.13 \pm 0.12$ & \citet[][]{sdss_dr17} \\
  & $A_{V}$ & $0.49 \pm 0.28$ & \citet[][]{green2019} \\
  \hline
  & age & $20.0 \pm 9.5$ & same prior as ASCC 19 \\
 Gulliver 6 & [Fe/H] & $-0.15 \pm 0.11$ & \citet[][]{sdss_dr17} \\
  & $A_{V}$ & $0.64 \pm 0.36$ & \citet[][]{green2019} \\
  \hline
   & age & $20.0 \pm 4.6$ & \citet[][]{jeffries2017_gammavel, jeffries2009_gammavel} \\
 Pozzo 1 & [Fe/H] & $-0.04 \pm 0.05$ & \citet[][]{jeffries2014_gammavel,spina2014_gammavel} \\
  & $A_{V}$ & $0.131 \pm 0.055$ & \citet[][]{jeffries2009_gammavel} \\
  \hline
  & age & $17.4 \pm 4.0$ & \citet[][]{kharchenko2005_bh_56} \\
 BH 56 & [Fe/H] & $-0.12 \pm 0.11$ & \citet[][]{sdss_dr17} \\
  & $A_{V}$ & $0.2 \pm 0.2$ & \citet[][]{kharchenko2005_bh_56} \\
  \hline
   & age & $34.7 \pm 4.0$ & \citet[][]{jeffries2005_ngc_2547} \\
 NGC 2547 & [Fe/H] & $-0.03 \pm 0.06$ & \citet[][]{magrini2015_ngc_2264} \\
  & $A_{V}$ & $0.37 \pm 0.16$ & \citet[][]{naylor2006_ngc_2547} \\
  \hline
   & age & $63.1 \pm 7.3$ & \citet[][]{basri1999_alpha_per} \\
 Alpha Per & [Fe/H] & $-0.08 \pm 0.14$ & \citet[][]{sdss_dr17} \\
  & $A_{V}$ & $0.017 \pm 0.062$ & \citet[][]{green2019} \\
  \hline
   & age & $115 \pm 11$ & \citet[][]{gaia2018_hr_diagrams} \\
 Blanco 1 & [Fe/H] & $-0.09 \pm 0.13$ & \citet[][]{galah_dr3} \\
  & $A_{V}$ & $0.05 \pm 0.05$ & \citet[][]{schlafly2011} \\
  \hline
   & age & $155 \pm 18$ & \citet[][]{cummings2018_ngc_2422} \\
 NGC 2422 & [Fe/H] & $-0.08 \pm 0.21$ & \citet[][]{bailey2018_ngc2422_vsini} \\
  & $A_{V}$ & $0.284 \pm 0.097$ & \citet[][]{green2019} \\
  \hline
   & age & $447 \pm 52$ & \citet[][]{barnes2015_m48} \\
 NGC 2548 & [Fe/H] & $-0.03 \pm 0.10$ & \citet[][]{cui2012,hamer2021_lamost_data} \\
  & $A_{V}$ & $0.145 \pm 0.062$ & \citet[][]{green2019} \\
  \hline
\hline
\end{tabular}
\end{center}
\end{table*}

\subsection{Rotation Periods}
To determine rotation periods, we used a combination of TESS light curves from the PATHOS project \citep[][]{nardiello2019_pathos1, nardiello2020_pathos2, nardiello2020_pathos3, nardiello2021_pathos4} and \texttt{eleanor} software \citep[][]{feinstein2019}. As in \paperone, we generated each light curve's autocorrelation function \citep[ACF, e.g.][]{mcquillan2013, mcquillan2014} and periodgram \citep[e.g.][]{nielsen2013} to provide period estimates. Combining the information from these analysis tools with a Gaia-based list of nearby blended stars and prior expectations of color-period relations in open clusters \citep[e.g.][]{barnes2003, kovacs2015, angus2019}, we manually vetted each star in the sample. We did not apply a membership probability threshold for this part of the analysis in order to maximize the number of reported periods for stars that are potentially cluster members.

Our evaluations yielded periods for some stars in each cluster that demonstrate light curve modulation consistent with rotating starspots (as opposed to oscillations, pulsations, or eclipses). The stars passing the vetting process are also not blended with other stars of similar color or brightness, which would confuse the assignment of a periodic signal to the correct star. Finally, we applied extra scrutiny to signals near harmonics of the $\sim$ 13.7-day TESS satellite's orbital period due to the influence of scattered light at those periods.

\subsection{Projected Rotation Velocities}
\label{subsec:analysis_vsini}
After accessing the various data sources listed in Table \ref{tab:vsini_sources}, we referenced the source papers to determine an appropriate threshold for $v\sin i$ values. We do not include measurements below the threshold in our inclination analysis. Section 3.3 of \papertwo\ further discusses the motivation of excluding projected rotation velocity measurements too close to the predicted macroturbulent velocity of a star \citep[see also][]{kamiaka2018}.

For SDSS APOGEE-2 data, we established a 10 km s$^{-1}$ $v\sin i$ threshold and assumed uncertainties of $\sim$ 2 km s$^{-1}$ or 10\%, whichever was greater \citep[e.g.][]{tayar2015, simonian2020}. With the above threshold, this meant that every APOGEE $v\sin i$ value we analyzed had a 10\% uncertainty. To further exclude poor-quality data, we removed stars having the following SDSS quality flags: (\texttt{ASPCAPFLAGS}): \texttt{VSINI\_BAD}, \texttt{SN\_BAD}, \texttt{STAR\_BAD}, \texttt{CHI2\_BAD}, \texttt{VMICRO\_BAD}; (\texttt{STARFLAGS}): \texttt{BAD\_PIXELS}, \texttt{VERY\_BRIGHT\_NEIGHBOR}, \texttt{LOW\_SNR}.  

In two other cases, we adopted a $v\sin i$ threshold greater than suggested by source papers. We set thresholds of 5 km s$^{-1}$ for Blanco 1 data \citep[][]{mermilliod2009} (consistent with our analysis of the same source in \papertwo) and NGC 2422 \citep[][]{bailey2018_ngc2422_vsini}. 

For Pozzo 1 data \citep[][]{jeffries2014_gammavel}, we estimated $v\sin i$ uncertainties by comparing signal-to-noise ratios (S/N) for the cluster's spectroscopy with those reported by \citet[][]{jackson2016} for NGC 2264, NGC 2547, and NGC 2516. These latter data were collected with the same GIRAFFE spectrograph that was used for the majority of Pozzo 1 spectroscopy by \citet[][]{jeffries2014_gammavel}. We computed the mean reported uncertainty for \citet[][]{jackson2016} $v\sin i$ data within 18 bins of S/N. We then performed linear interpolation to estimate the $v\sin i$ uncertainties for Pozzo 1 as a function of S/N using the binned data. Above the cluster's reported threshold of 10 km s$^{-1}$, we estimated a mean uncertainty of $\sim 20\%$ in $v\sin i$.

\subsection{Accounting for Systematic Error} \label{subsec:systematic_errors}

\begin{figure*}
    \centering
    \includegraphics[scale=0.35]{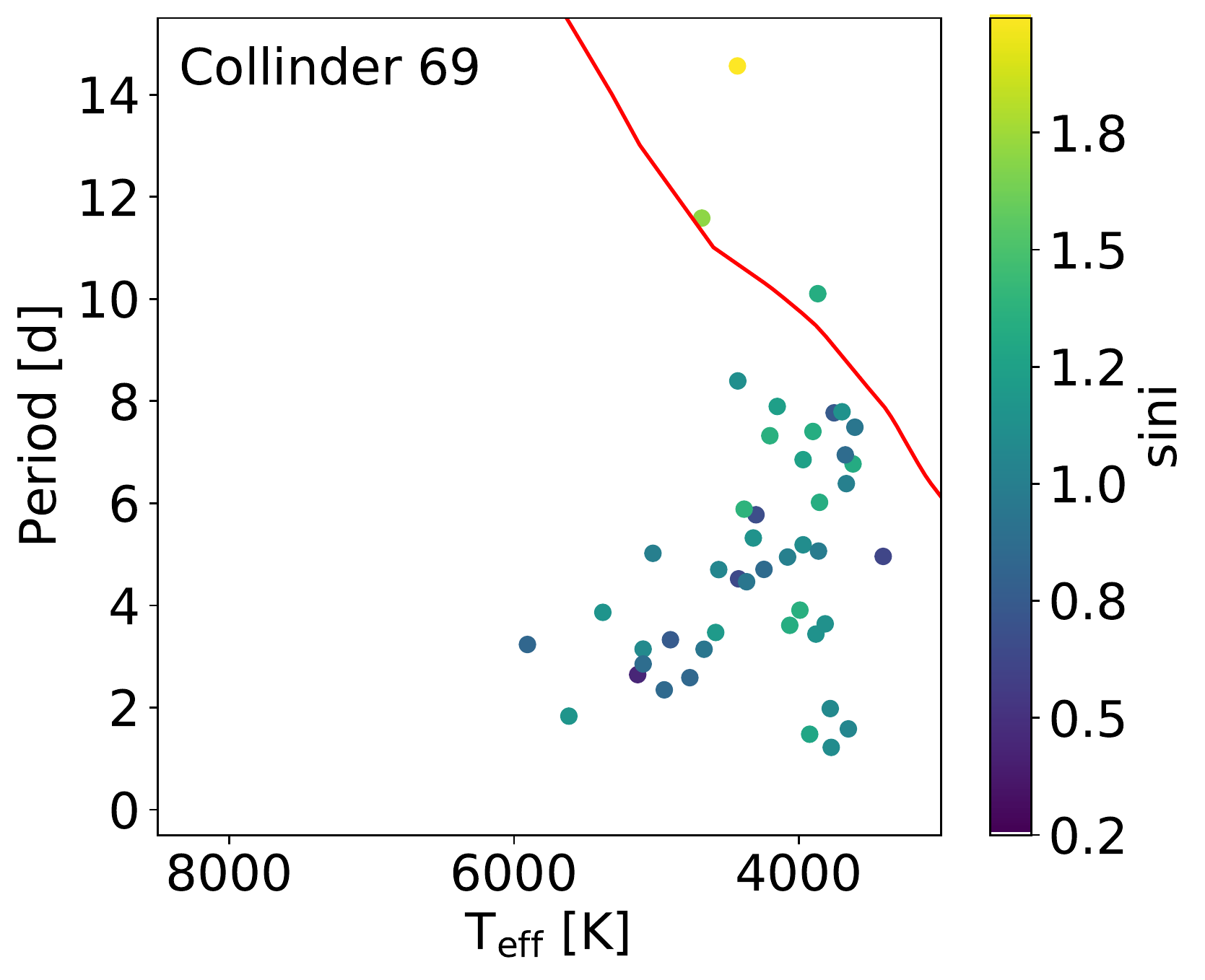}
    \includegraphics[scale=0.35]{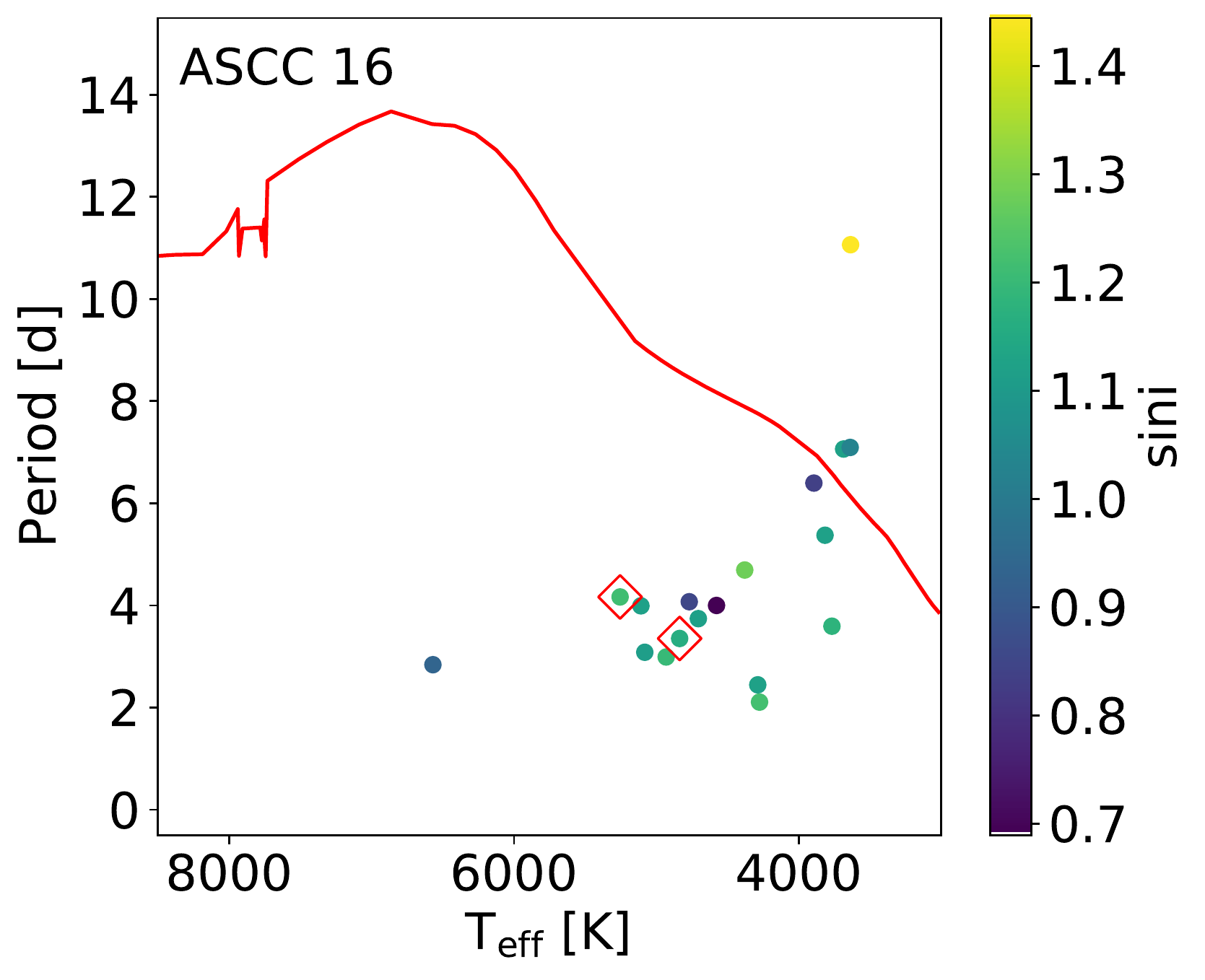}
    \includegraphics[scale=0.35]{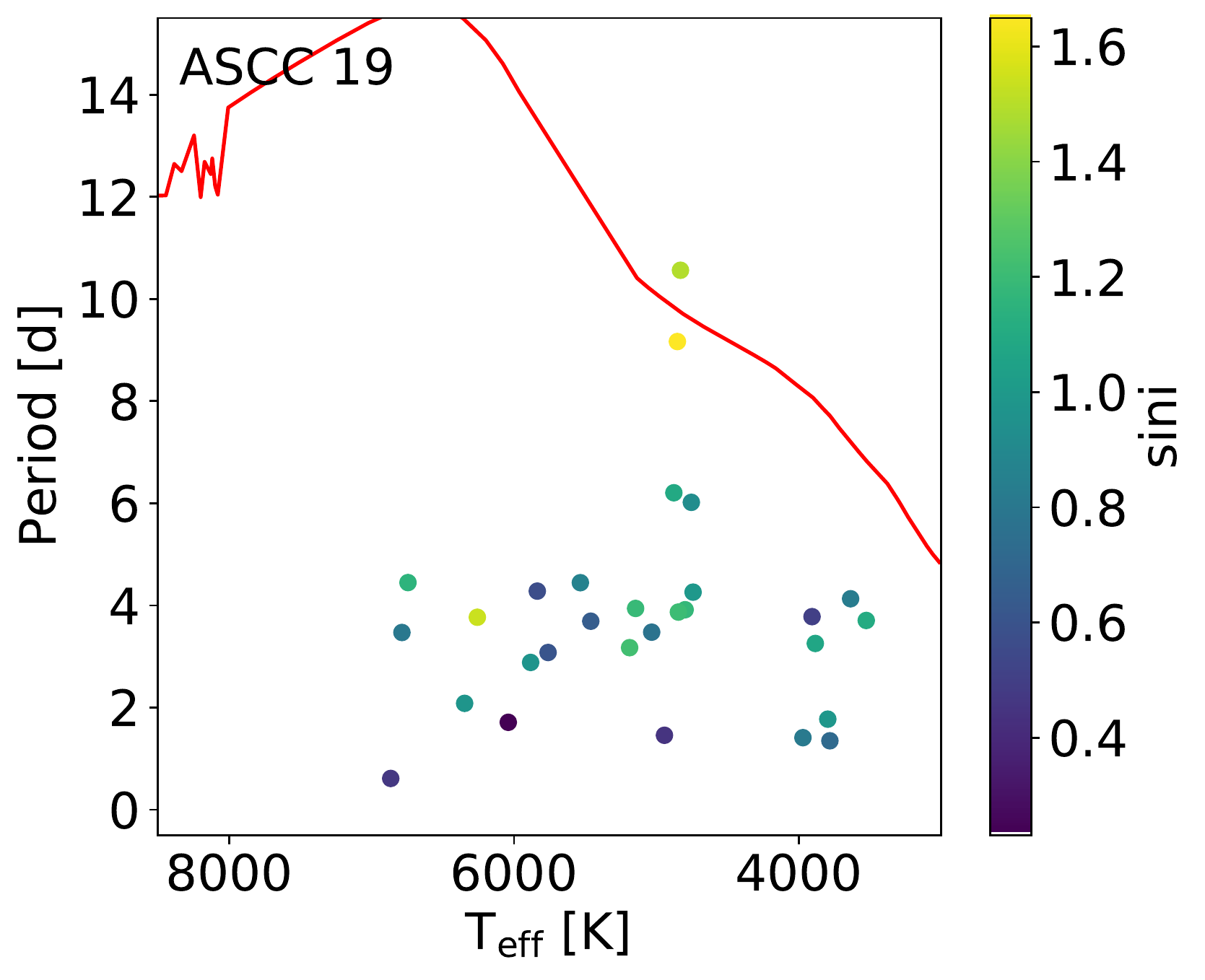}
    \includegraphics[scale=0.35]{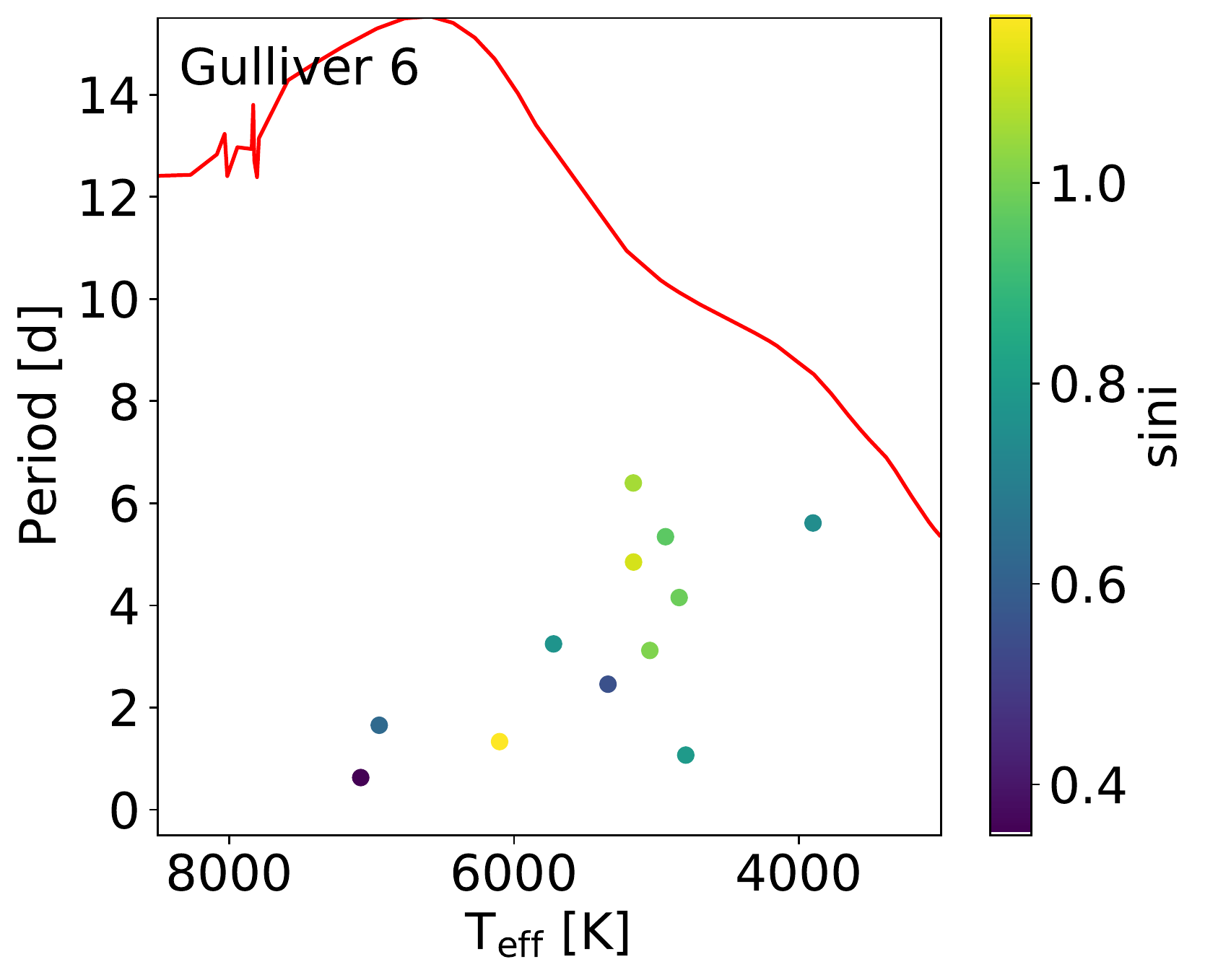}
    \includegraphics[scale=0.35]{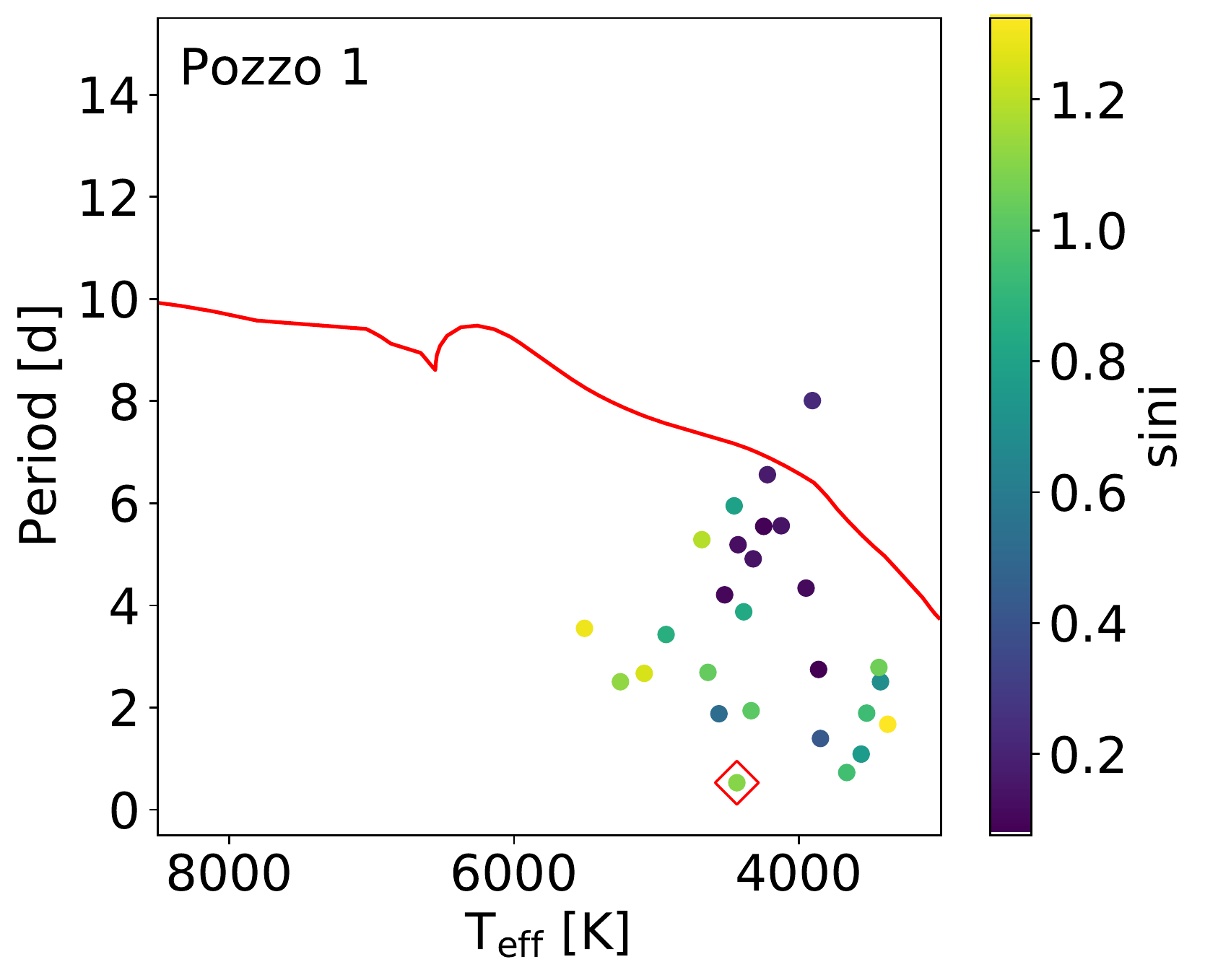}
    \includegraphics[scale=0.35]{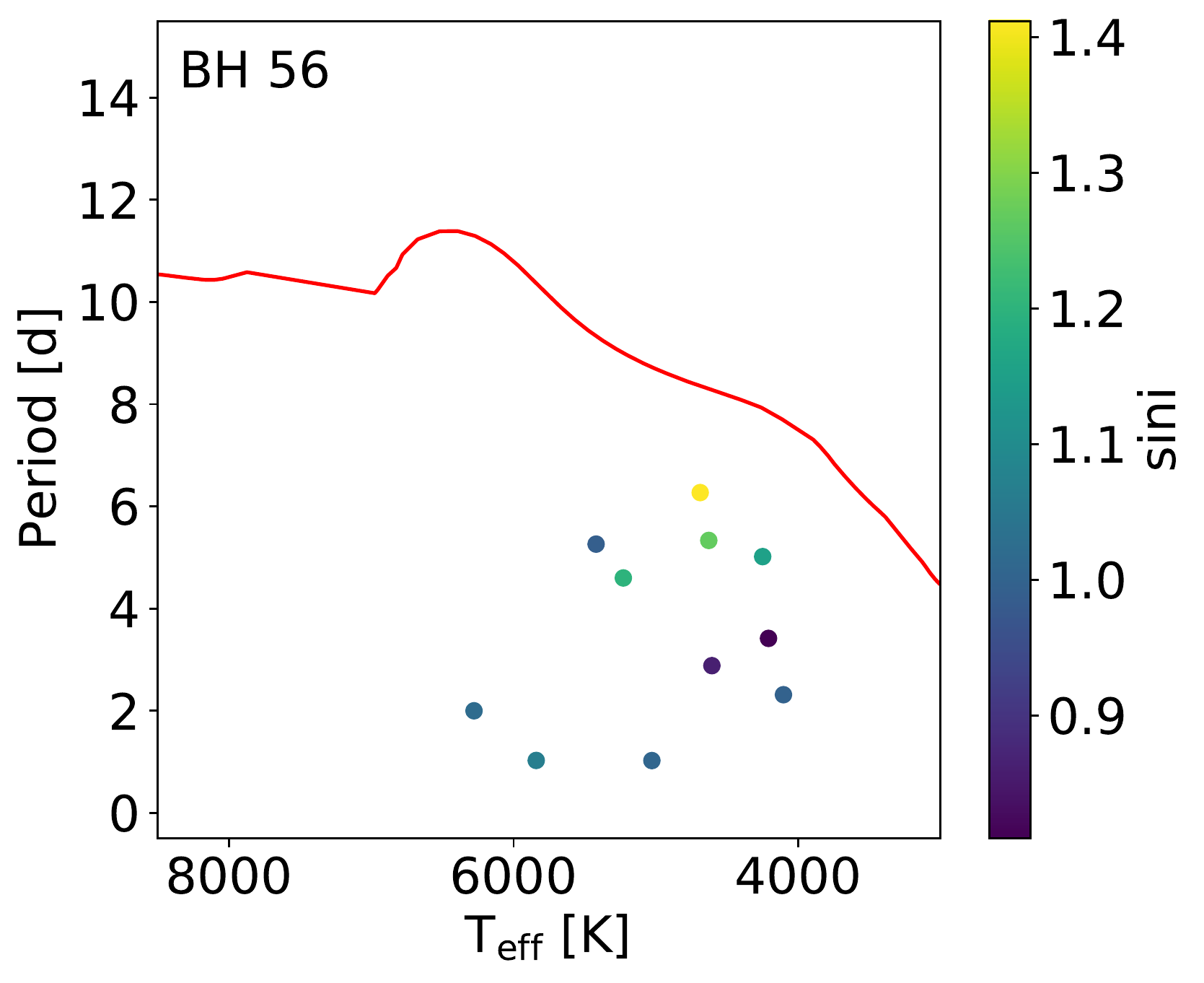}
    \includegraphics[scale=0.35]{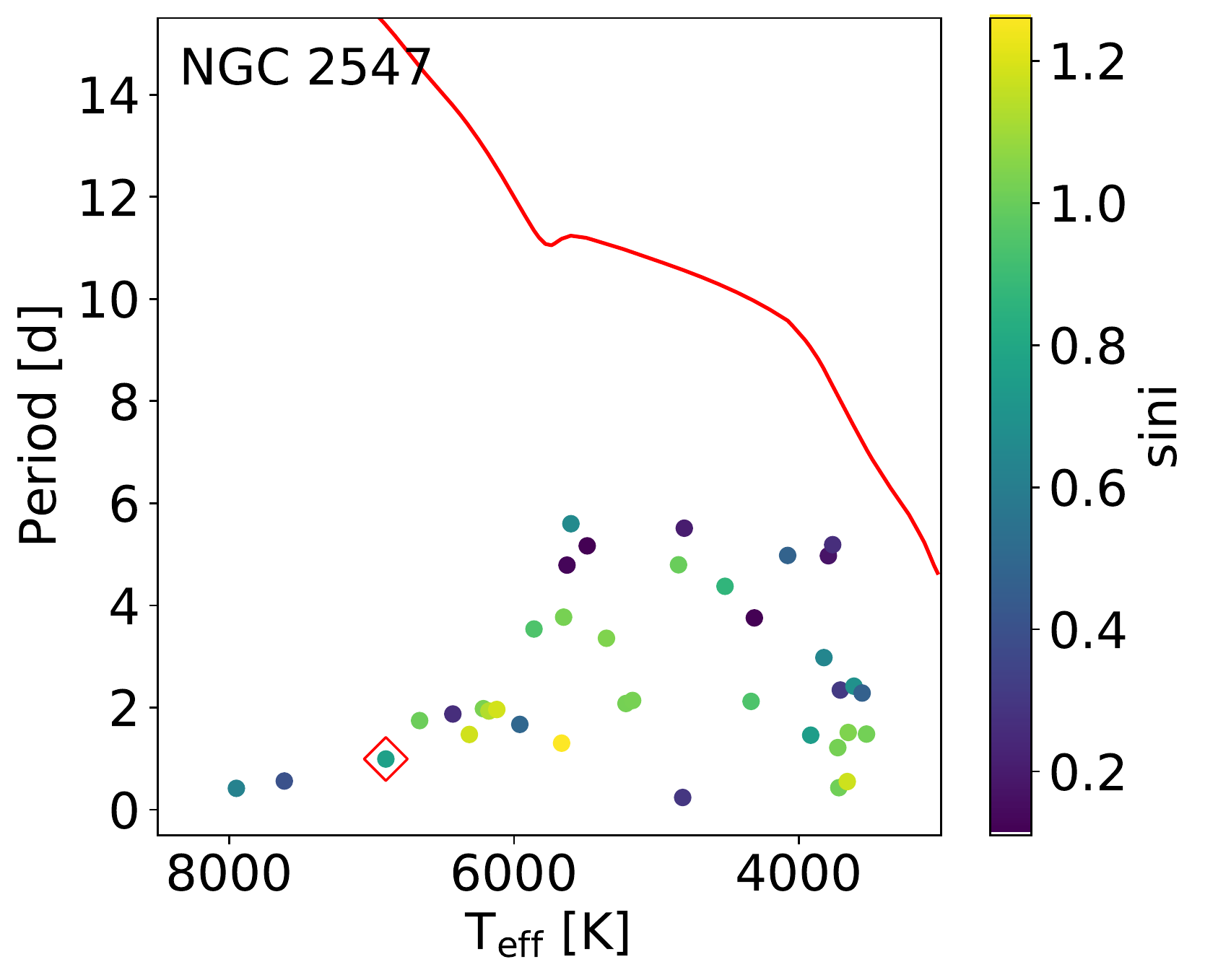}
    \includegraphics[scale=0.35]{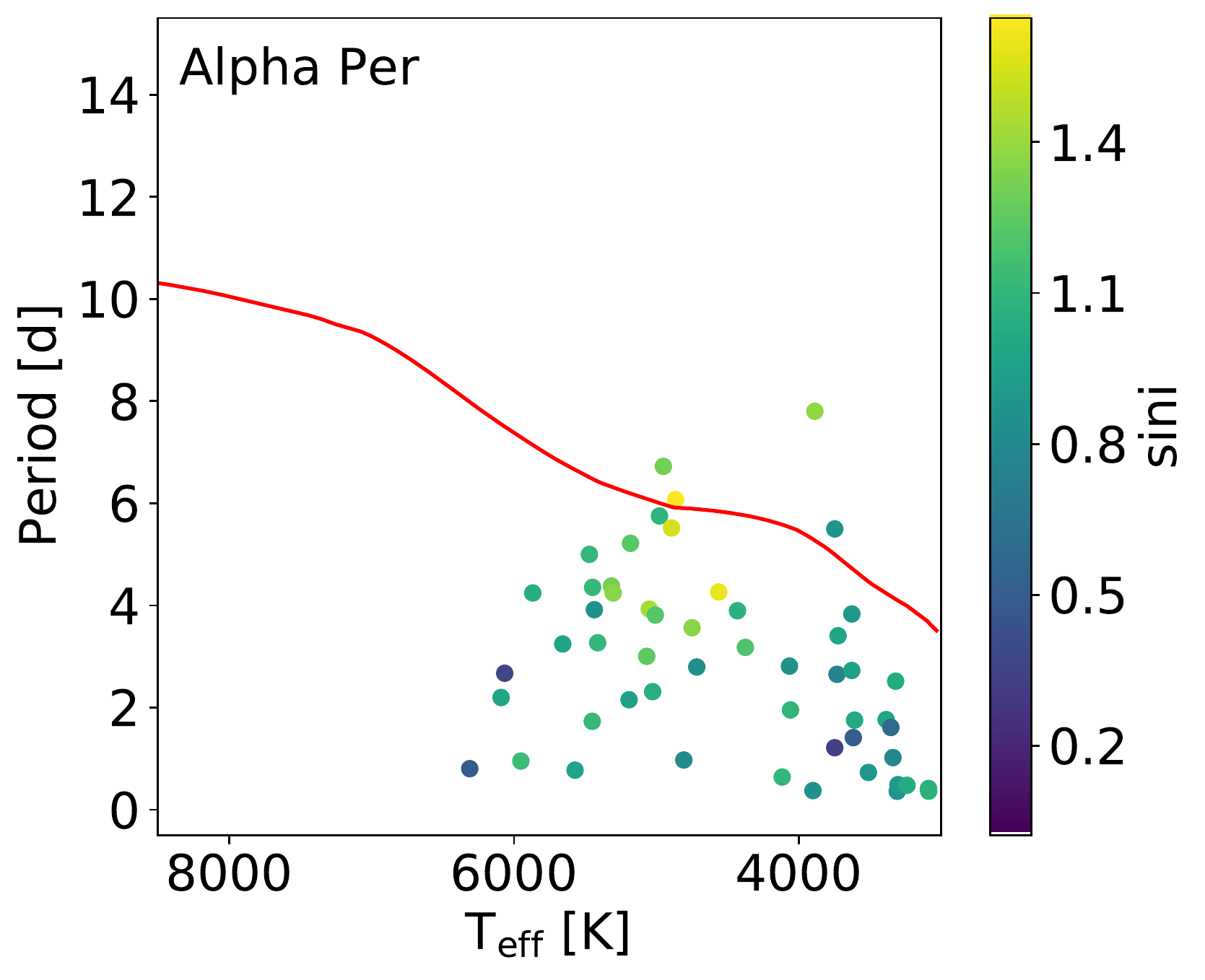}
    \includegraphics[scale=0.35]{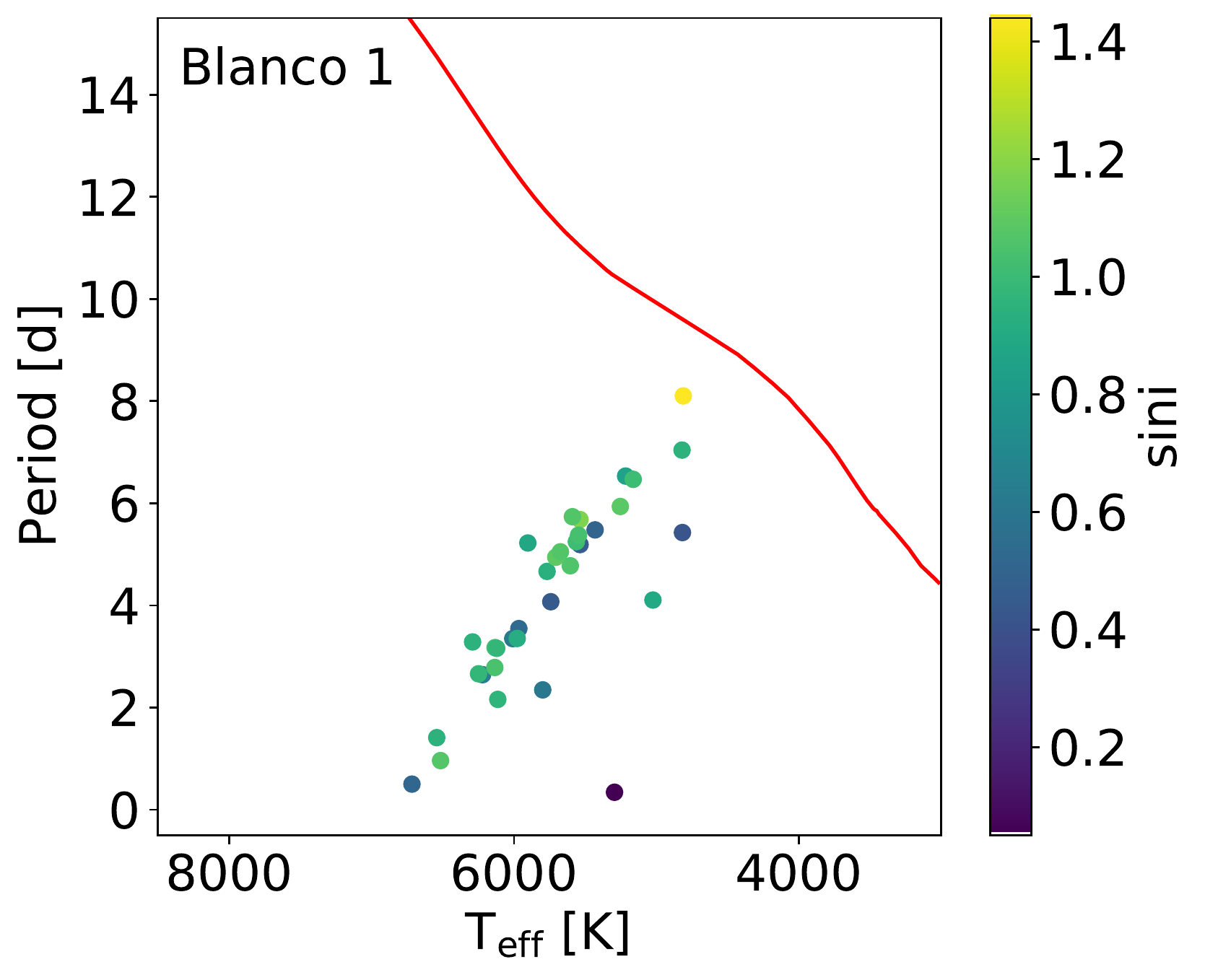}
    \includegraphics[scale=0.35]{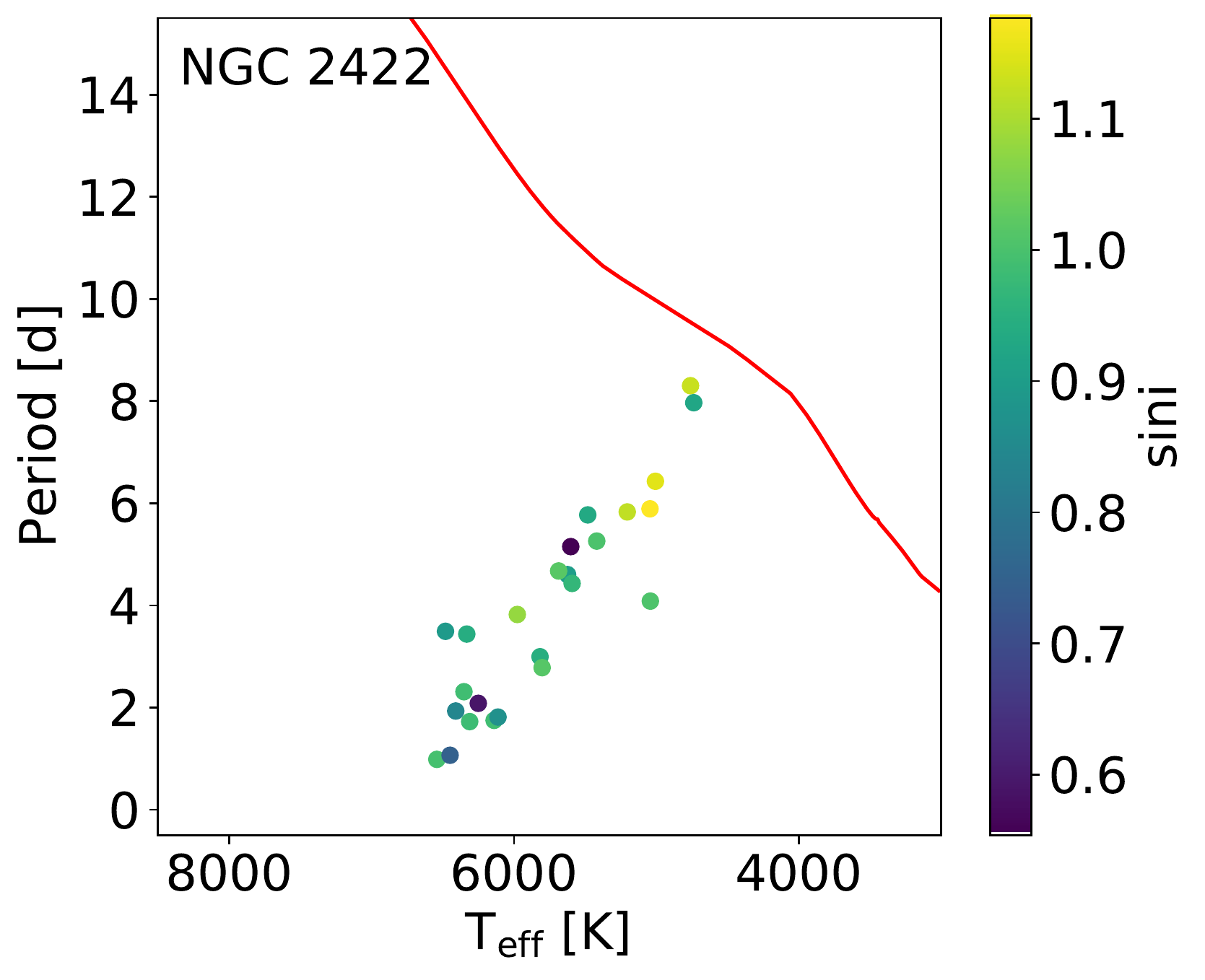}
    \includegraphics[scale=0.35]{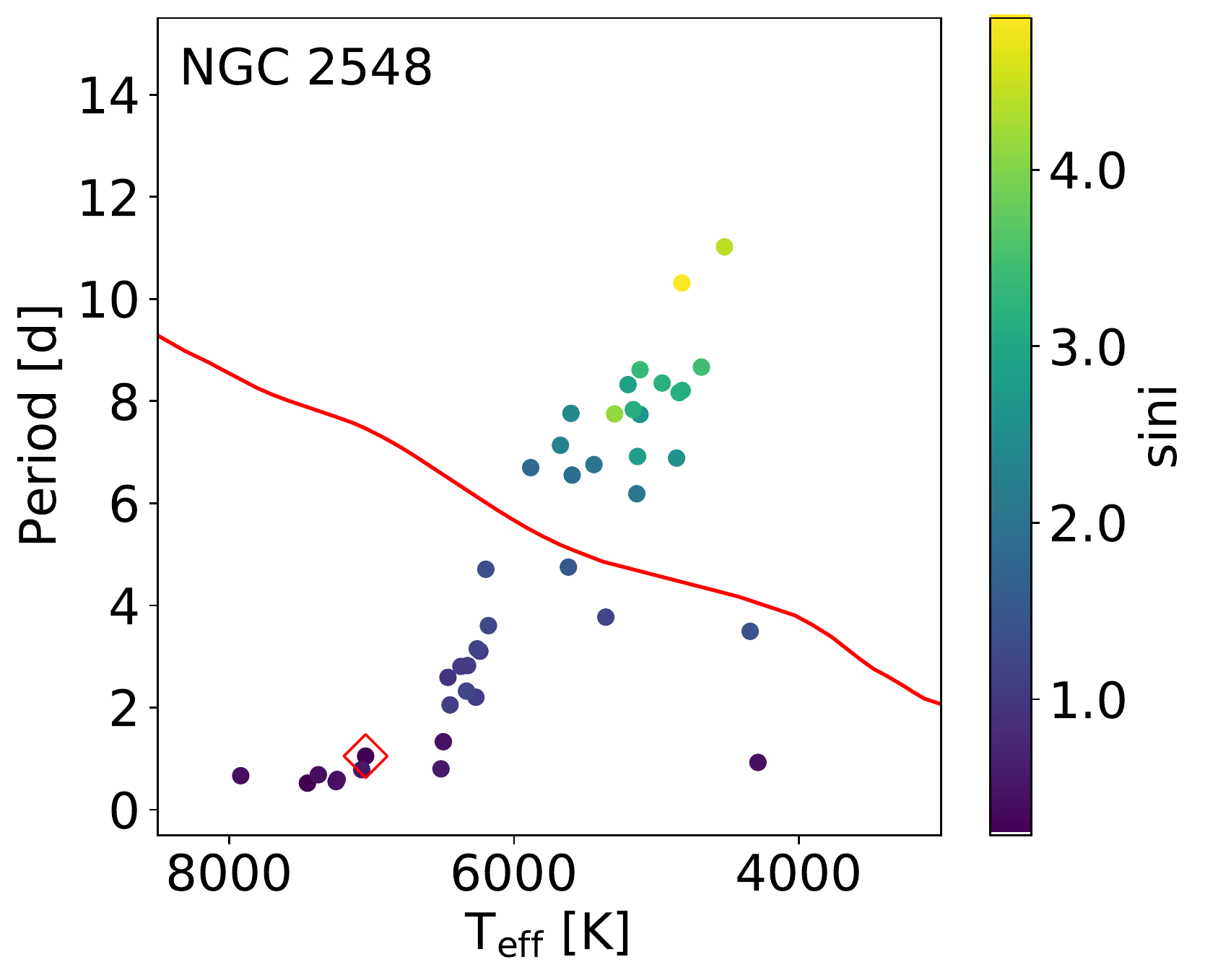}
    \caption{Period-effective temperature plots for the clusters newly analyzed in this paper. Stars are color-coded by the $\sin i$ value their parameters yield. Red curves show the contour on which a star having a threshold $v\sin i$ value yields $\sin i = 1$, incorporating scatter in the $P$ and $R$ values. Above this contour, stars have nonphysical inclinations and are discarded for their potential contribution of systematic error to a cluster's inclination distribution. Red diamonds indicate stars removed from the sample due to having an outlying radial velocity value indicative of a potential binary system.}
    \label{fig:per_teff_plots}
\end{figure*}

For each cluster, we used the technique described in Section 3.3 of \papertwo\ to remove $v\sin i$ values suffering from potential systematic error. Figure \ref{fig:per_teff_plots} shows the period and effective temperature $T_{\rm eff}$ values we determine for stars having $v\sin i$ values passing the above criteria for selection. For each cluster, we used a MIST isochrone generated from the cluster's mean parameters to assign radius values to a grid of $P$ and $T_{\rm eff}$ values spanning the plotted parameter space. For each point on the grid, we calculated an equatorial rotation velocity (i.e. a maximum $v\sin i$) from the associated $P$ and $R$ values. The red contour in each plot represents combinations of parameters that result in a maximum $v\sin i$ equal to our reliability threshold for that cluster.

Since all stars in Figure \ref{fig:per_teff_plots} already survived the threshold process, points located above the contour have a $v\sin i$ measurement greater than the limit set by the star's rotation period and radius (after allowing for observed scatter in these two parameters). The color-coding of the panels indicates that these stars' preliminary $\sin i$ values are often greater than unity, emphasizing the potential presence of a systematic error in $v\sin i$. We removed stars falling above their cluster's contour to decrease the influence of systematic errors on the resulting inclination distributions. In addition, we removed a small number of stars having a highly discrepant radial velocity compared to the typical values for their cluster (marked by a red diamond).

We also estimated the effect of magnetic radius inflation due to spots on the surface of the stars in our sample. The fraction of a star's surface that is covered by spots can range from a few to tens of percent, depending on age and mass \citep[e.g.][]{nicholsfleming2020_spotfraction, jackson2014a}. Starspots inhibit convection, reducing a star's effective temperature. Under the simplifying assumption that the star's luminosity can be described by the Stefan-Boltzmann equation, a reduction in effective temperature at the same luminosity will increase the stellar radius (see Sections 5.2 of \papertwo\ and \ref{subsec:discussion_radius_inflation} for further discussions of radius inflation). Since MIST isochrones do not account for the presence of starspots, our radius values from SED fitting could be underestimated, resulting in an overestimation of $\sin i$ (See Equation \ref{eq:spectrophot}). This bias could be amplified by the requirement of our analysis that stars display light curve modulation assumed to be caused by starspots.

We accessed SPOTS isochrones \citep[][]{somers2020_spots}, which incorporate starspots into their solar-metallicity stellar models. The isochrones assume that the effective temperature of starspots are 80\% that of the non-spotted stellar surface. To obtain an estimate of radius inflation's effect without overcompensating, we selected isochrones corresponding to the lowest nonzero fractional spot coverage modeled, $f_{\rm spot} = 0.17$. For each cluster, we compared MIST-derived radii at the same age and solar metallicity interpolated onto the mass values of SPOTS. We computed the mean fractional difference between the two sets of radius estimates, finding that the radii we report may be underestimated by $\sim 2-13\%$. To incorporate this prediction into our analysis without excessively modifying our results, we treated the mean fractional difference in radius as a systematic error and added it in quadrature to the lower error bar of each $\sin i$ value. We only did this for stars up to 1.3\msun\ to avoid extrapolating the SPOTS isochrones above their available range of masses.

Finally, following Section 5.6 of \papertwo, we estimated the effect of differential rotation on $\sin i$ values, which we summarize here. Starspots at nonzero latitudes on the stellar surface will rotate more slowly that spots at the equator, leading to overestimated rotation periods \citep[e.g.][]{hirano2014}. In \papertwo, we predicted that if starspots exist at solar-like latitudes ($l = 20 \pm 20^{\circ}$) with a solar-like rate of differential rotation (a fractional difference between equatorial and polar rotation rates between 0.1 and 0.2), the resulting rotation periods will be overestimated by an amount between $\sim$ 1-2\% on average.

Projected rotation velocities will be underestimated in the presence of differential rotation, since some parts of the star are rotating more slowly than the equatorial rotation rate. This effect will manifest on the scale of $\sim$ 5-10\% in solar-like rotators \citep[][]{hirano2014}. Averaging the combination of opposing systematic effects due to differential rotation on rotation periods and $v\sin i$ values, we compute a mean systematic underestimate of $\sin i$ by 5.7\%. We added this fractional error in quadrature to the upper error bar of our reported inclination values.

\subsection{Inclination Distributions}
We combined $R$, $P$, and $v\sin i$ values to determine the sine of each star's inclination and its uncertainty via the Bayesian method described by \citet[][]{masuda2020}. We assumed that each parameter was represented by a Gaussian uncertainty distribution except for the $v\sin i$ values of NGC 2547. For this cluster, we instead modeled each projected rotation velocity as a t-distribution with two degrees of freedom to better describe the extended tails of each measurement's distribution according to \citet[][]{jackson2015}.

We modeled the resulting inclination distributions using the cone model created by \citet[][]{jackson2010}. This framework imagines stellar spin-axes uniformly distributed within a cone having a spread half-angle $\lambda$ and mean inclination $\alpha$. Isotropic spin-axis distributions ($\lambda \sim 90^{\circ}$) reduce the distribution to one that is uniform in $\cos i$, while tightly aligned distributions will have $\lambda$ values closer to $0^{\circ}$. Mean inclinations range from pole-on orientations ($\alpha \sim 0^{\circ}$) to edge-on ($\alpha \sim 90^{\circ}$). We also employ our modification of this framework to include a third parameter representing a fraction of stars aligned amid an otherwise isotropic distribution. More details about this three-parameter model are offered in Section 3.5 of \papertwo.

We forward-modeled the effect of establishing a threshold in $v\sin i$ on the inclination distributions. This selection effect will generally eliminate stars that are either intrinsically rotating slowly or have low projected rotation velocity due to a nearly pole-on inclination. This second category of stars is especially important to consider, since the preferential elimination of data as an implicit function of the quantity of interest will directly affect our overall results. As described in Section 3.4 of \papertwo, we used a Monte Carlo simulation to create a population of $\sim 10^5$ stars matching the color distribution for each cluster's stars. 
We used a Gaussian Mixture Model clustering algorithm to identify rotation sequences in the color-period space of our data, and we designated a proportionate fraction of simulated stars to each sequence based on the results of the clustering.

We fit polynomial splines to each identified color-period sequence to assign rotation periods to simulated stars. We used a MIST isochrone to assign radius values as a function of color. Combining the simulated radius and period values, we computed rotation velocities for our simulated population. We paired each rotation velocity with a $\sin i$ value generated from the \citet[][]{jackson2010} cone model with the desired $\lambda$ and $\alpha$ values. Thus, we created a collection of model $v\sin i$ values for which we set the threshold chosen for each cluster. By inducing a change in the associated $\sin i$ distribution after discarding simulated stars below the $v\sin i$ threshold, we modeled the expected selection effect on our data after setting the same threshold. 

To compare our $\sin i$ determinations for each cluster with the simulated values generated using the cone model, we computed cumulative distribution functions (CDFs) for both. We modified the resulting model CDFs to account for random uncertainties in $R$, $P$ and $v\sin i$. As in \papertwo, we assumed normal distributions of fractional errors in $P$ and $R$, and log-normal distributions of fractional errors in $v\sin i$ \citep[e.g.][]{kovacs2018}.\footnote{While we assumed that each individual $v\sin i$ measurement was quantified by either a normal or t-distribution depending on the measurement source, we found that the overall ensemble of fractional $v\sin i$ uncertainties for each cluster was better described by a log-normal distribution.} Setting a $v\sin i$ threshold increases the $\sin i$ value associated with a given cumulative probability in the model CDFs, shifting them rightward compared to a model without a threshold. At any given threshold level, the magnitude of the shift is proportional to a $\sin i$ value's proximity to $0^{\circ}$. Using these models, we performed a 10,000-step MCMC analysis of parameter space using the same Gaussian likelihood function as Equation 2 of \papertwo:

\begin{equation}
    \ln{\rm{LF}} = -0.5 * \sum_{i=1}^{\rm{n}} \bigg[\ln{(2\pi \sigma_{i}^2)} + \frac{(x_{i} - \hat{x_{i}})^2}{\nu \sigma_{i}^2} \bigg].
    \label{eq:likelihood}
\end{equation}

Here, $\rm{n}$ is the number of $\sin i$ values for a cluster, $x_{i}$ is a determination of $\sin i$, $\sigma_{i}$ is the uncertainty, and $\hat{x_{i}}$ is the model's predicted inclination at the same cumulative probability as $x_{i}$. The number of degrees of freedom, $\nu = \rm{n} - \rm{m}$, is used to calculate the reduced chi-squared value $\chi_{\rm red}^{2}$, having either $\rm{m} = 2$ or $\rm{m} = 3$ fitted parameters ($\alpha$, $\lambda$, and $f$ for the three-parameter model). For the two-parameter model, as in \papertwo, we discarded a 1000-step burn-in before marginalizing the posteriors to obtain probability distributions (PPDs) for $\alpha$ and $\lambda$. To facilitate the use of asymmetric error bars, we used either the upper or lower value of $\sigma_{i}$ depending on the sign of the difference between data and model $\sin i$ values. For $x_{i} - \hat{x_{i}} > 0$, we used the lower value of $\sigma_{i}$, and vice-versa. We also explored a three-parameter model when two parameters did not provide a satisfactory fit to the data (see Section \ref{subsec:threeparam_model_runs}).

\subsection{Cluster Kinematics}
To search for connections between stellar rotation and bulk motion within a cluster, we again computed plane-of-sky and line-of-sight (LOS) internal kinematics based on \citet[][]{kamann2019} and Paper \citetalias{paper1_healy} (Section 3.7) and Paper \citetalias{paper2_healy} (Section 3.6).

To calculate plane-of-sky kinematics in multiple bins of radial distance from cluster center, we accounted for bias in the Gaia EDR3 proper motions of bright stars \citep[][]{cantatgaudin2021_pm_correction}, transformed to Cartesian coordinates \citep[][]{gaiaclusters2018}, subtracted contributions due to each cluster's motion along the LOS \citep[][]{vandeven2006}, and transformed again to polar coordinates \citep[][]{vanleeuwen2000}. In this coordinate system, a star's radial unit vector points in the direction opposite that of cluster center, while the tangential unit vector is oriented perpendicular to the radial direction and is positive in the clockwise direction on the sky (see Footnote 5 of \papertwo). 

To analyze LOS kinematics, we used Gaia EDR3 LOS velocities for Blanco 1 and NGC 2422. For all other clusters lacking a sufficient quantity of Gaia LOS velocities, we consulted the targeted spectroscopic studies listed in Table \ref{tab:vsini_sources}. After discarding likely spectroscopic binary outliers from the primary LOS velocity distribution, we determined mean LOS kinematics for each cluster. The smaller number of LOS measurements compared to proper motions prevented us from drawing significant insights from binning the results by radial distance.

\section{Results} \label{sec:results}
\begin{figure*}
    \centering
    \includegraphics[scale=0.3]{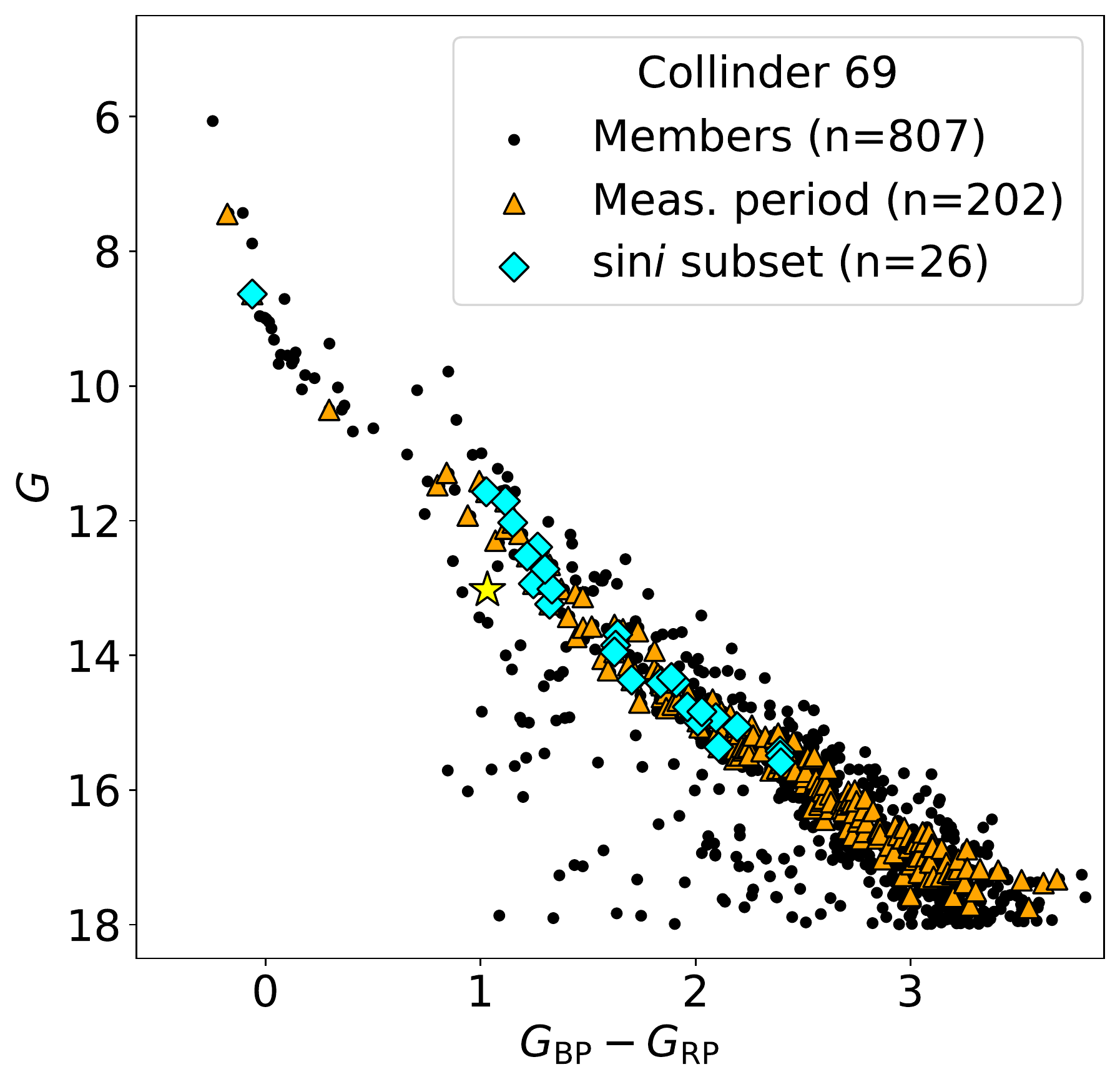}
    \includegraphics[scale=0.3]{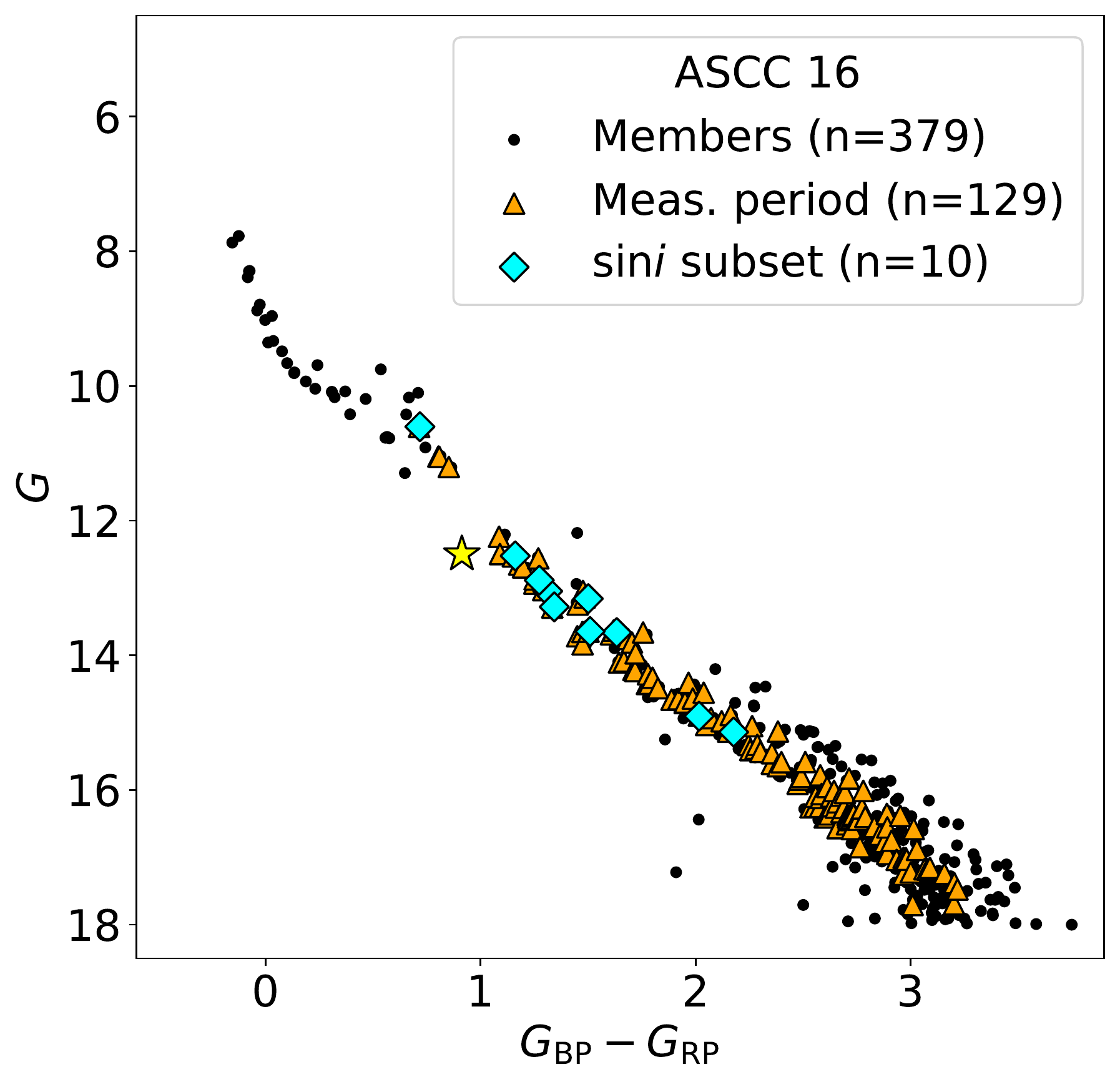}
    \includegraphics[scale=0.3]{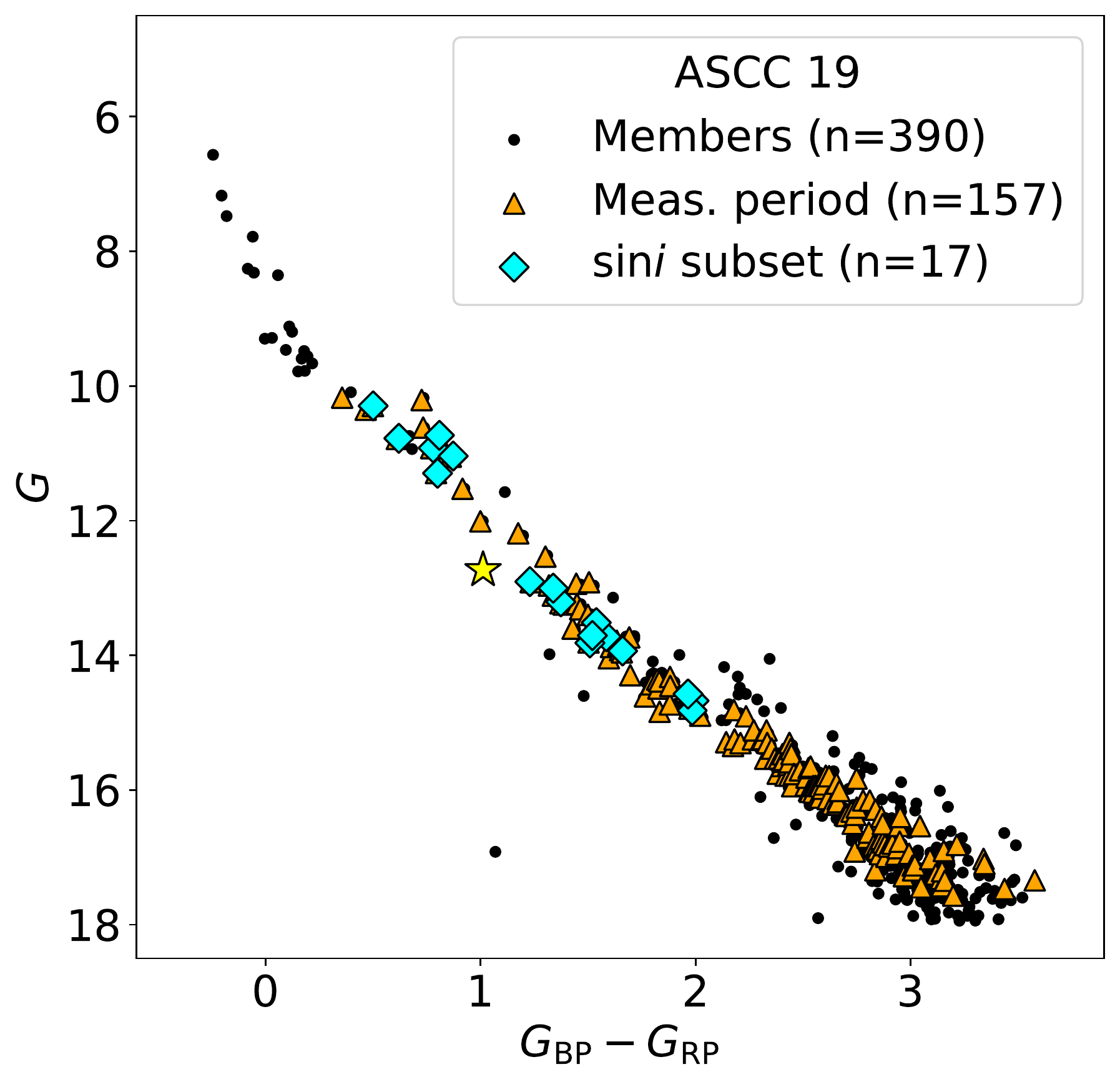}
    \includegraphics[scale=0.3]{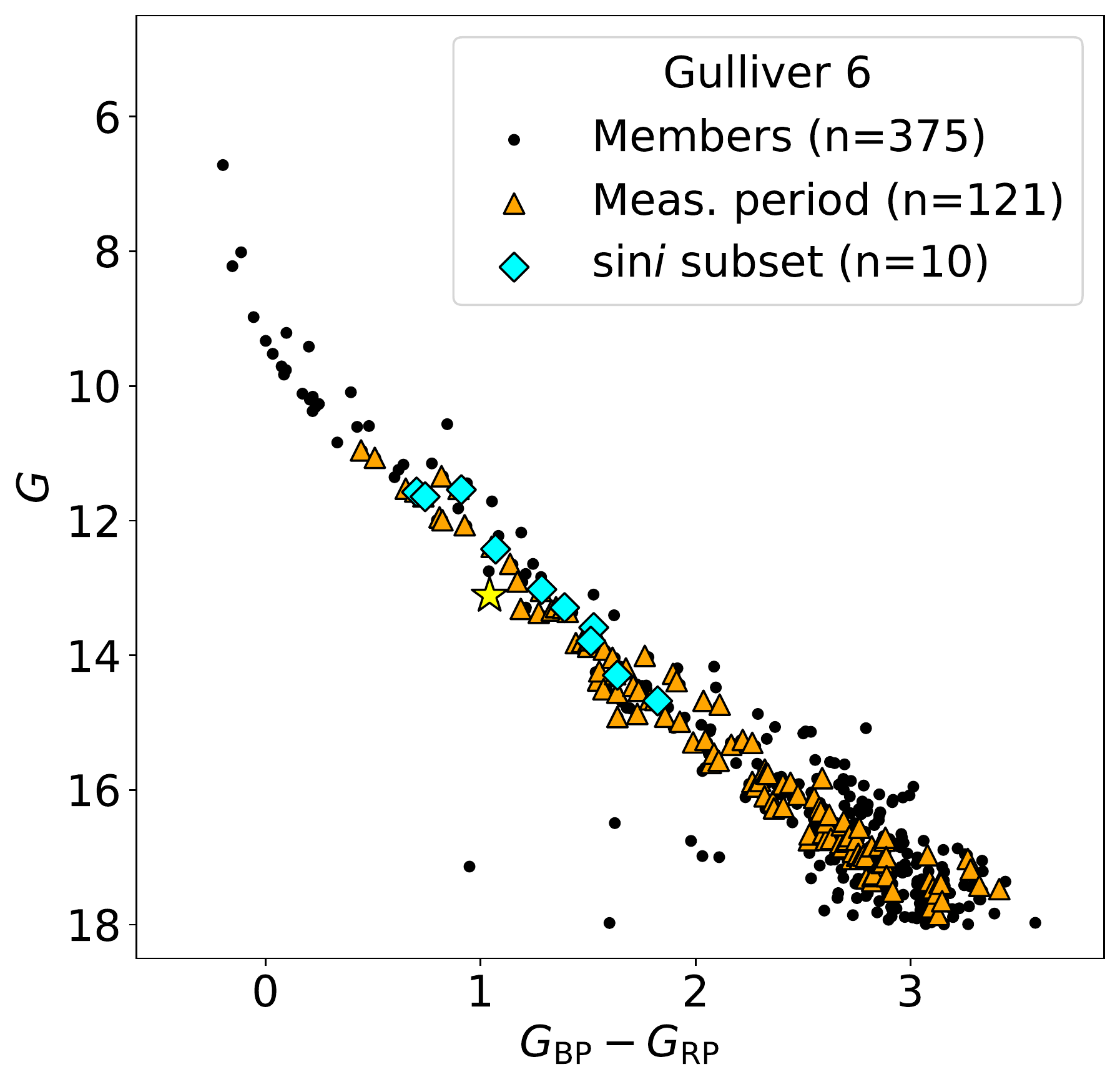}
    \includegraphics[scale=0.3]{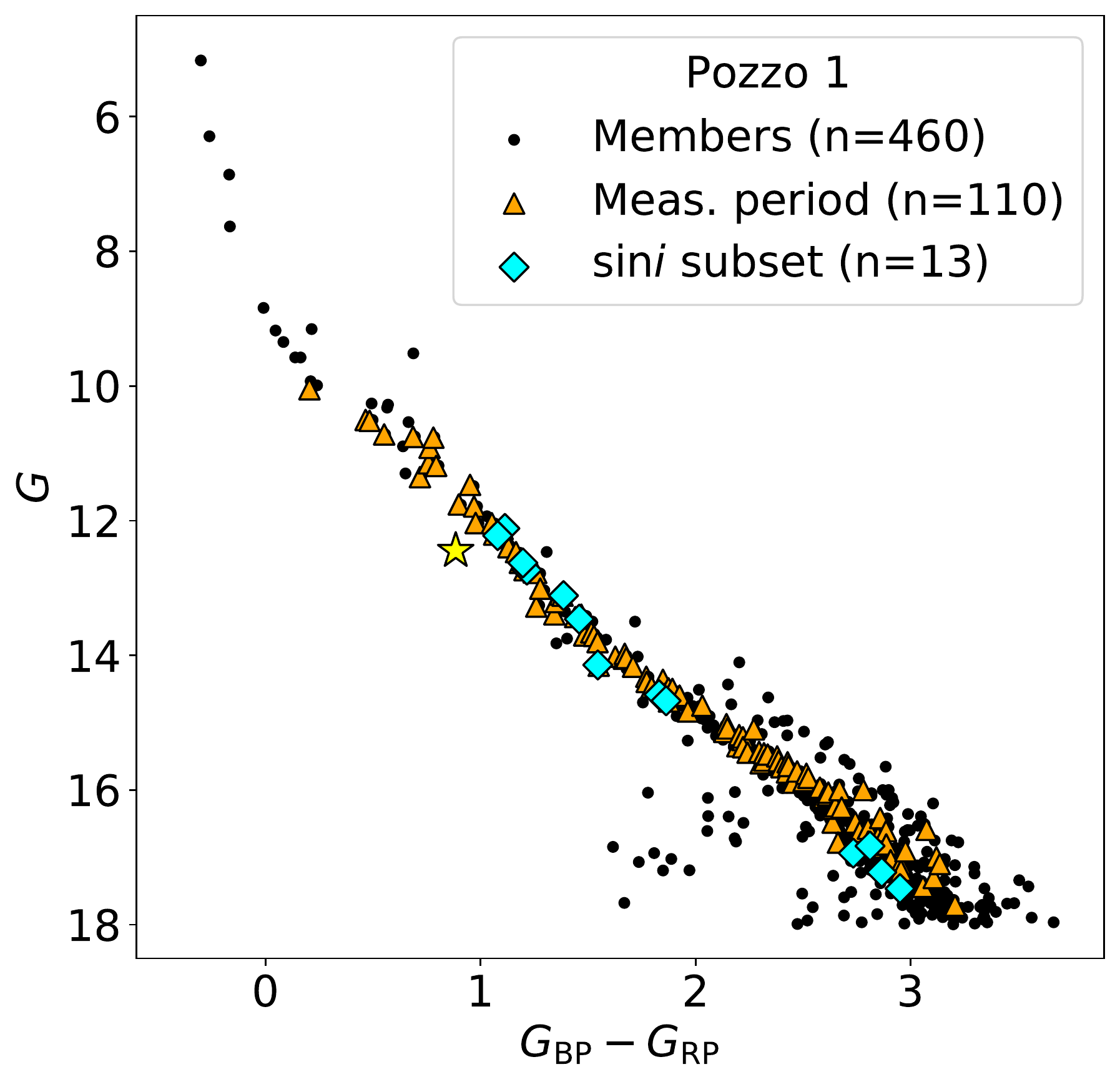}
    \includegraphics[scale=0.3]{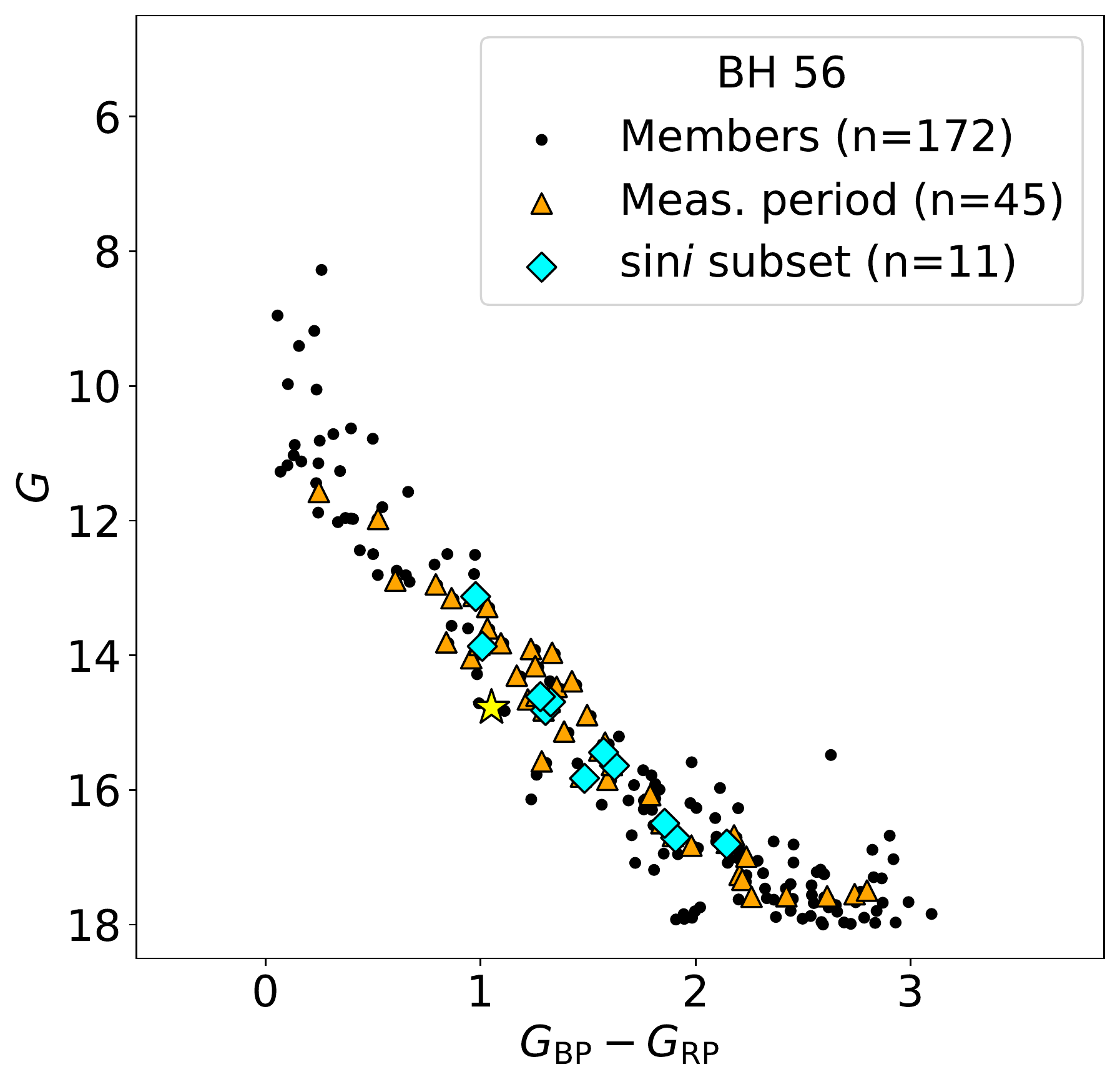}
    \includegraphics[scale=0.3]{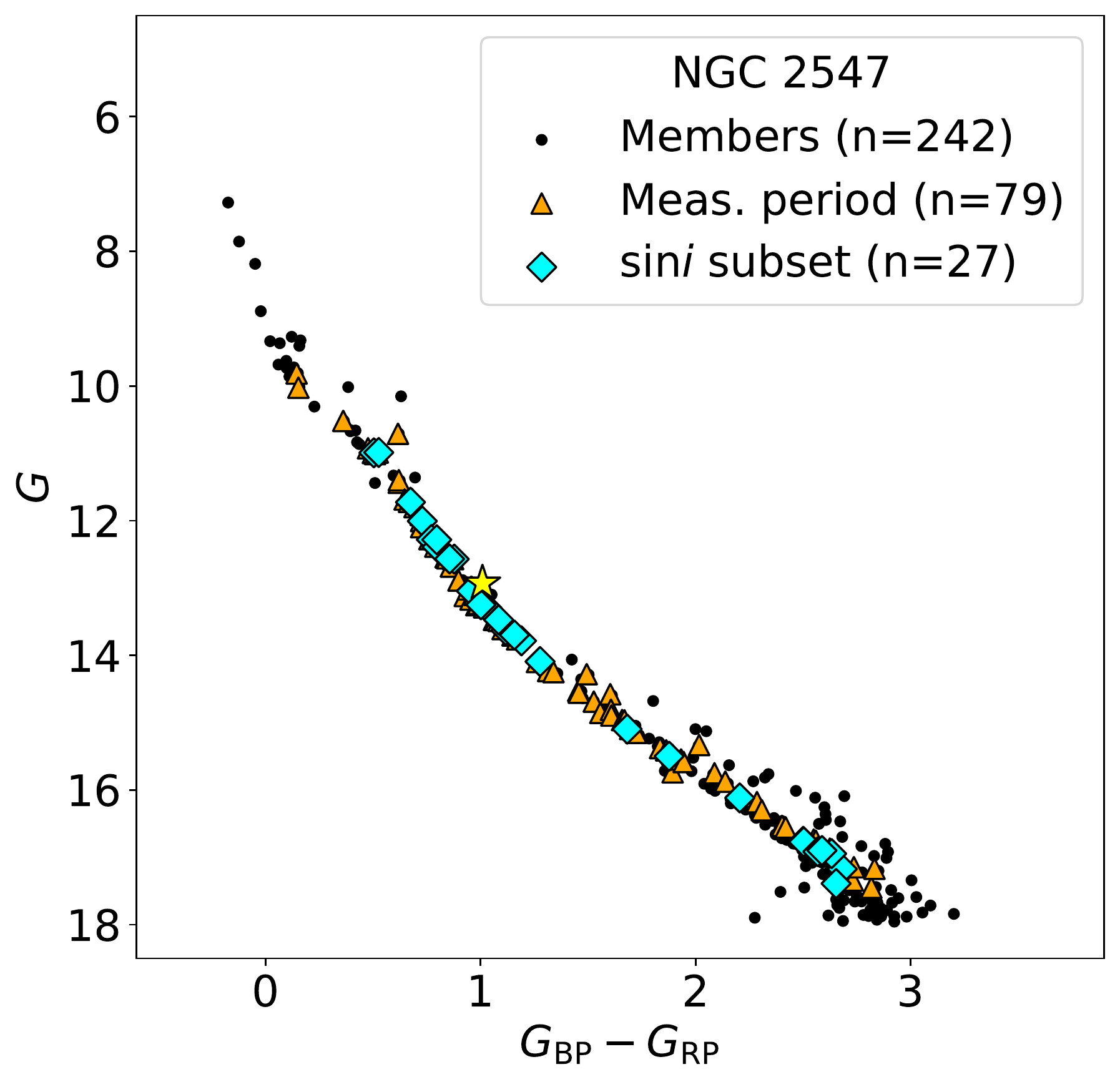}
    \includegraphics[scale=0.3]{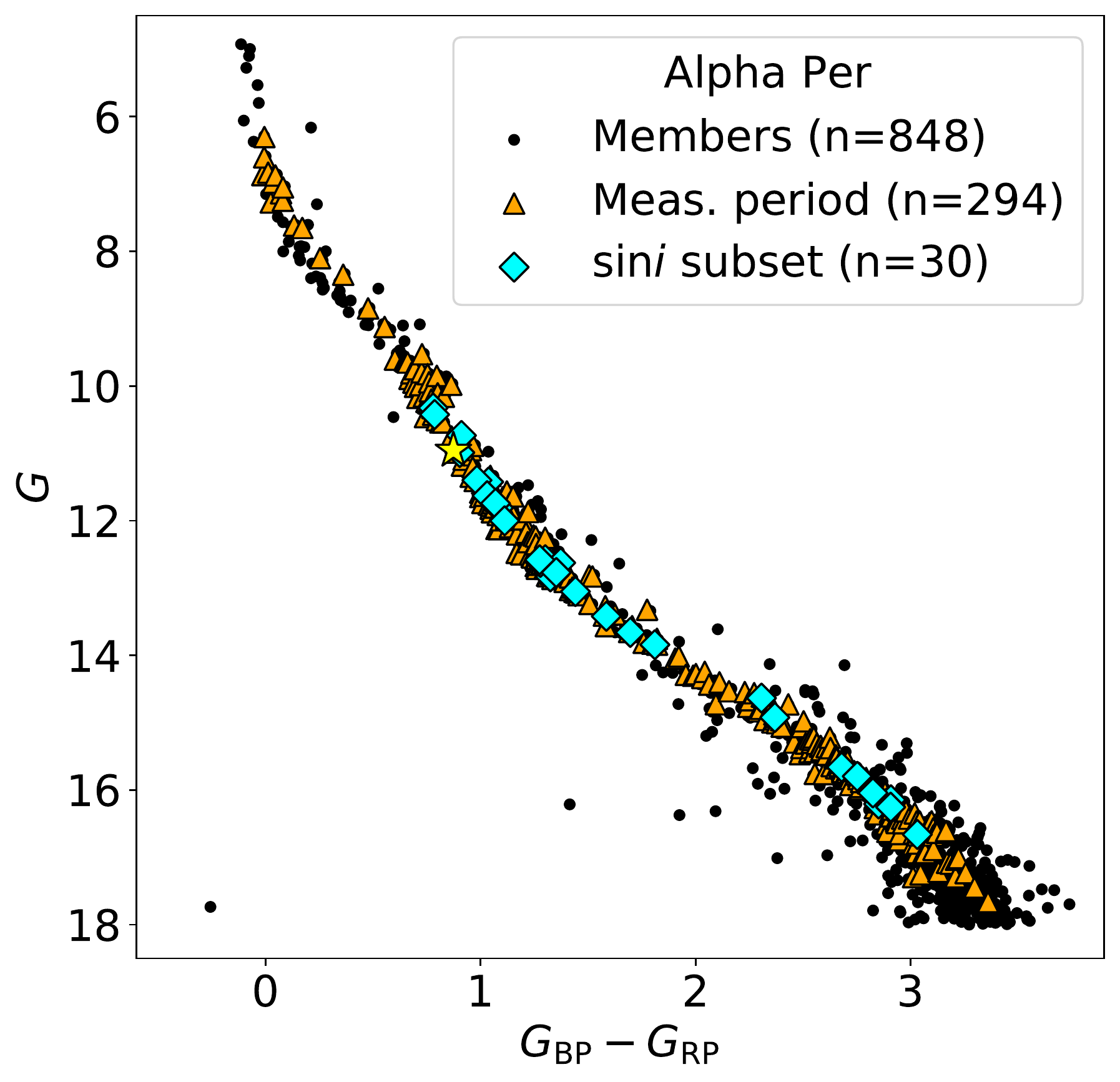}
    \includegraphics[scale=0.3]{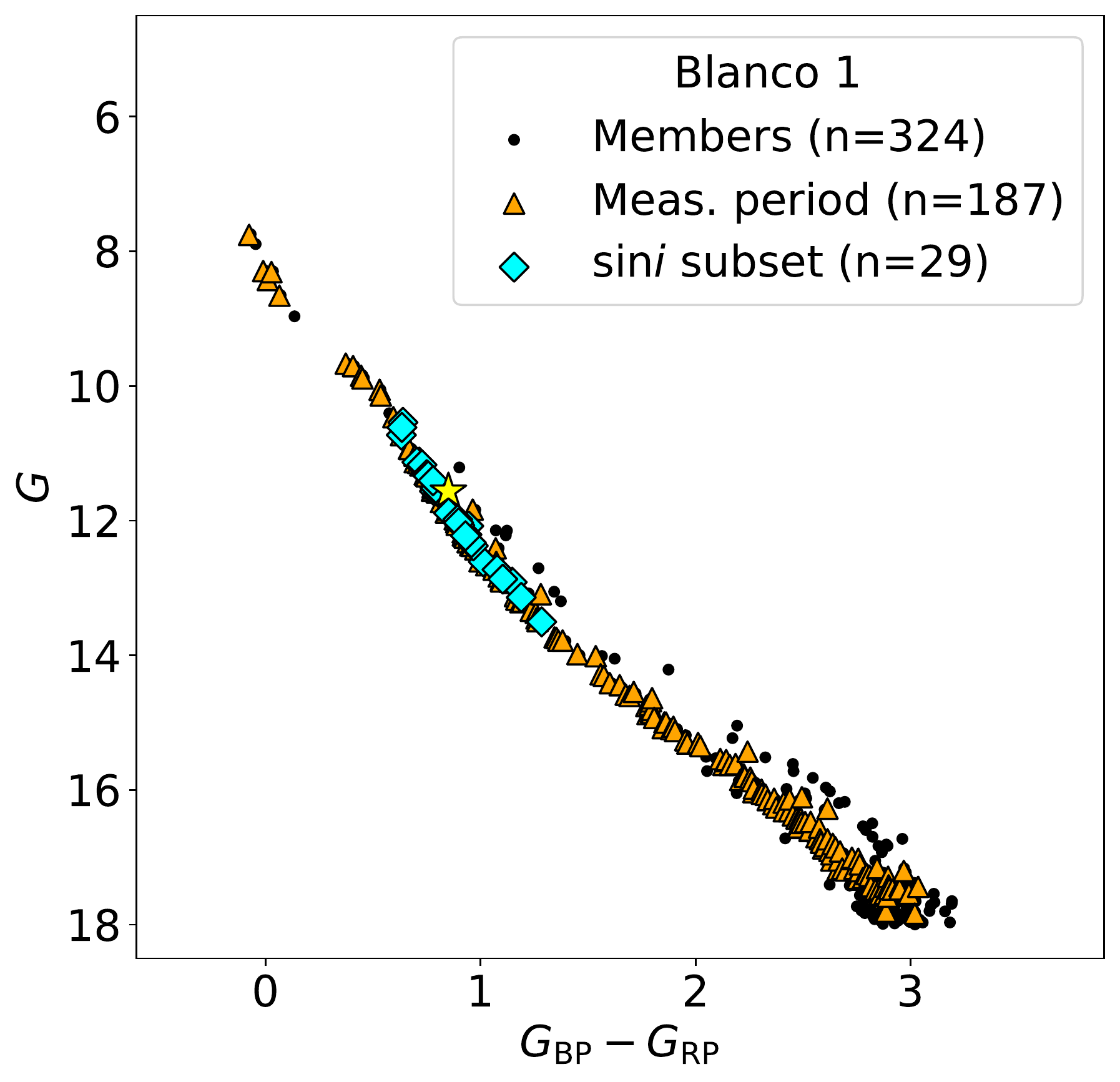}
    \includegraphics[scale=0.3]{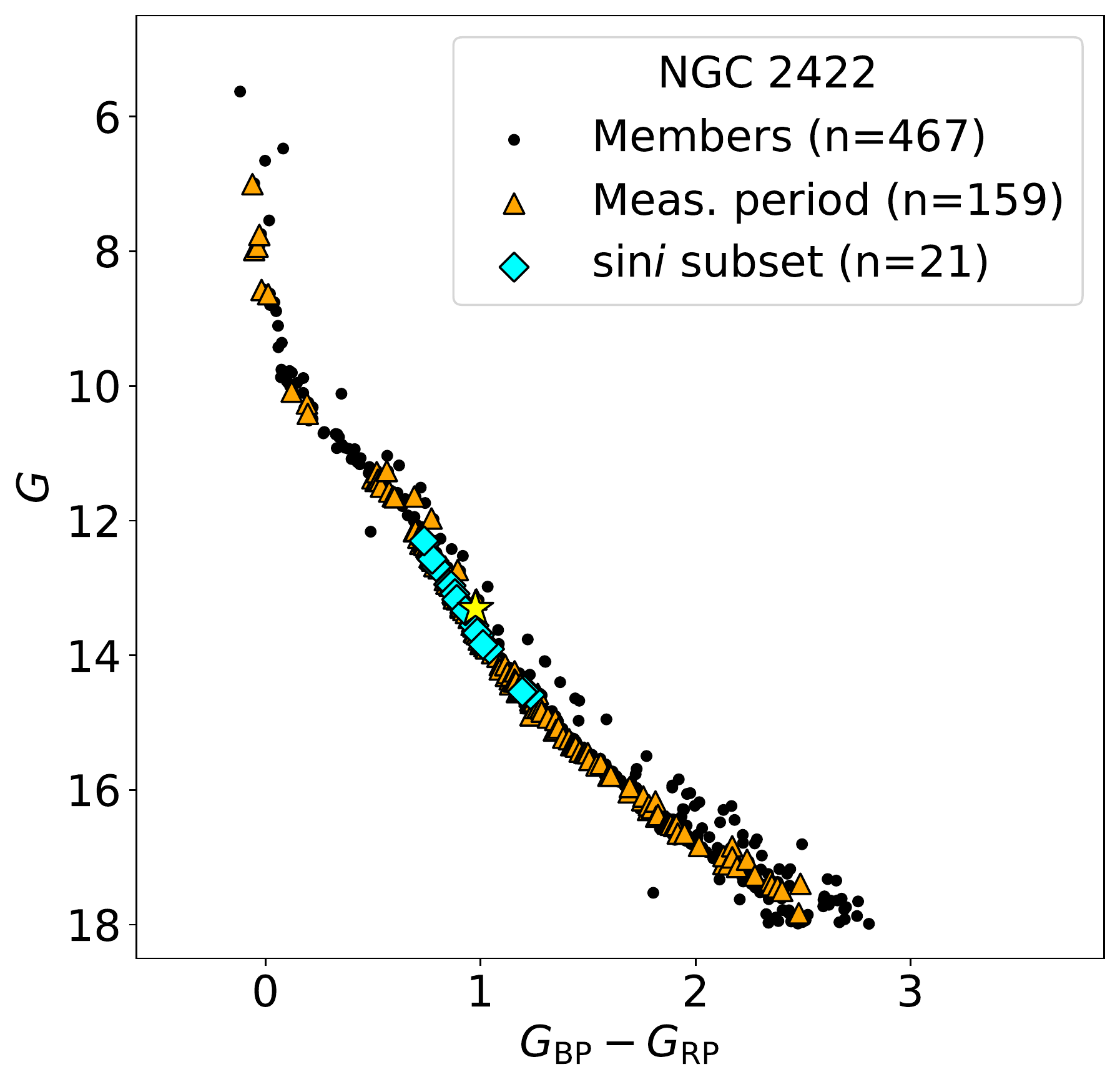}
    \includegraphics[scale=0.3]{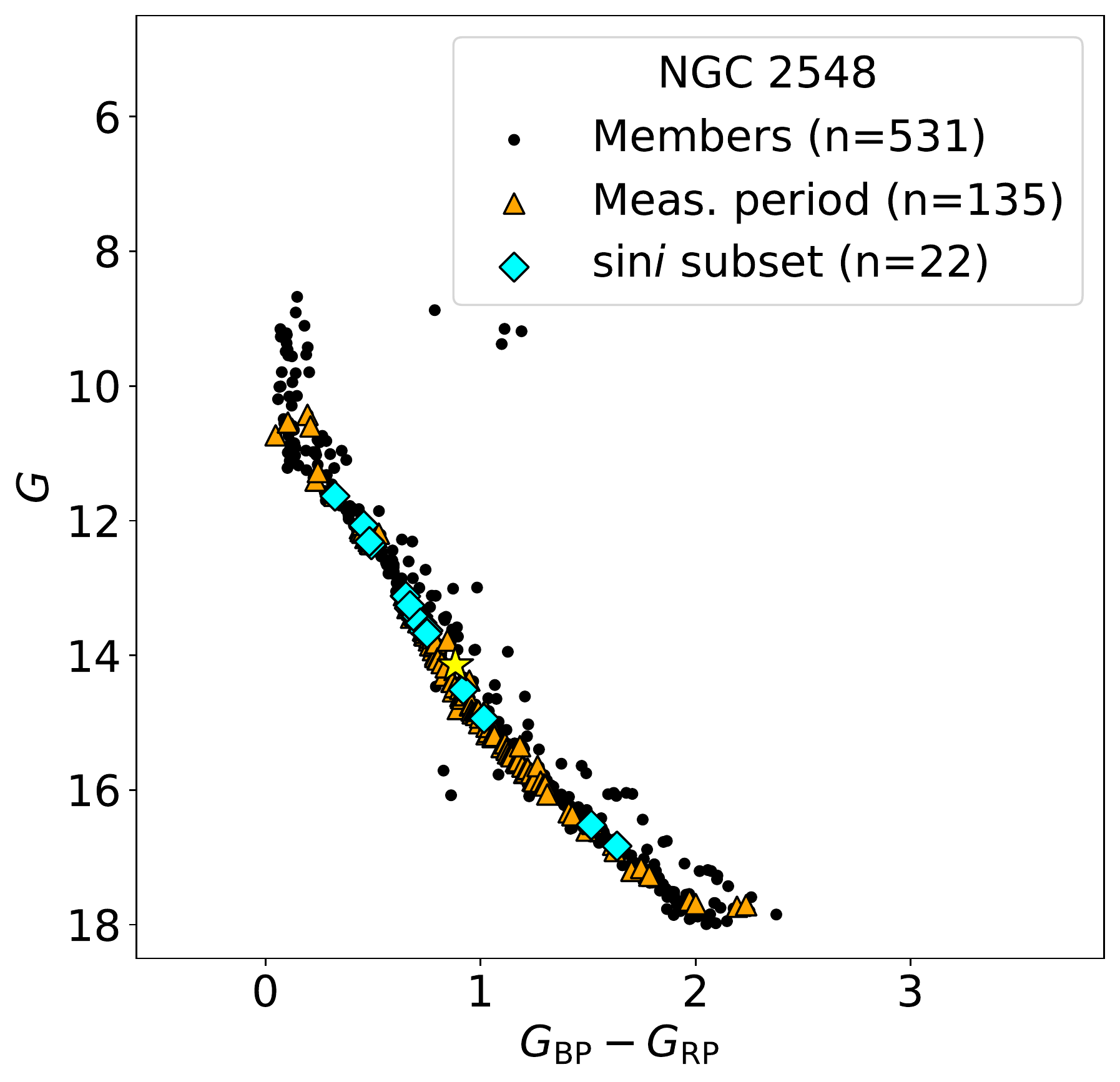}
    \caption{Color-magnitude diagram plots for all clusters. Orange triangles indicate stars for which we measured rotation periods with TESS. Cyan diamonds show stars with $\sin i$ values determined by this work. Yellow star symbols indicate reddened and extincted solar values at each cluster's distance.}
    \label{fig:CMD_plots}
\end{figure*}

We plot color-magnitude diagrams for each cluster in Figure \ref{fig:CMD_plots}. The black points in these figures represent all stars on the CG18 membership list. In each panel, we also plot the subsets of cluster members with rotation period measurements and $\sin i$ determinations. The following subsections and their associated tables and figures further describe our results for the 11 clusters we analyzed.

\subsection{Rotation Periods}
\begin{figure*}
    \centering
    \includegraphics[scale=0.325]{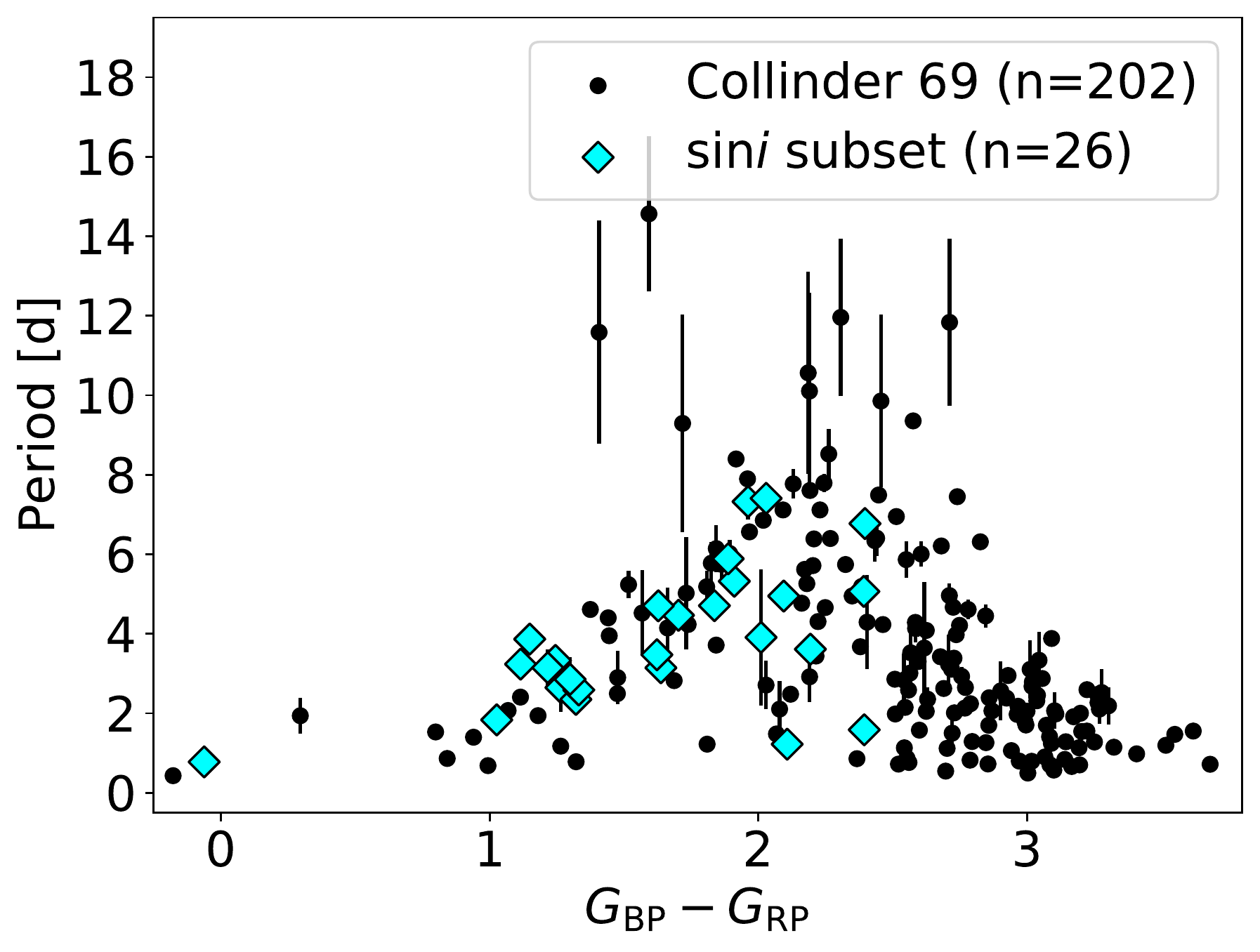}
    \includegraphics[scale=0.325]{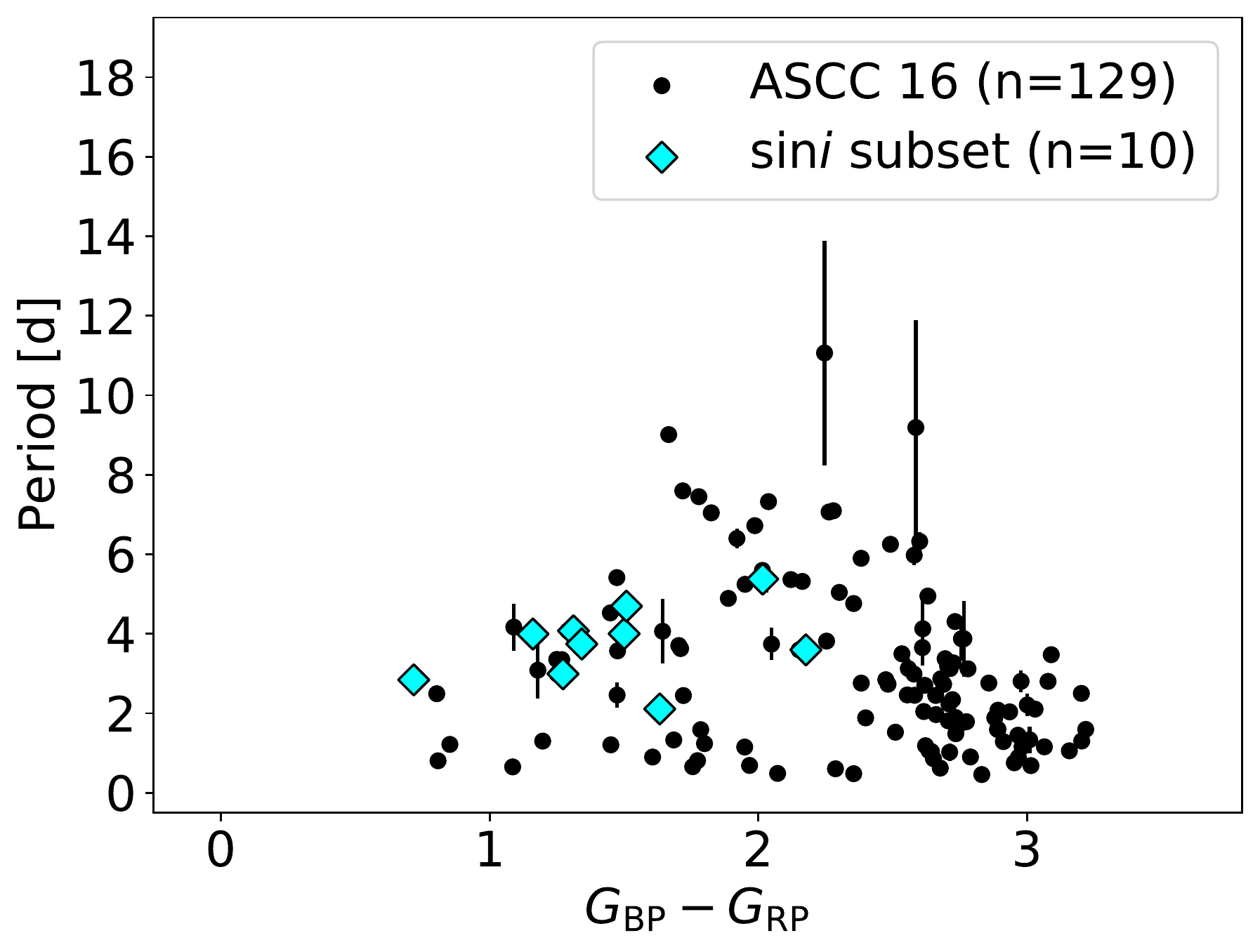}
    \includegraphics[scale=0.325]{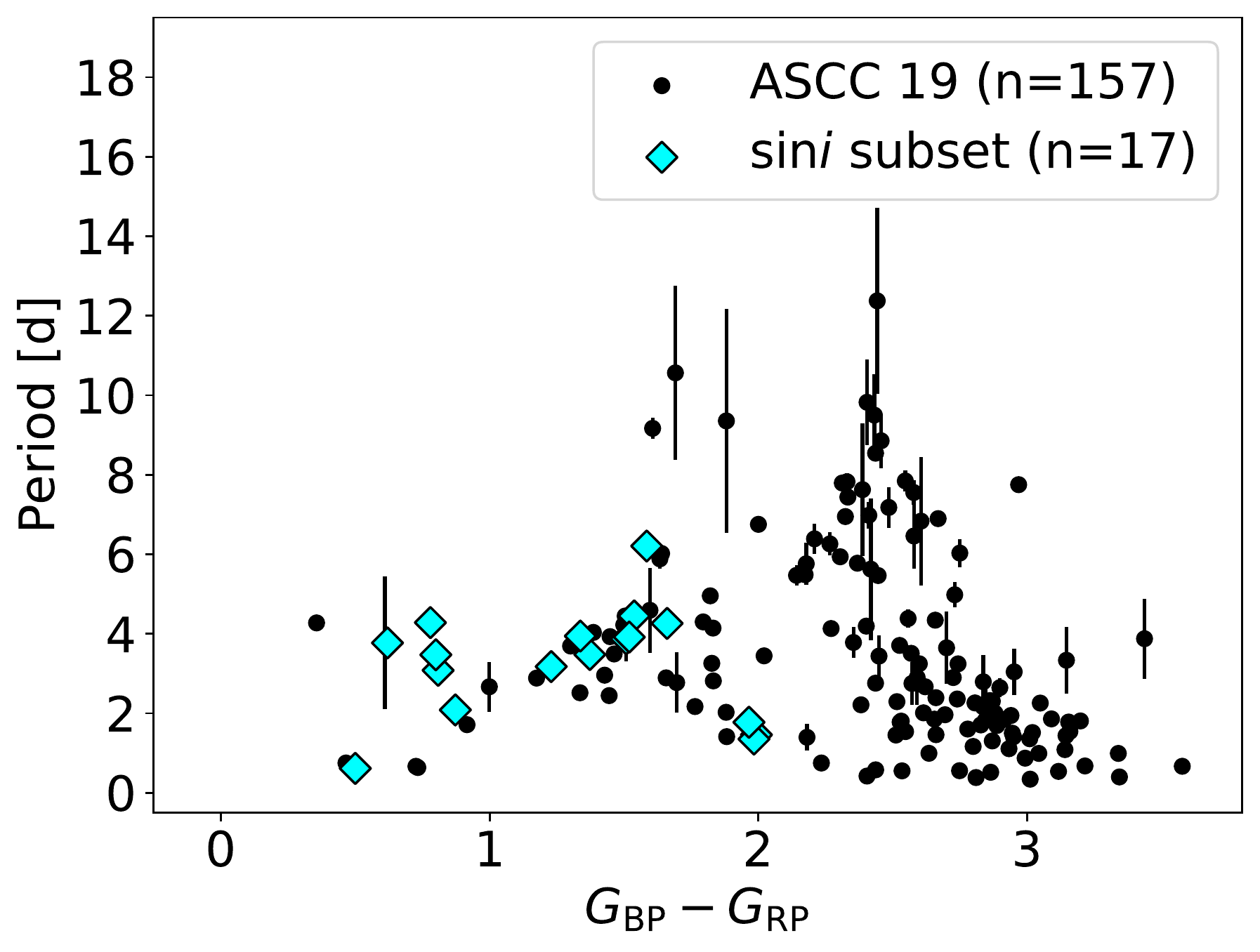}
    \includegraphics[scale=0.325]{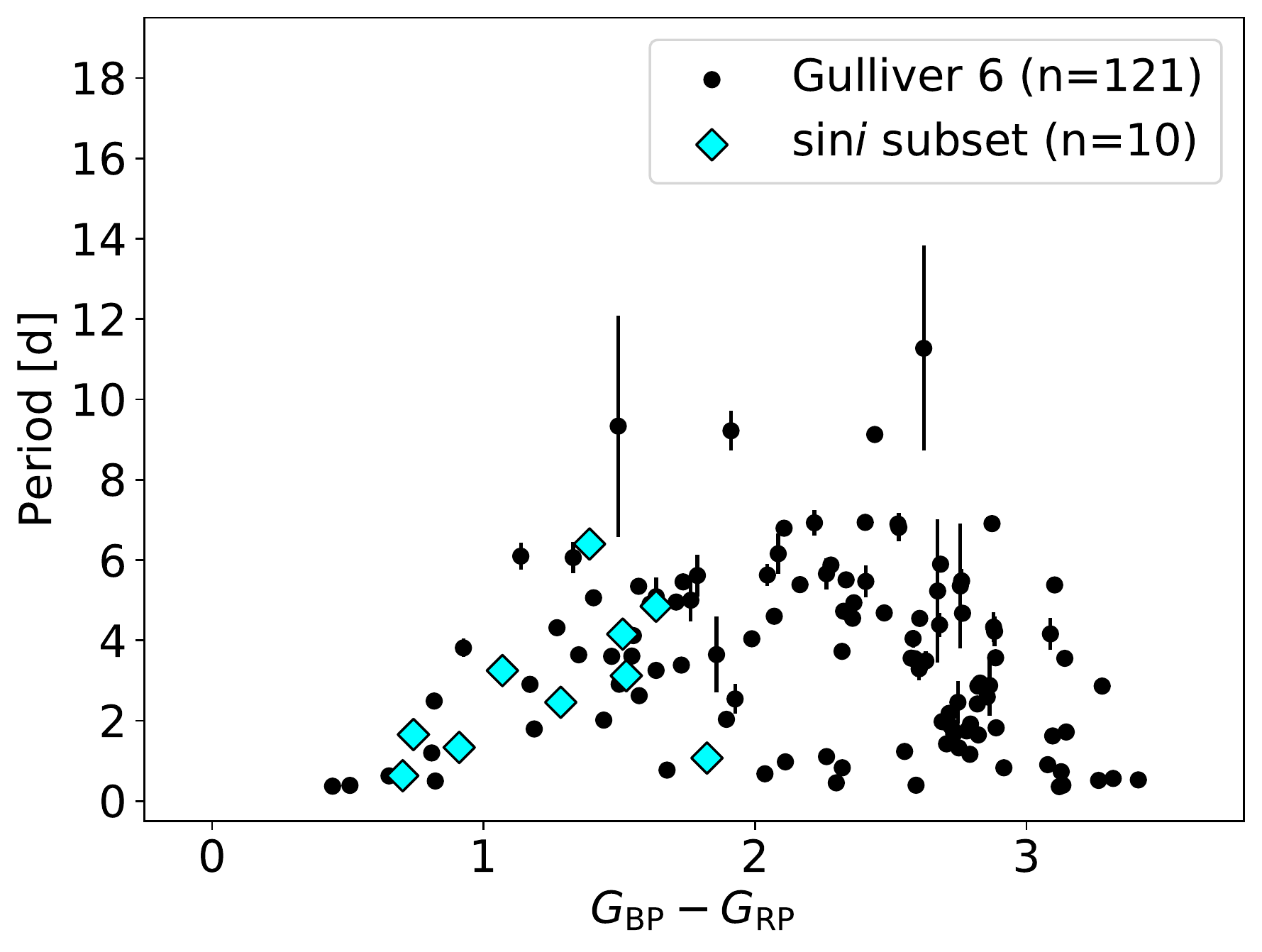}
    \includegraphics[scale=0.325]{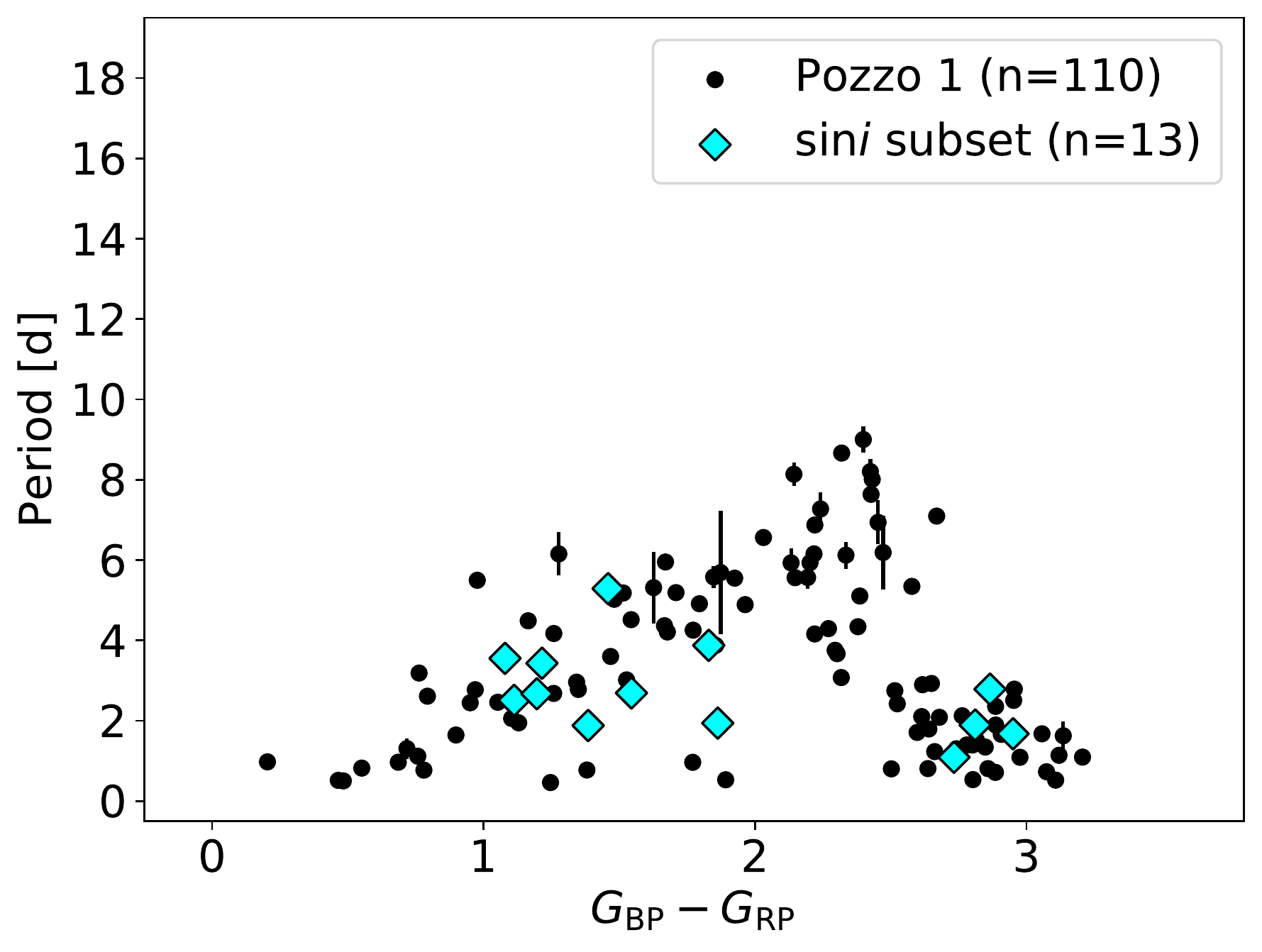}
    \includegraphics[scale=0.325]{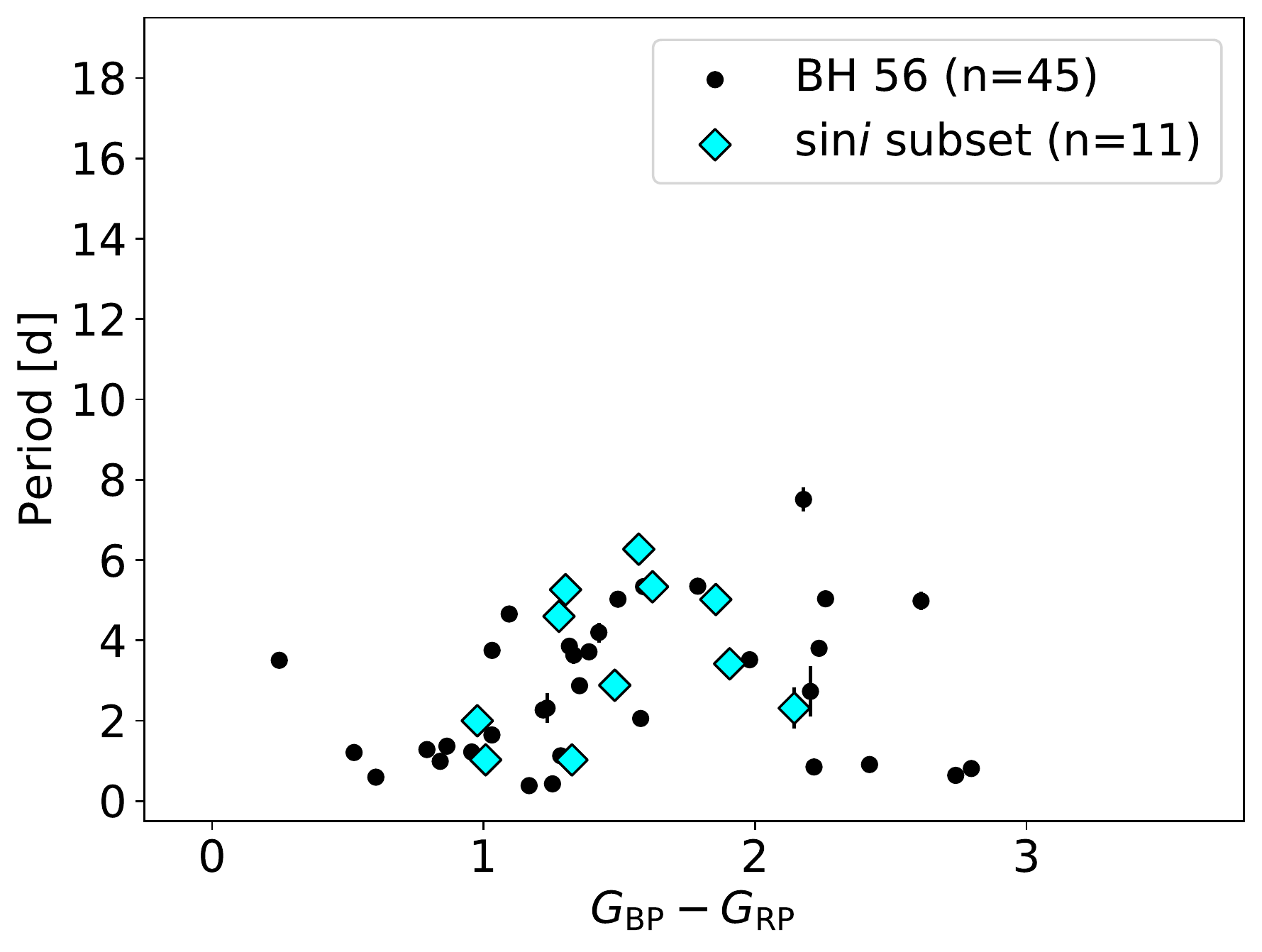}
    \includegraphics[scale=0.325]{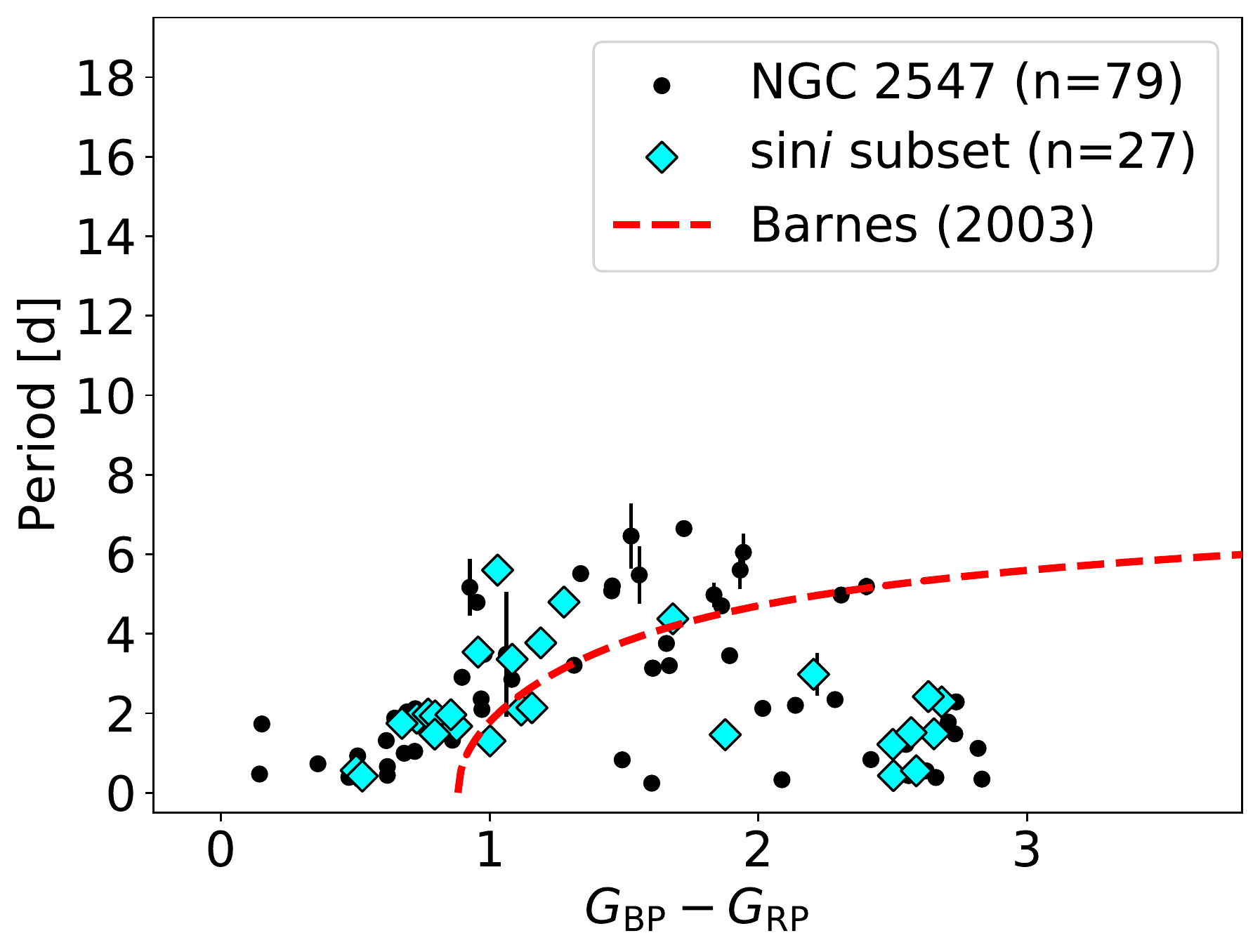}
    \includegraphics[scale=0.325]{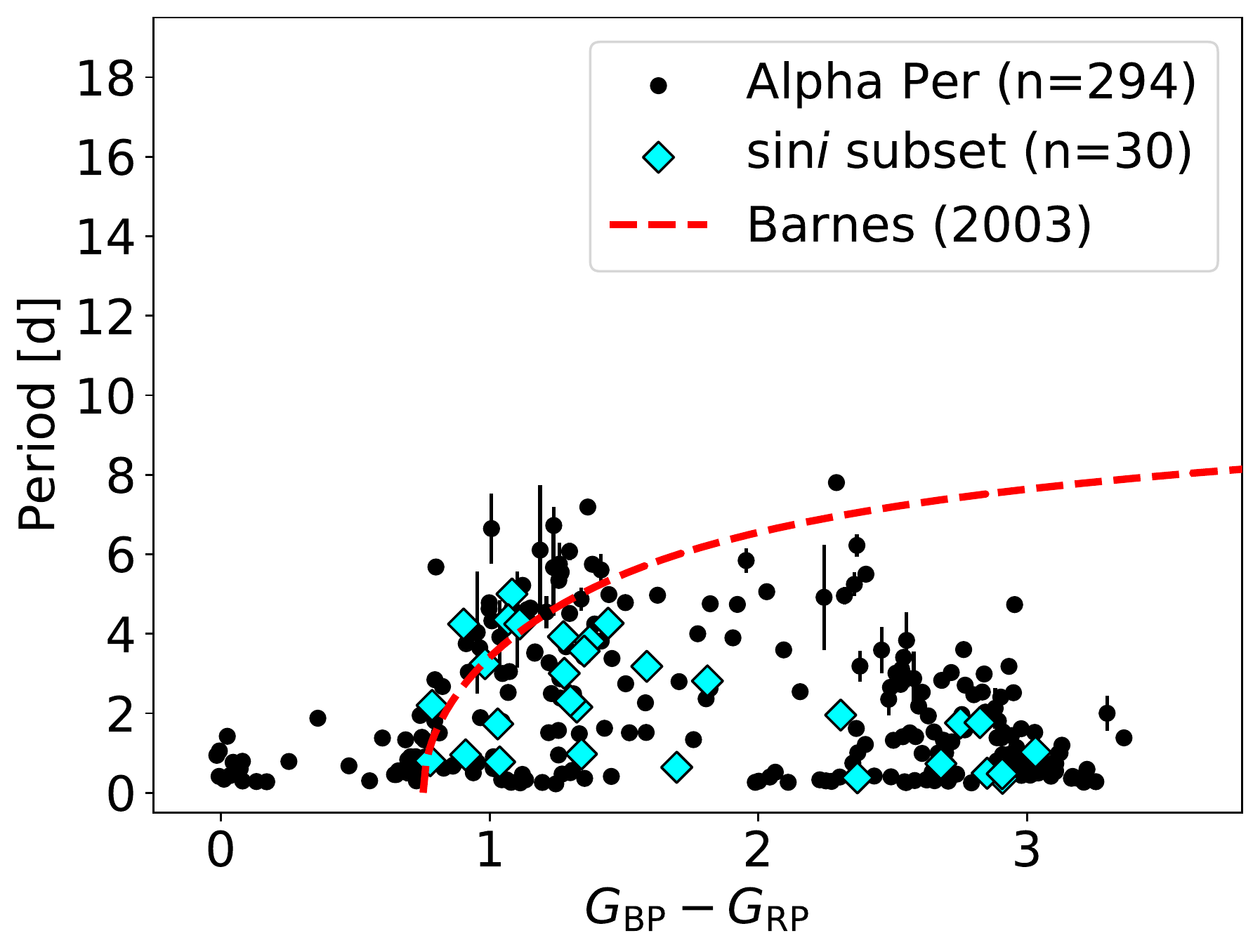}
    \includegraphics[scale=0.325]{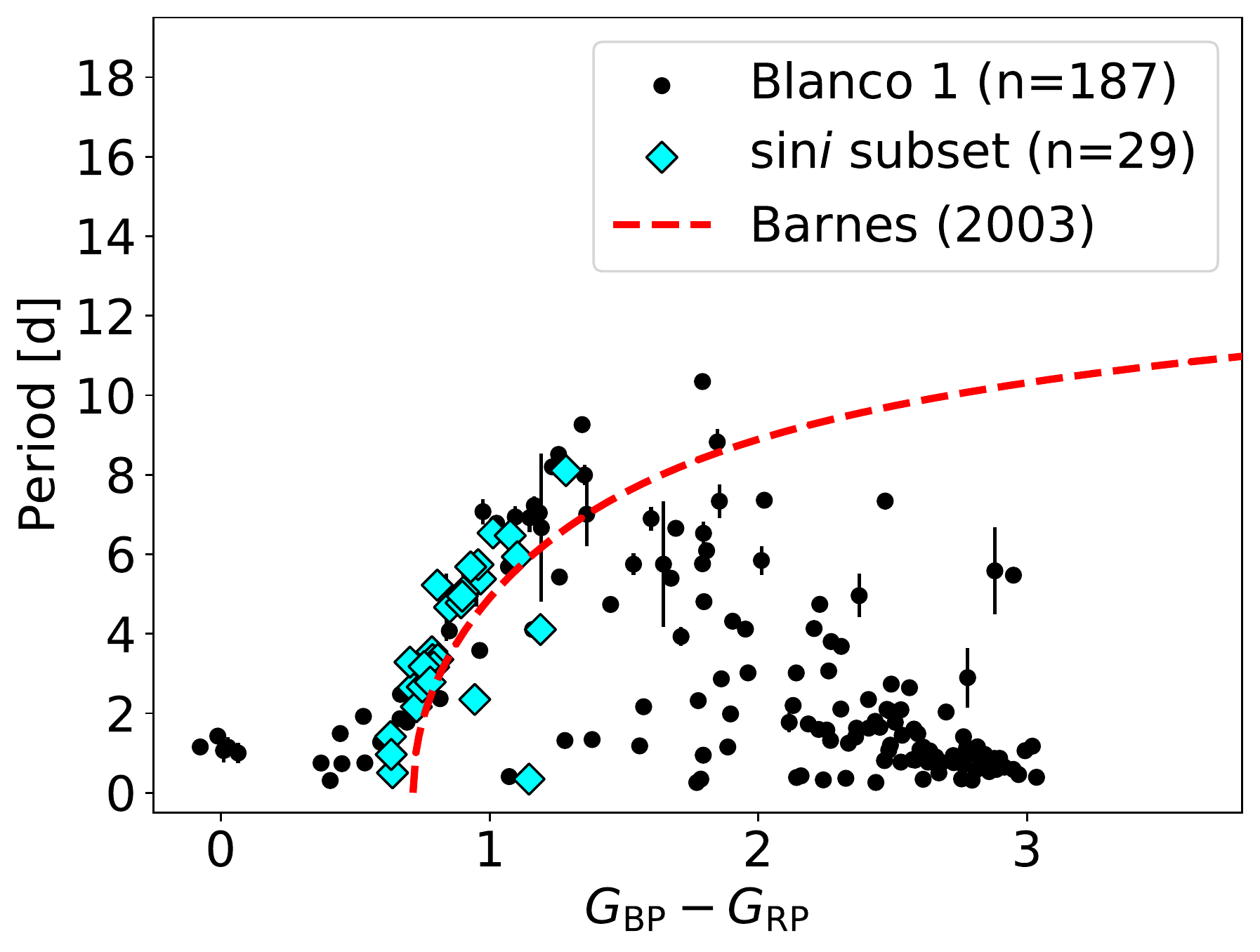}
    \includegraphics[scale=0.325]{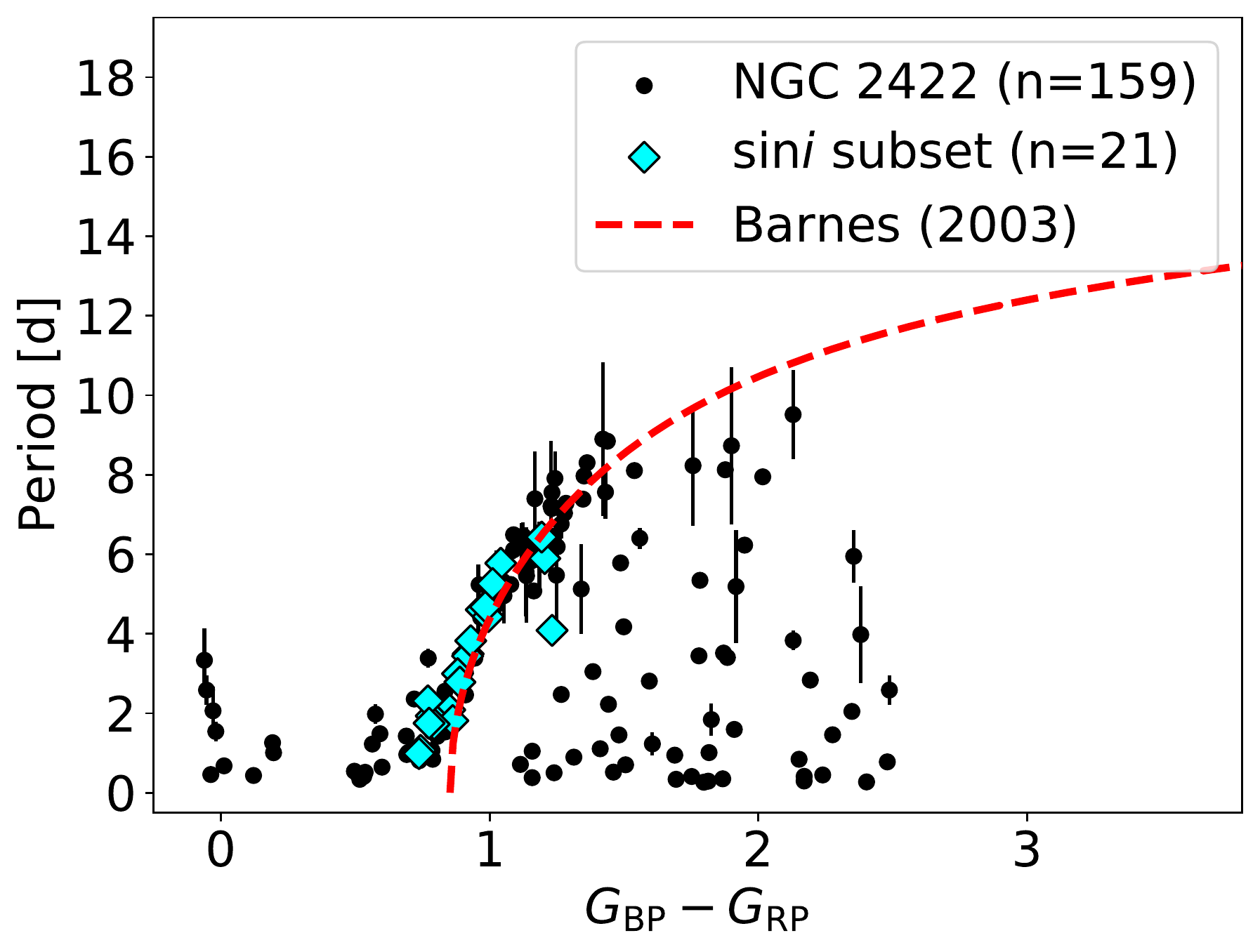}
    \includegraphics[scale=0.325]{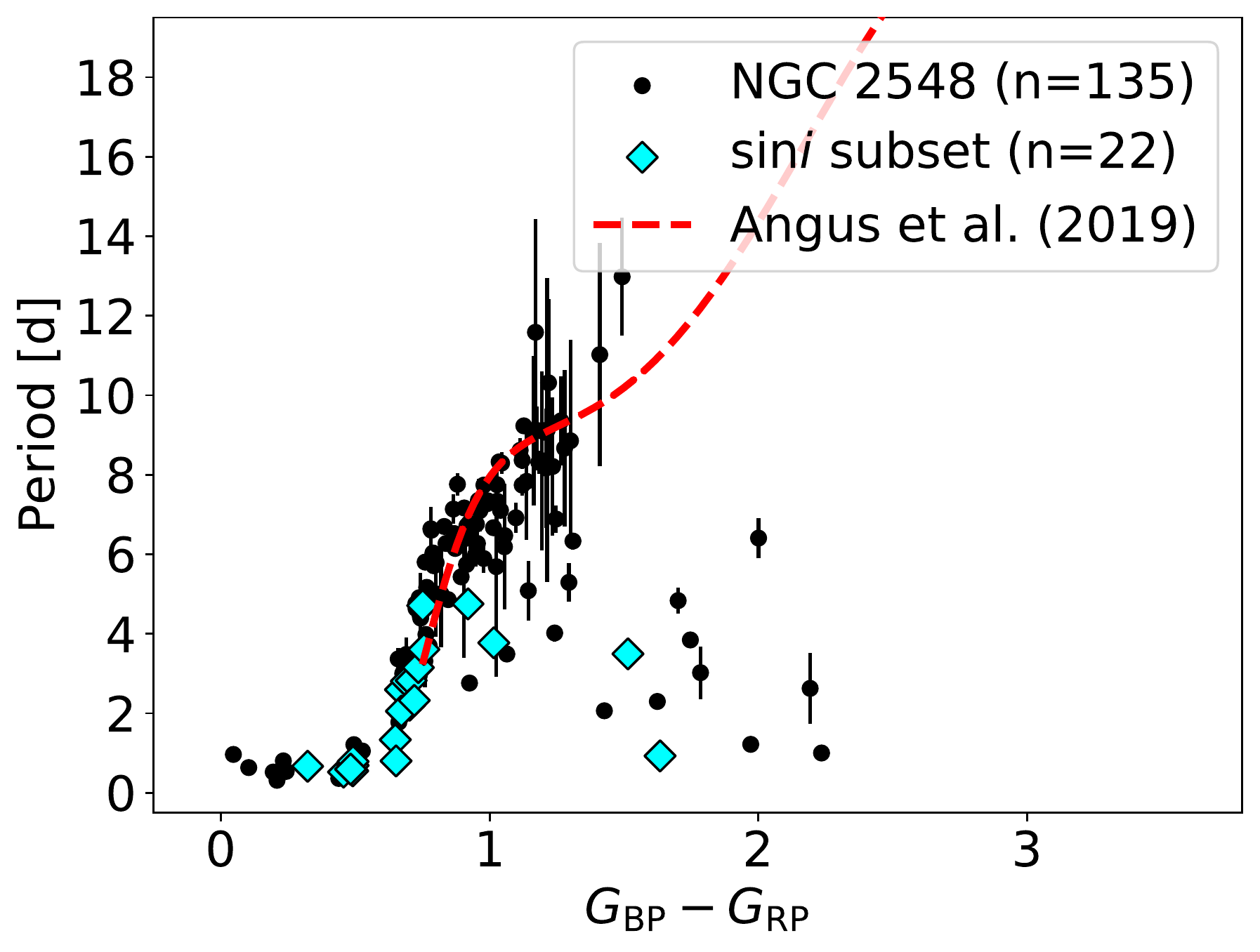}
    \caption{Color-rotation period plots for all clusters. All plotted periods are determined with TESS data by this work. For clusters older than 30 Myr, we plot an empirical sequence of slow rotators \citep[][]{barnes2003, angus2019} which validate the parameters we determined via SED fitting.}
    \label{fig:color_period_plots}
\end{figure*}

For the 11 clusters, we report 1618 rotation periods we determined with TESS in Table \ref{tab:periods}. We plot rotation periods versus $G_{\rm BP} - G_{\rm RP}$ color for all clusters in Figure \ref{fig:color_period_plots}. For clusters older than 30 Myr, we include an empirical isochrone for the slow-rotating sequence of stars (red dashed line) based on either \citet[][]{barnes2003} or \citet[][]{angus2019}. For the former, we estimated $G_{\rm BP} - G_{\rm RP}$ colors as a function of $B - V$ using Gaia documentation.\footnote{``Photometric relationships between Gaia
photometry and existing photometric systems,'' C. Jordi} The gyro isochrones serve to aid the visualization of each cluster's rotation sequence and support the ages and reddening values that result from our SED fitting analysis. We determined rotation periods for between $\sim$ 24-58\% of stars across the entire sample, with a mean of 33\%. 

\subsection{Inclinations}
\label{subsec:results_inclinations}
We determined $\sin i$ values for a total of 216 stars across the 11-cluster sample. Table \ref{tab:results} lists these values along with their upper and lower uncertainties, including the systematic errors described in Section \ref{subsec:systematic_errors}. The mean fractional uncertainty of the $\sin i$ determinations is $\sim$ 13\%. Table \ref{tab:results} also lists values for $T_{\rm eff}$ and $R$ from SED fitting, $P$ from our rotation period analysis, and $v\sin i$ from the corresponding data source (see Table \ref{tab:vsini_sources}).

\begin{figure*}
    \centering
    \includegraphics[scale=0.25]{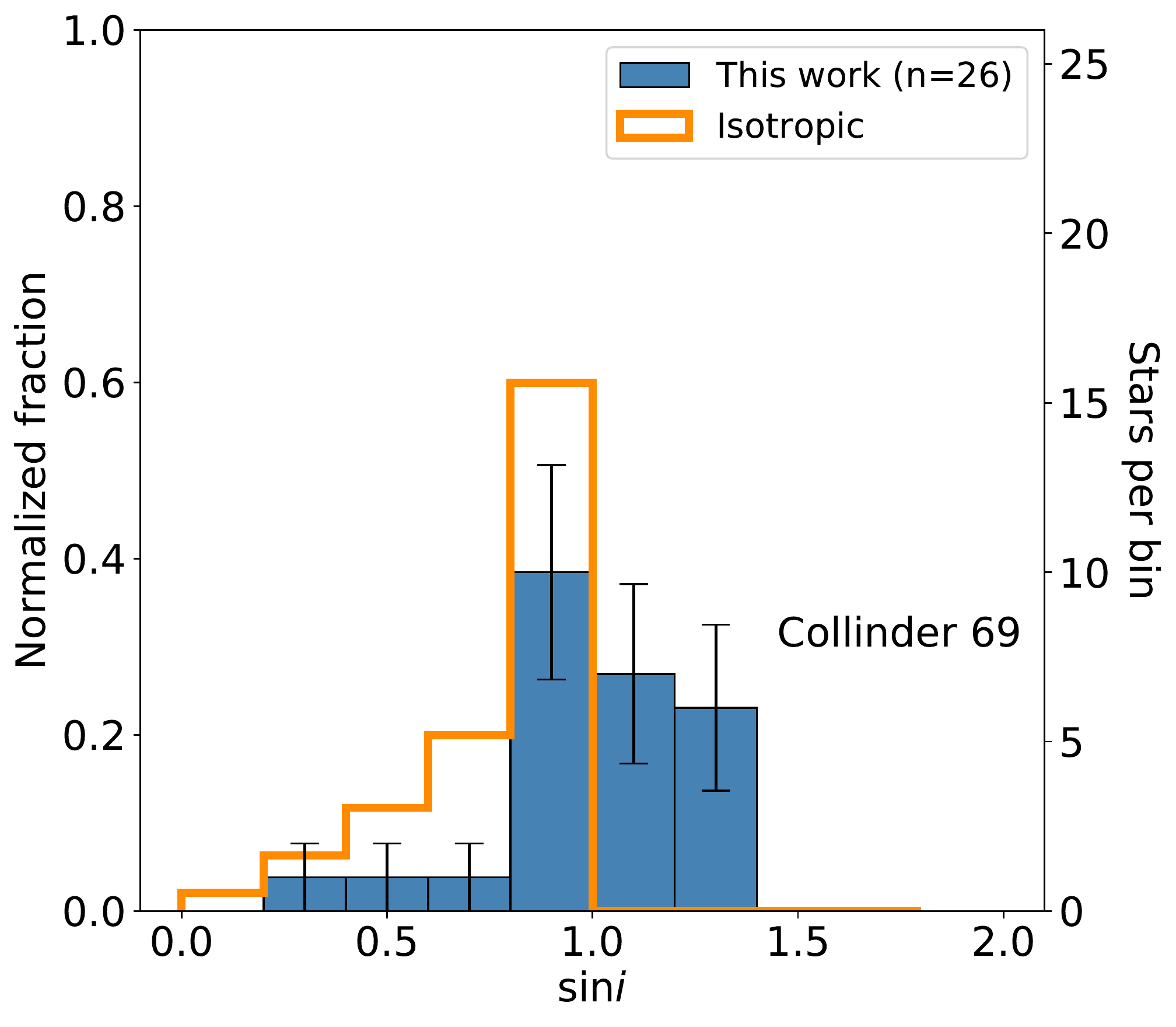}
    \includegraphics[scale=0.25]{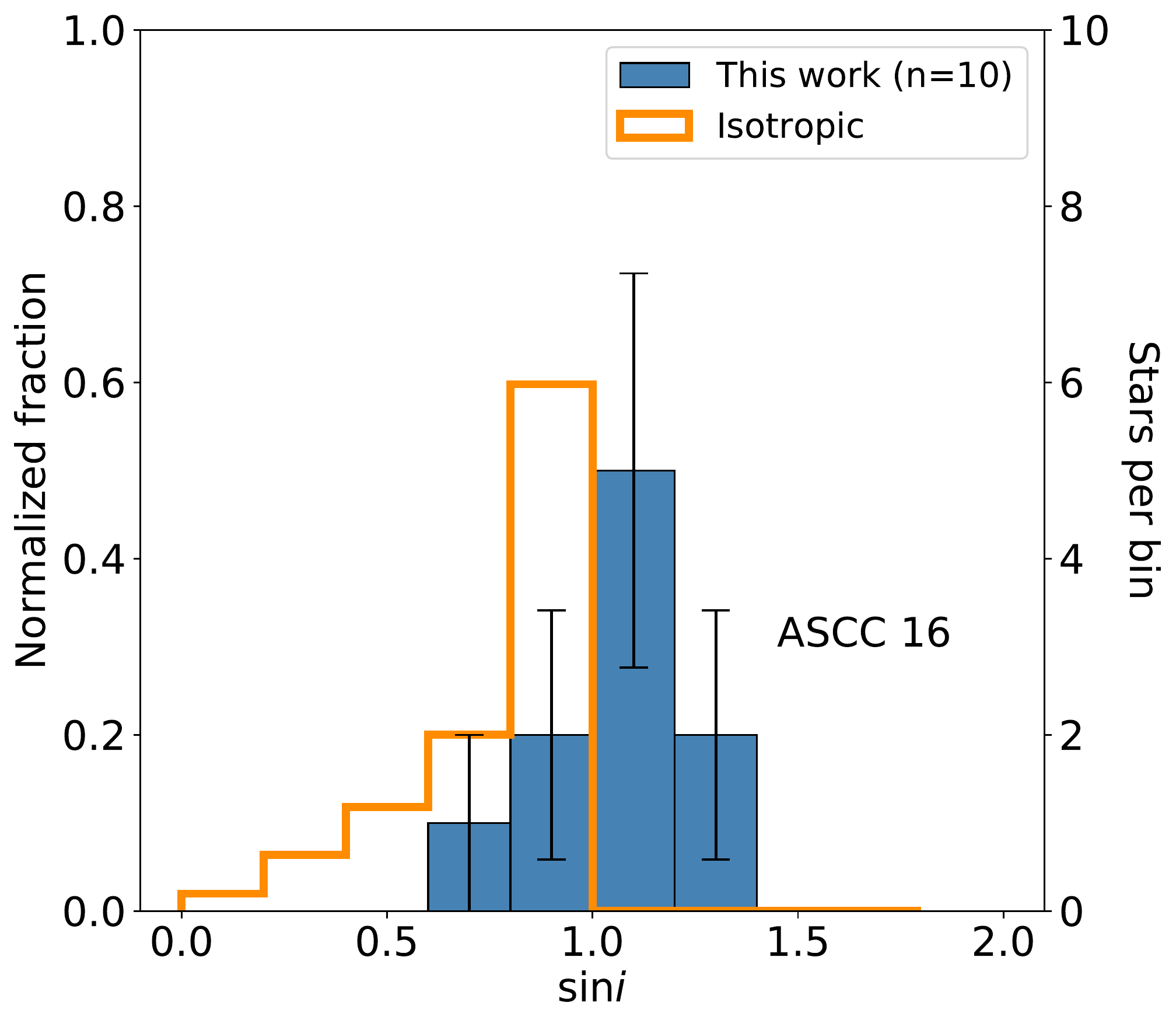}
    \includegraphics[scale=0.25]{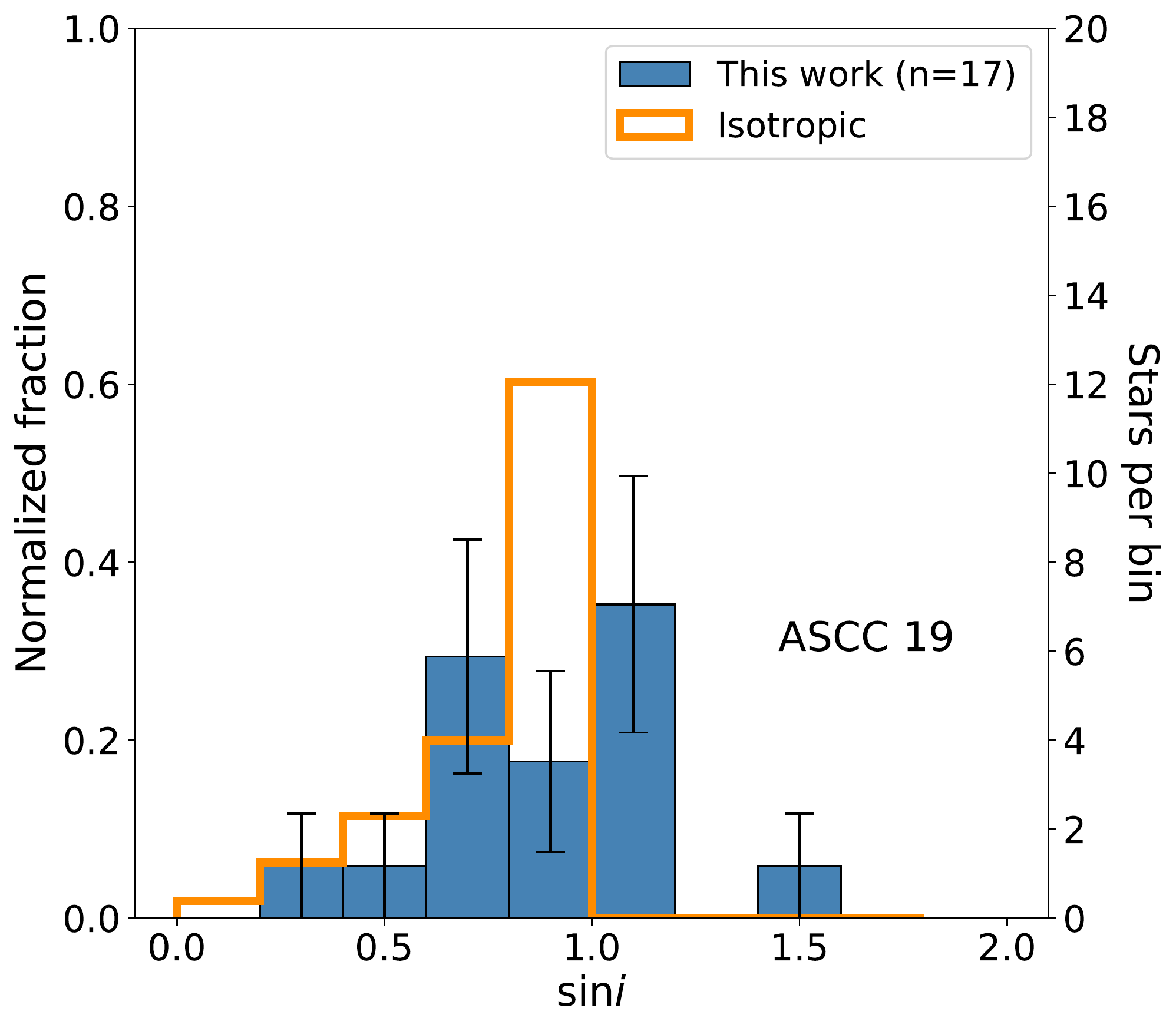}
    \includegraphics[scale=0.25]{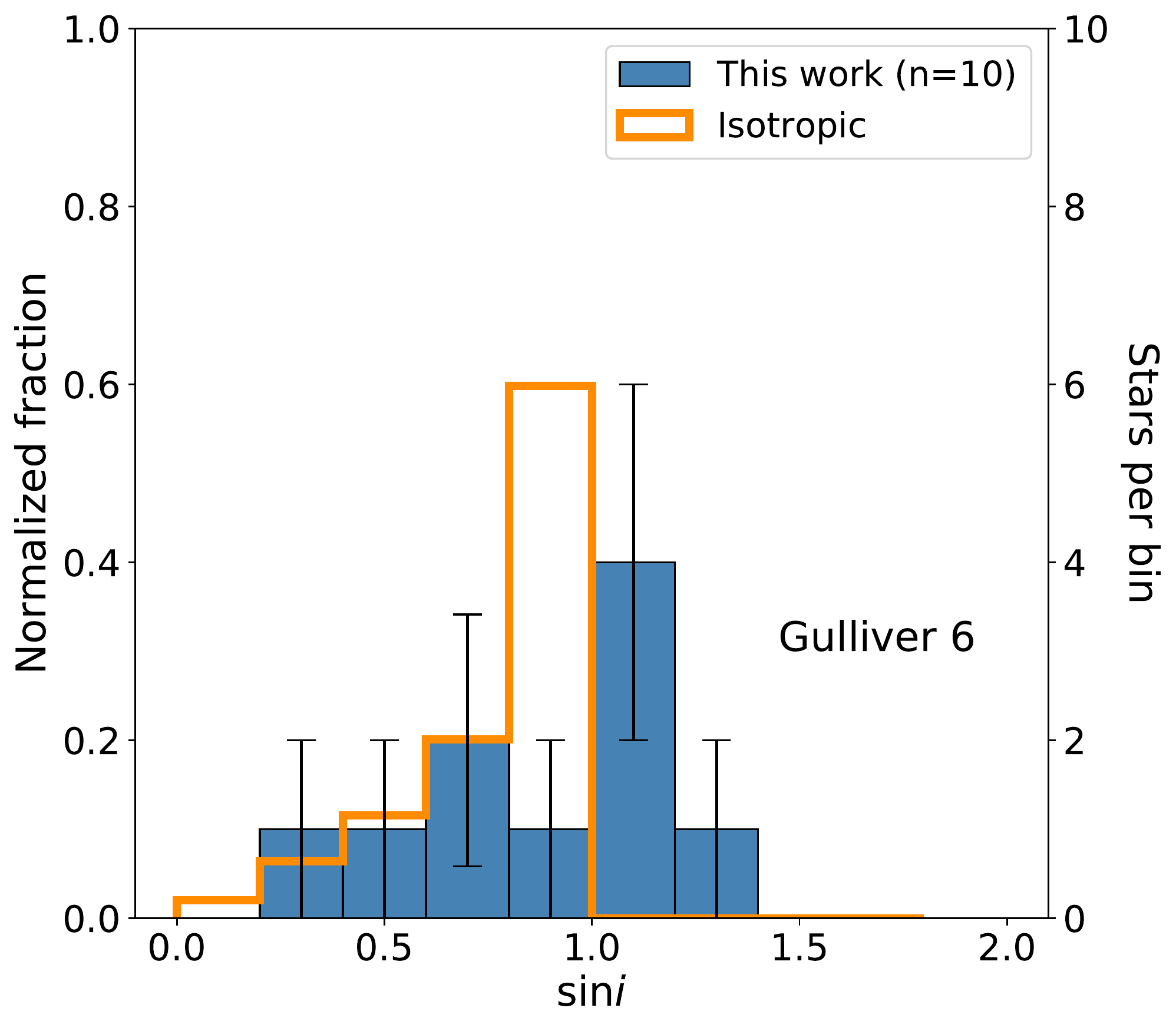}
    \includegraphics[scale=0.25]{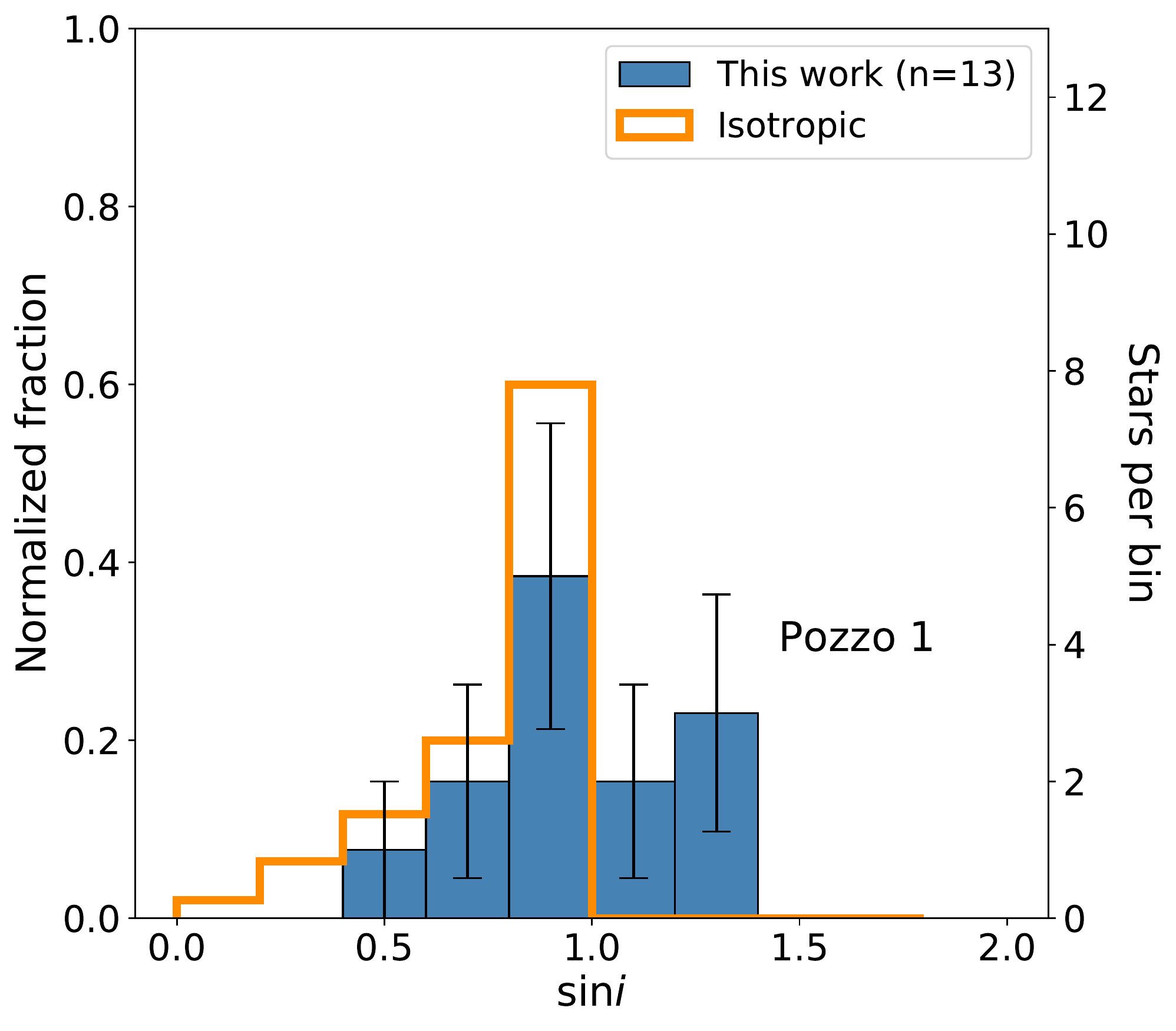}
    \includegraphics[scale=0.25]{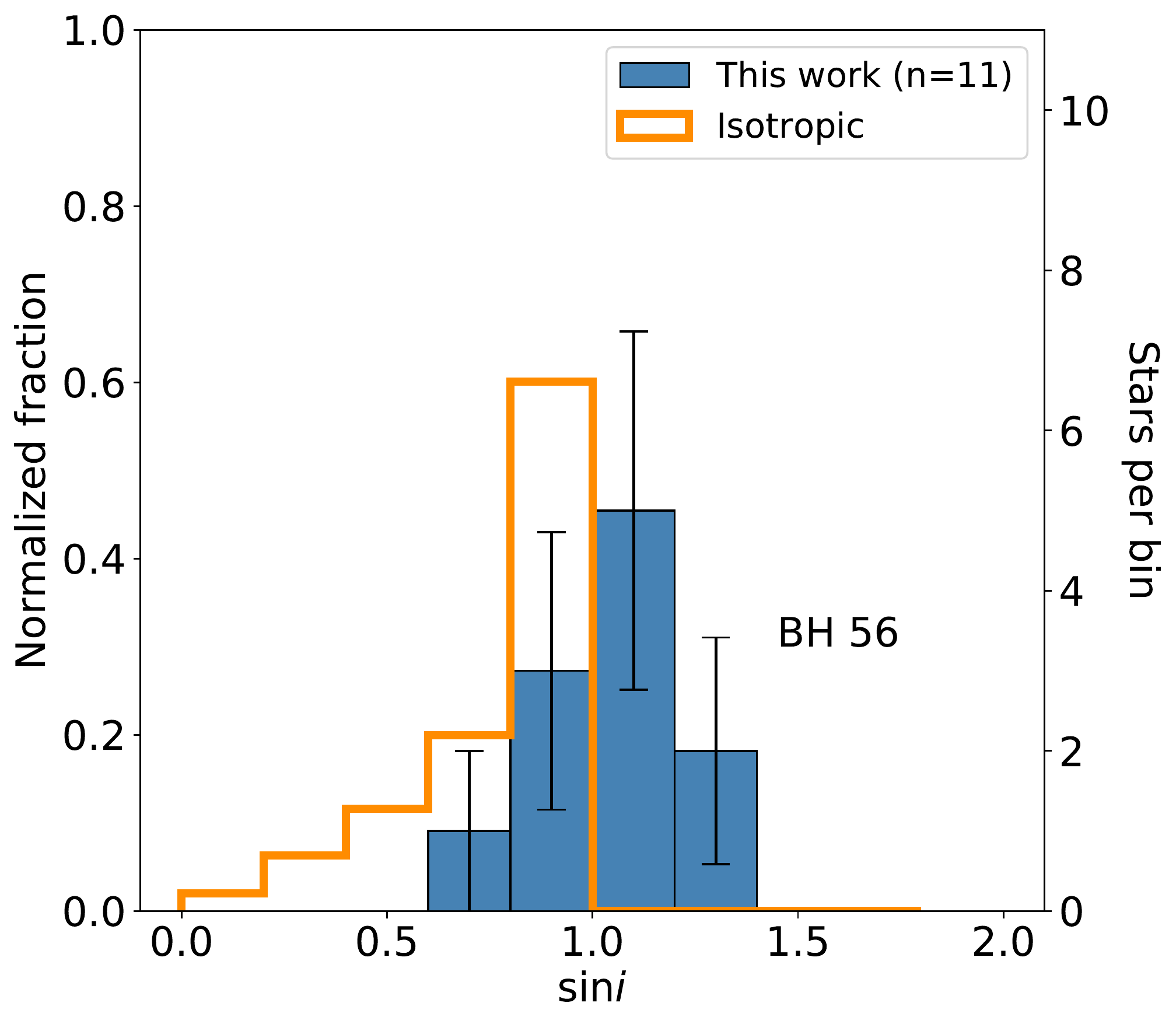}
    \includegraphics[scale=0.25]{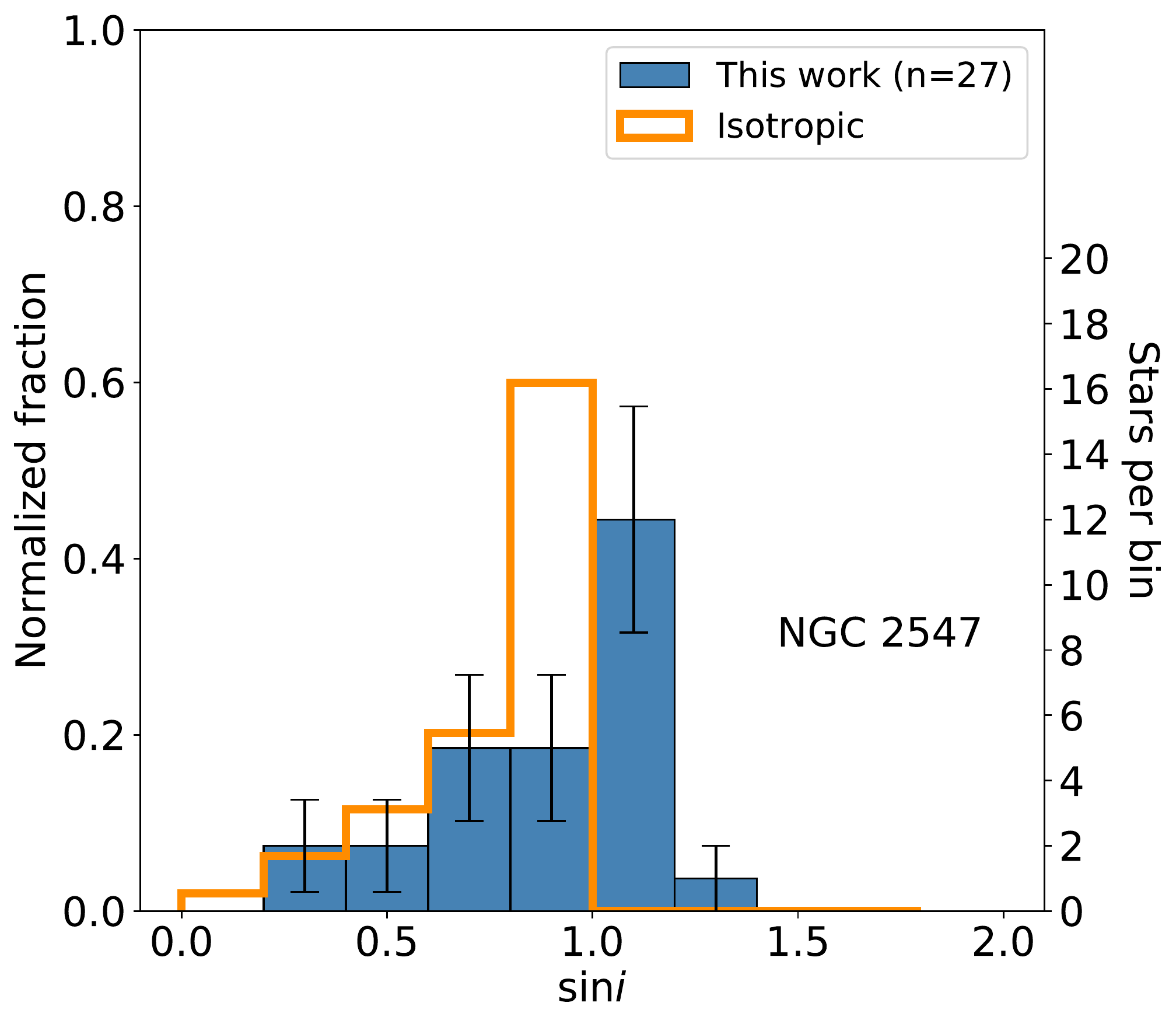}
    \includegraphics[scale=0.25]{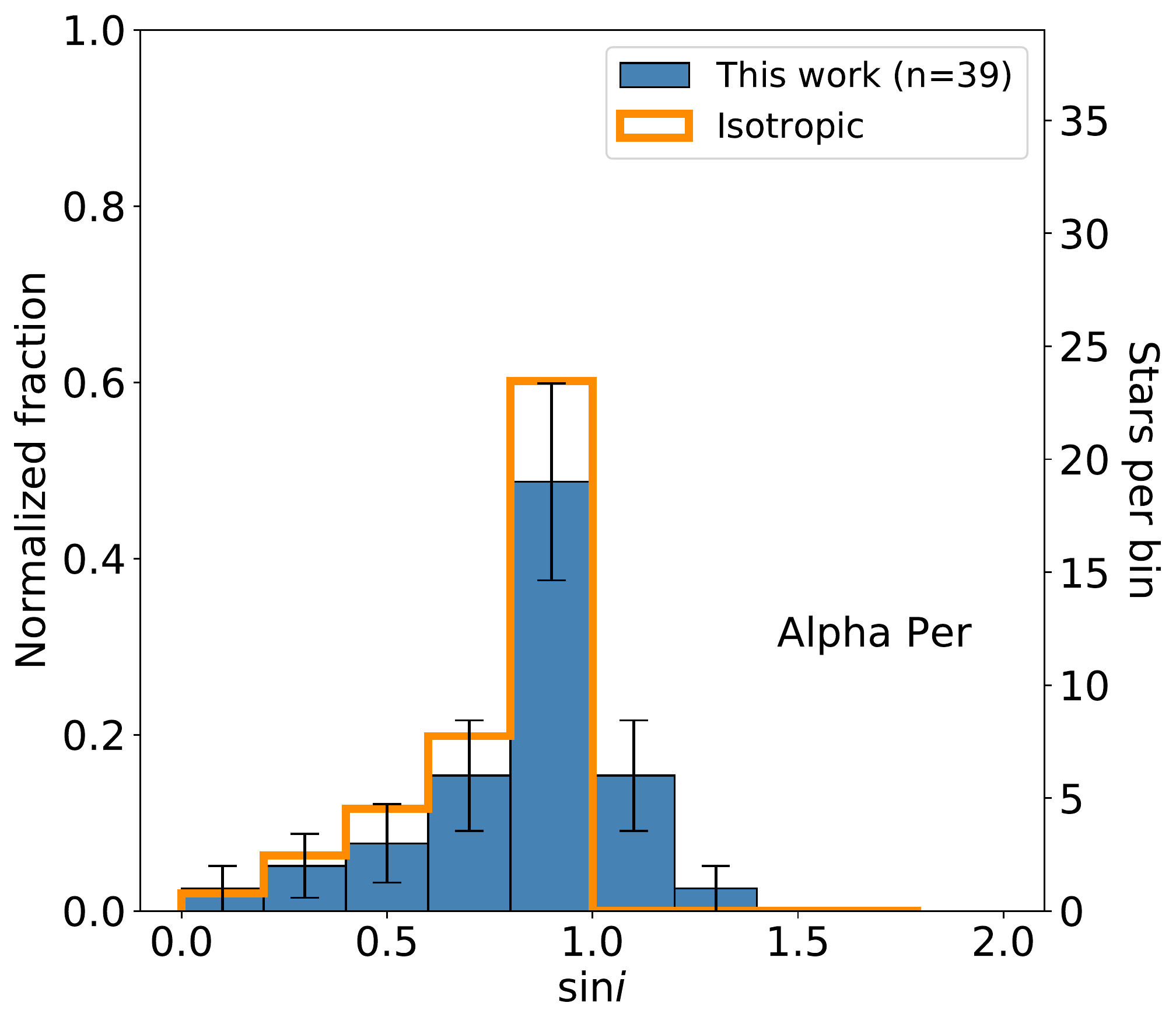}
    \includegraphics[scale=0.25]{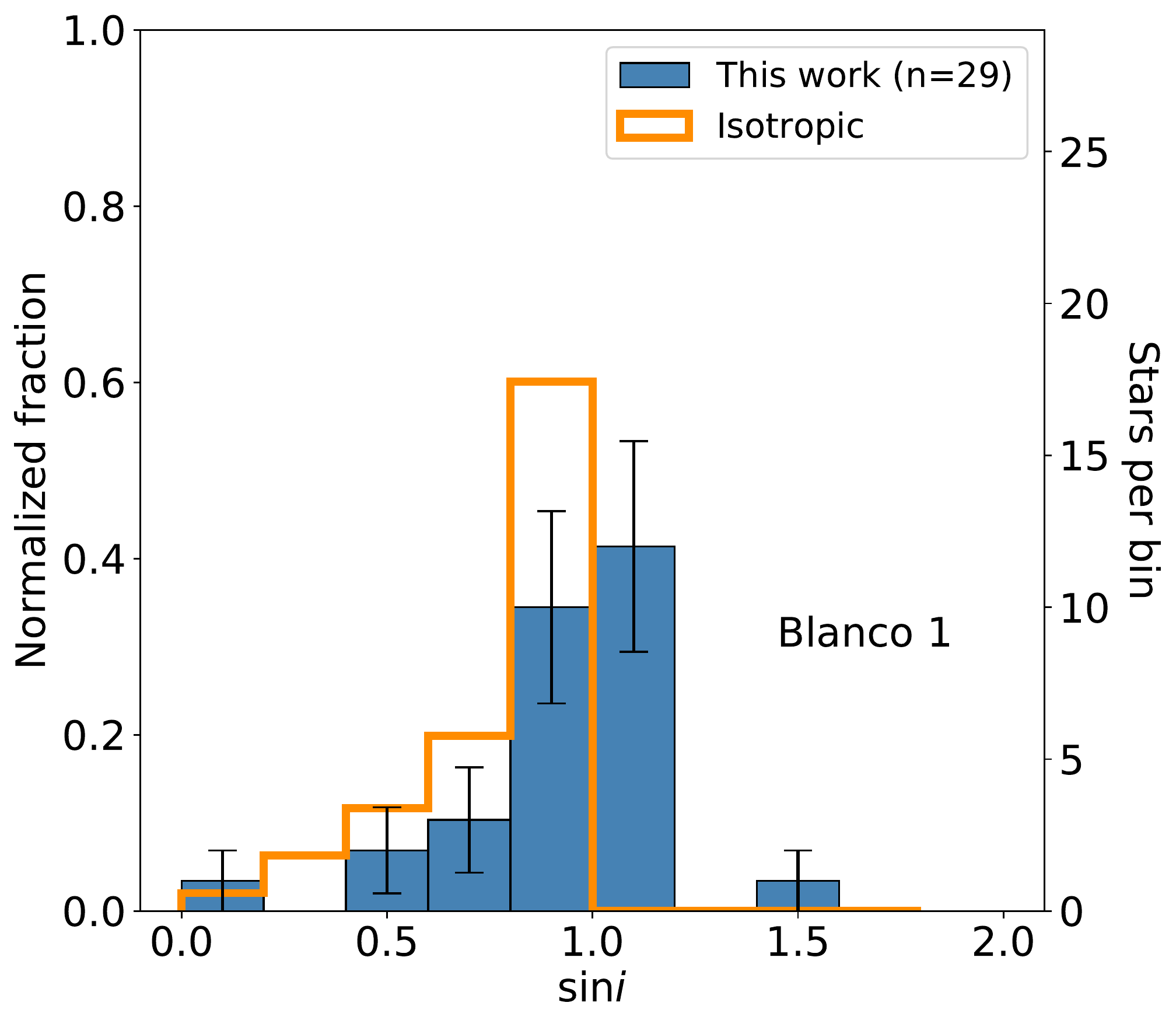}
    \includegraphics[scale=0.25]{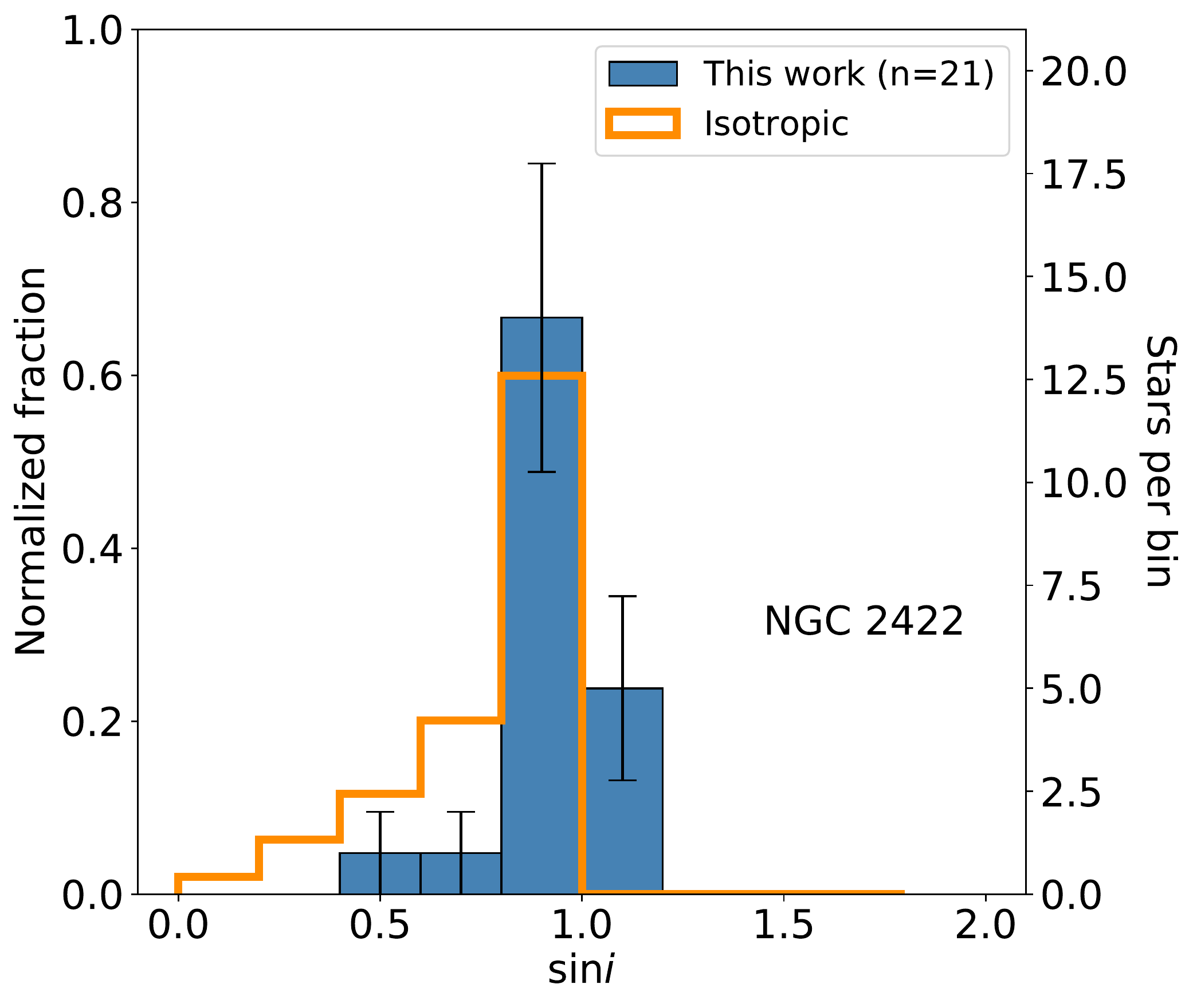}
    \includegraphics[scale=0.25]{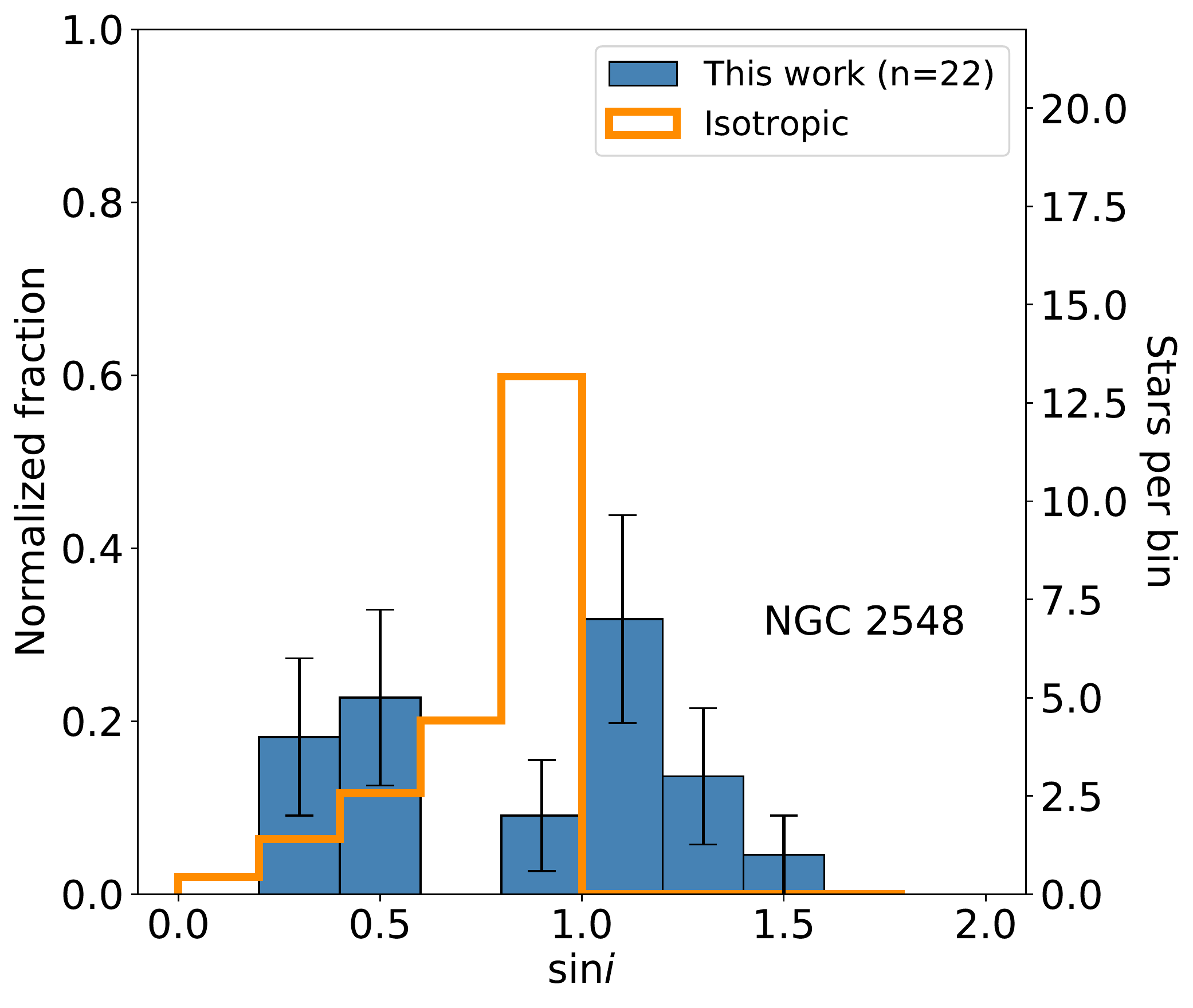}
    \caption{Histograms of projected inclinations for all clusters (blue), with isotropic distributions overlaid in orange. Error bars assume Poisson statistics.}
    \label{fig:histograms}
\end{figure*}

Figure \ref{fig:histograms} shows histograms for each cluster's inclinations compared to an isotropic distribution. Figure \ref{fig:likelihood_plots} shows plots of the cone model posteriors (normalized by the maximum value) in $\alpha$ and $\lambda$ parameter space along with 1D marginalized PPDs. Figure \ref{fig:ecdf_plots} plots the empirical CDFs for each cluster's inclinations, an isotropic distribution, a CDF corresponding to the PPD-peak values of $\lambda$ and $\alpha$, and CDFs representing model parameters drawn from the 68\% confidence intervals of the PPDs. Table \ref{tab:incl_distribs} summarizes the PPD-peak and 95\% confidence intervals in $\lambda$ and $\alpha$ for each cluster's two-parameter MCMC run, along with the reduced $\chi^{2}$ and maximum natural logarithm of the posterior values for the most probable results.

The numerical simulations of \citet[][]{corsaro2017} suggest that stars below 0.7 \msun\ do not show spin alignment even if more massive cluster members do. We ran a second iteration of our MCMC analysis on the stars greater than this mass threshold for each cluster, finding no significant differences in results compared to the entire-sample analyses.

\subsection{Three-parameter Model Runs}
\label{subsec:threeparam_model_runs}
We performed a grid search of 3-parameter model space for each cluster in order to identify significant improvements over two-parameter model fits. The $f$ parameter of this model quantifies the fraction of stars aligned in a cone described by a mean inclination $\alpha_{\rm aniso}$ and spread half-angle $\lambda_{\rm aniso}$. The remaining $1-f$ fraction of stellar spins are modeled as isotropic. We computed the natural log-posteriors for values of $\alpha$ and $\lambda$ between 10$^{\circ}$ and 90$^{\circ}$ (spaced in 10$^{\circ}$ intervals) and for $f$ between 0.1 and 0.9 (spaced by 0.1).

\begin{figure*}
    \centering
    \includegraphics[scale=0.25]{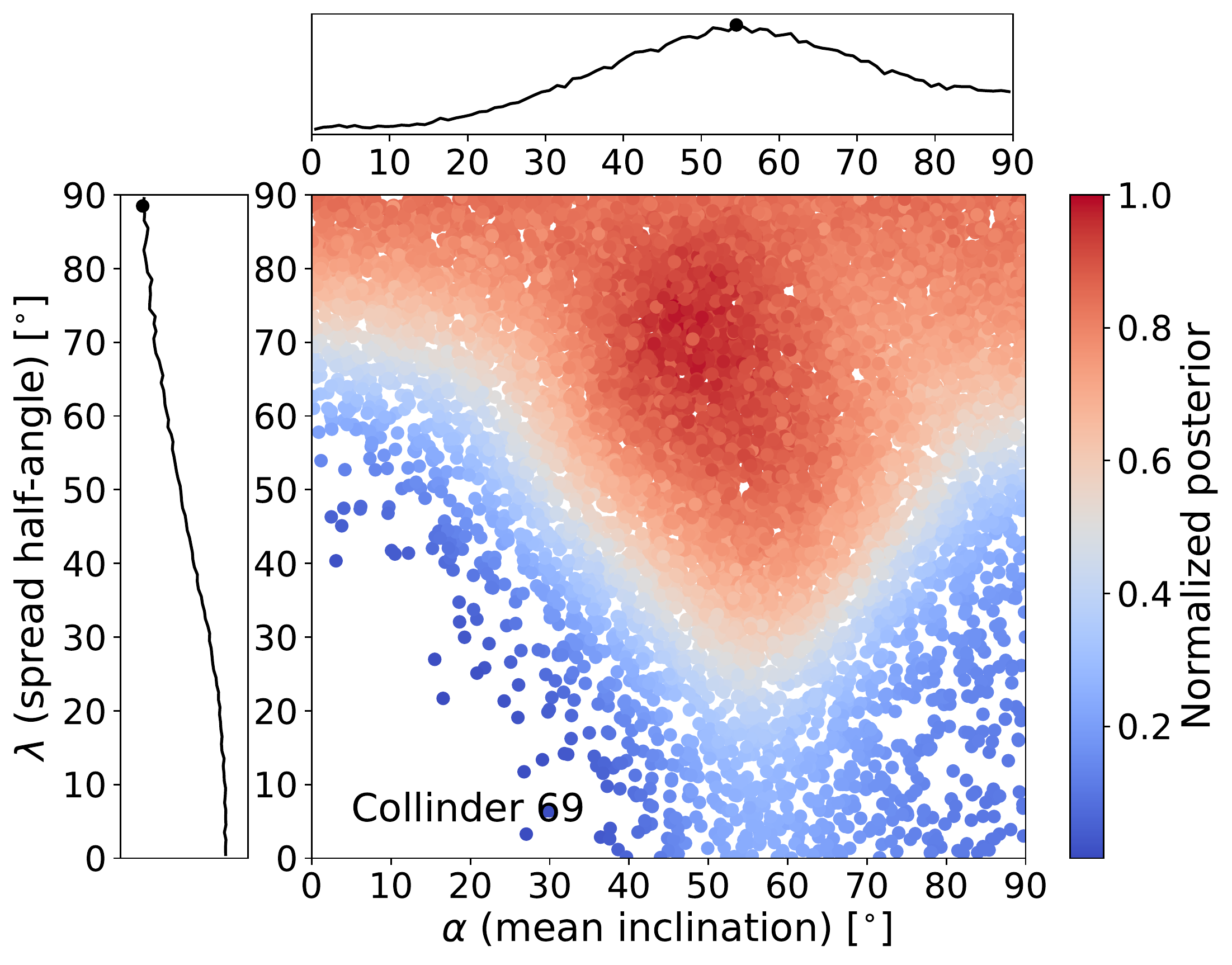}
    \includegraphics[scale=0.25]{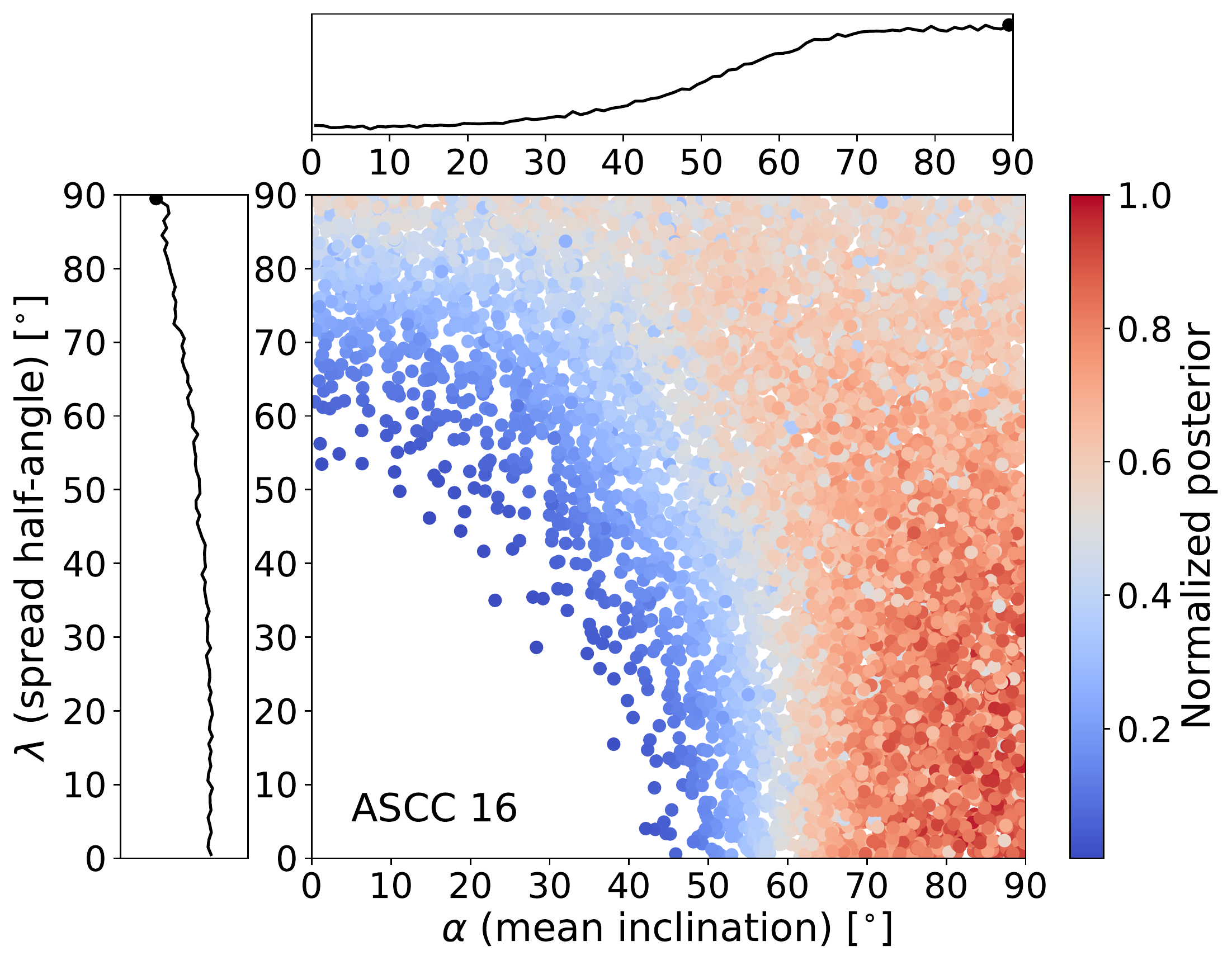}
    \includegraphics[scale=0.25]{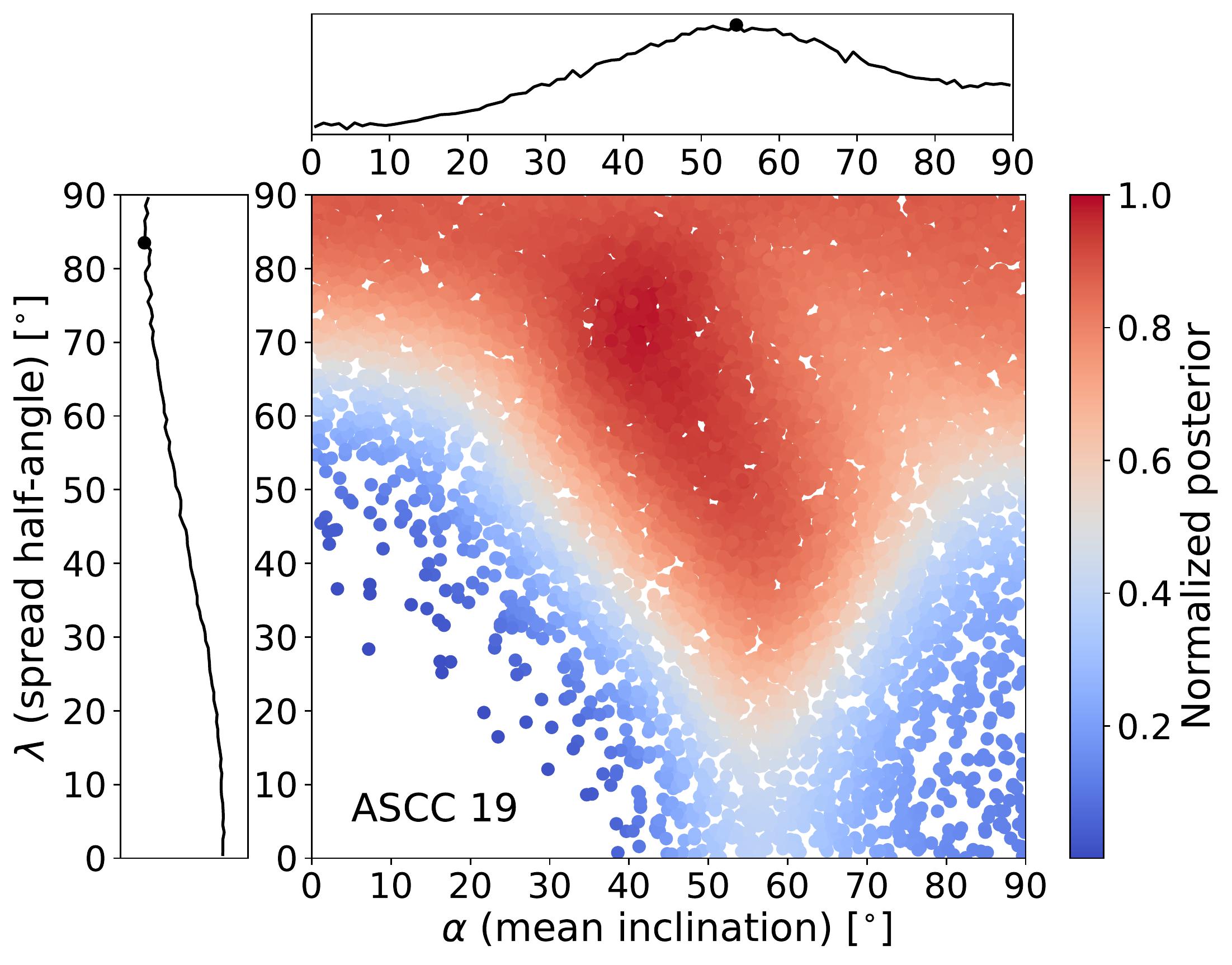}
    \includegraphics[scale=0.25]{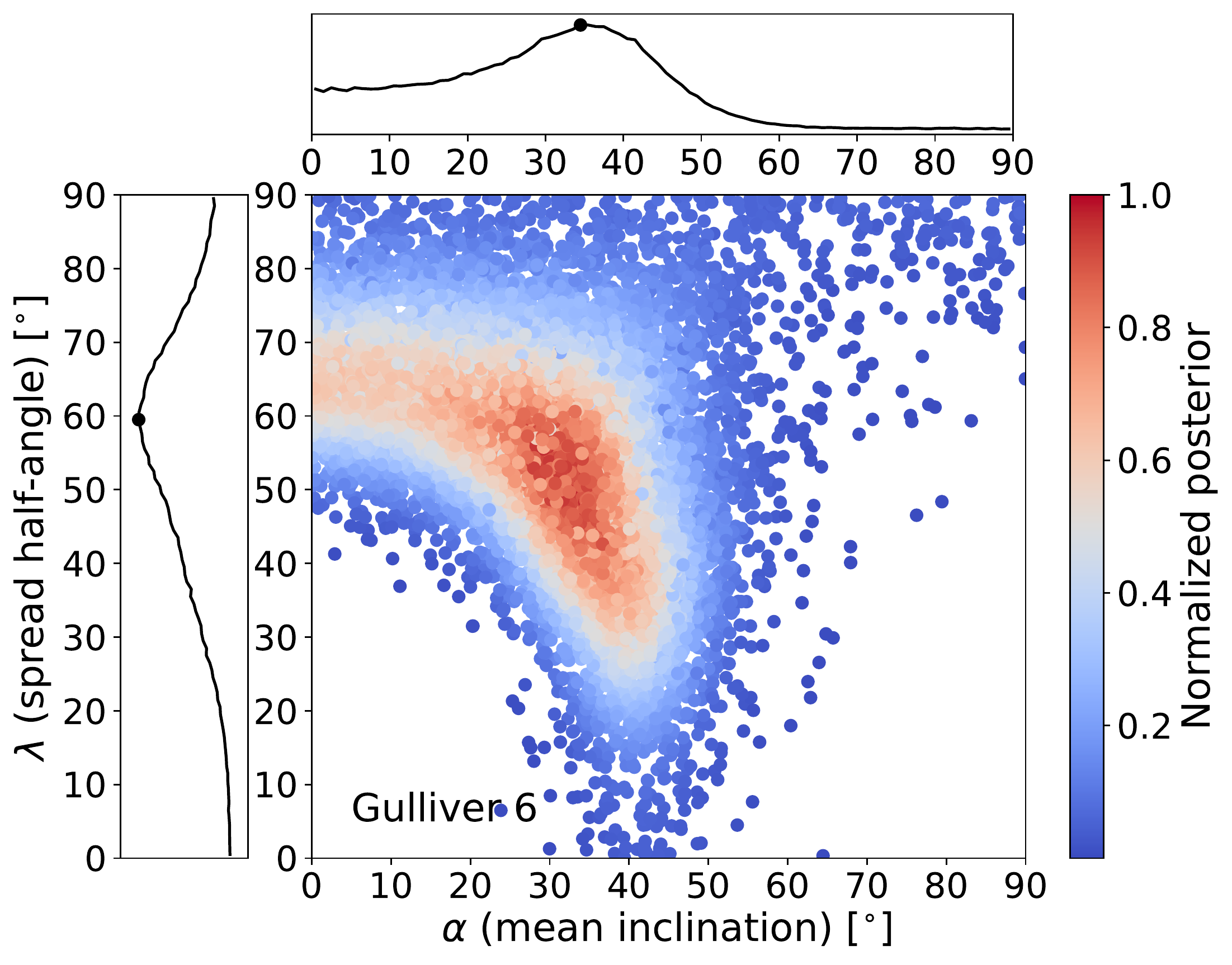}
    \includegraphics[scale=0.25]{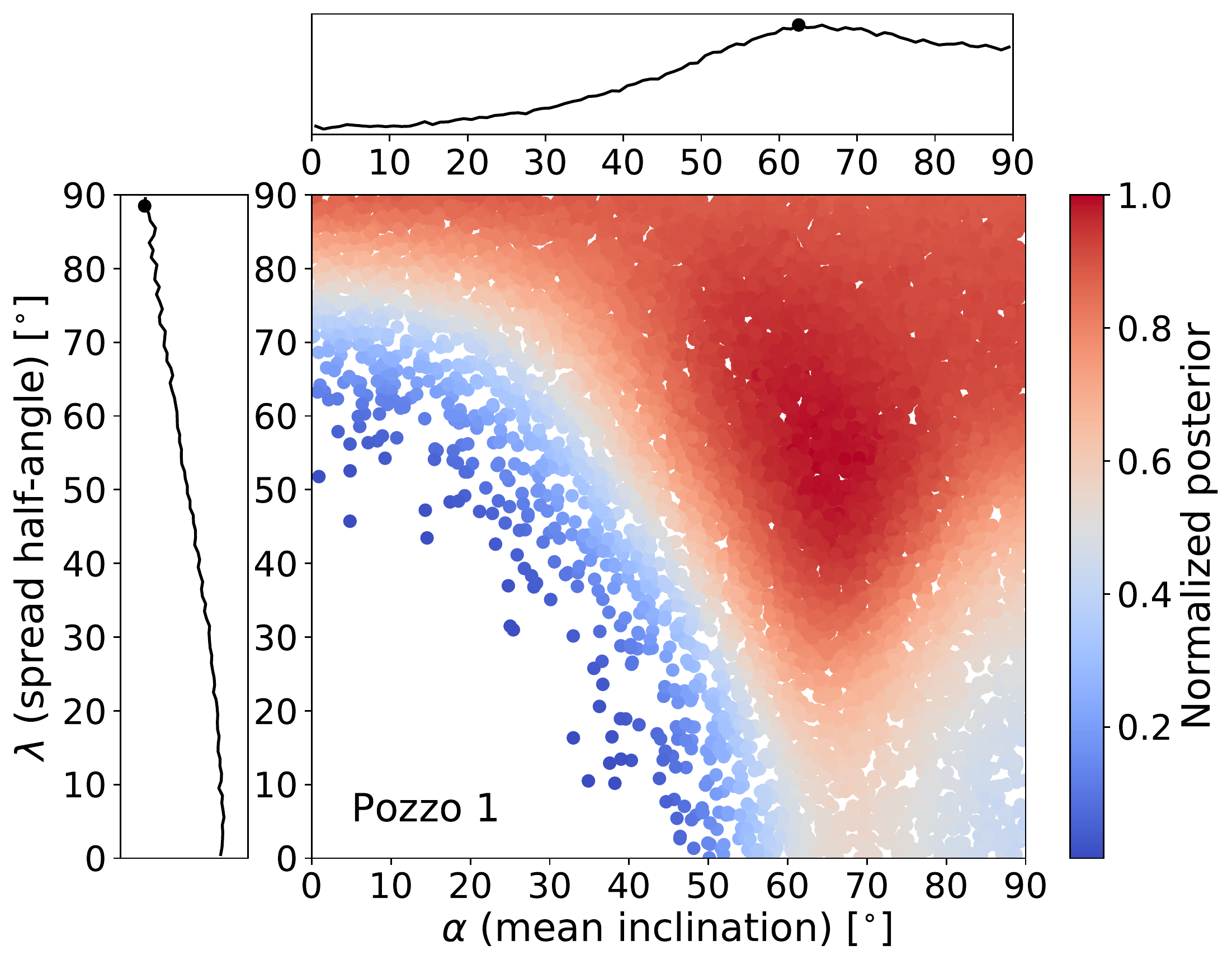}
    \includegraphics[scale=0.25]{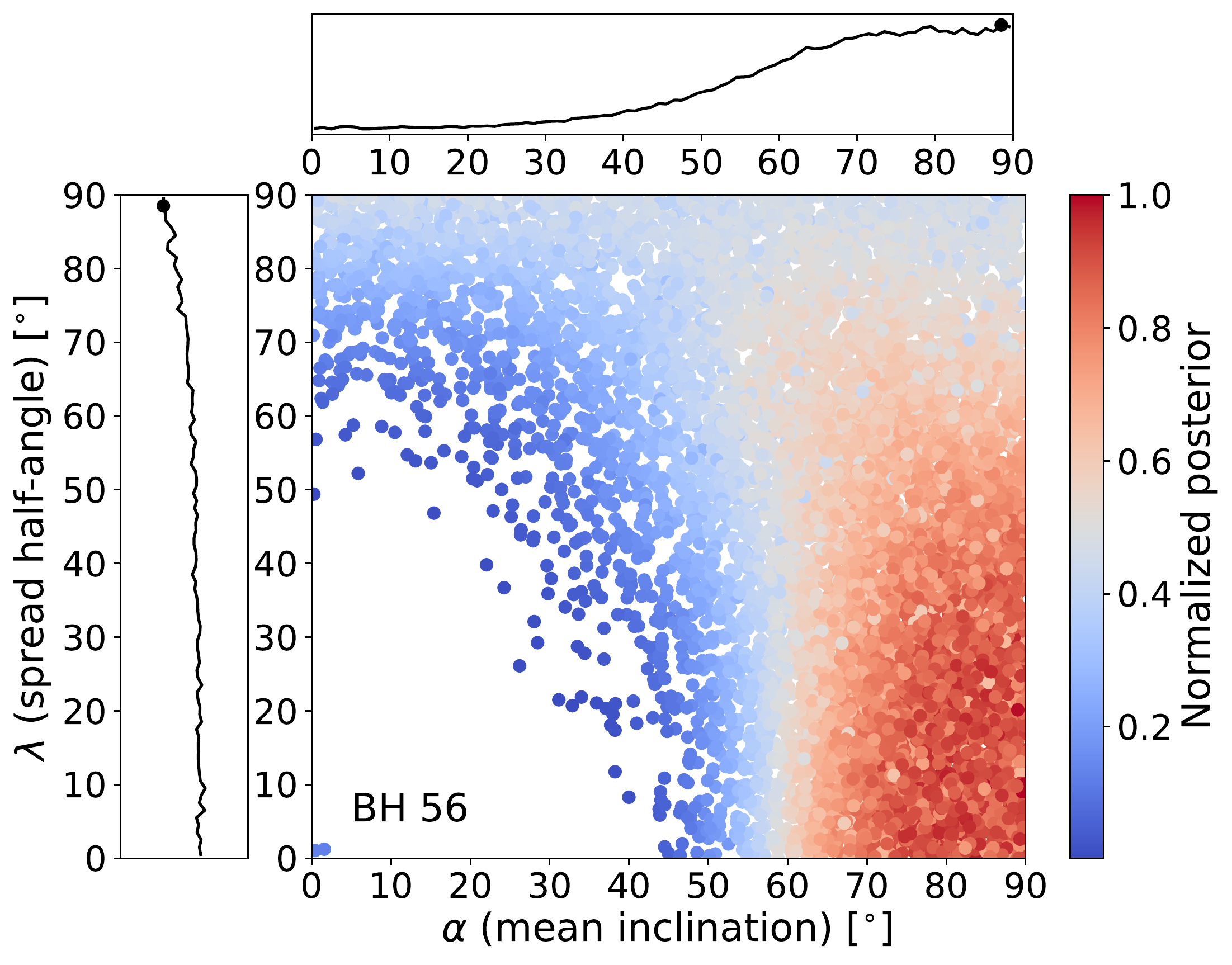}
    \includegraphics[scale=0.25]{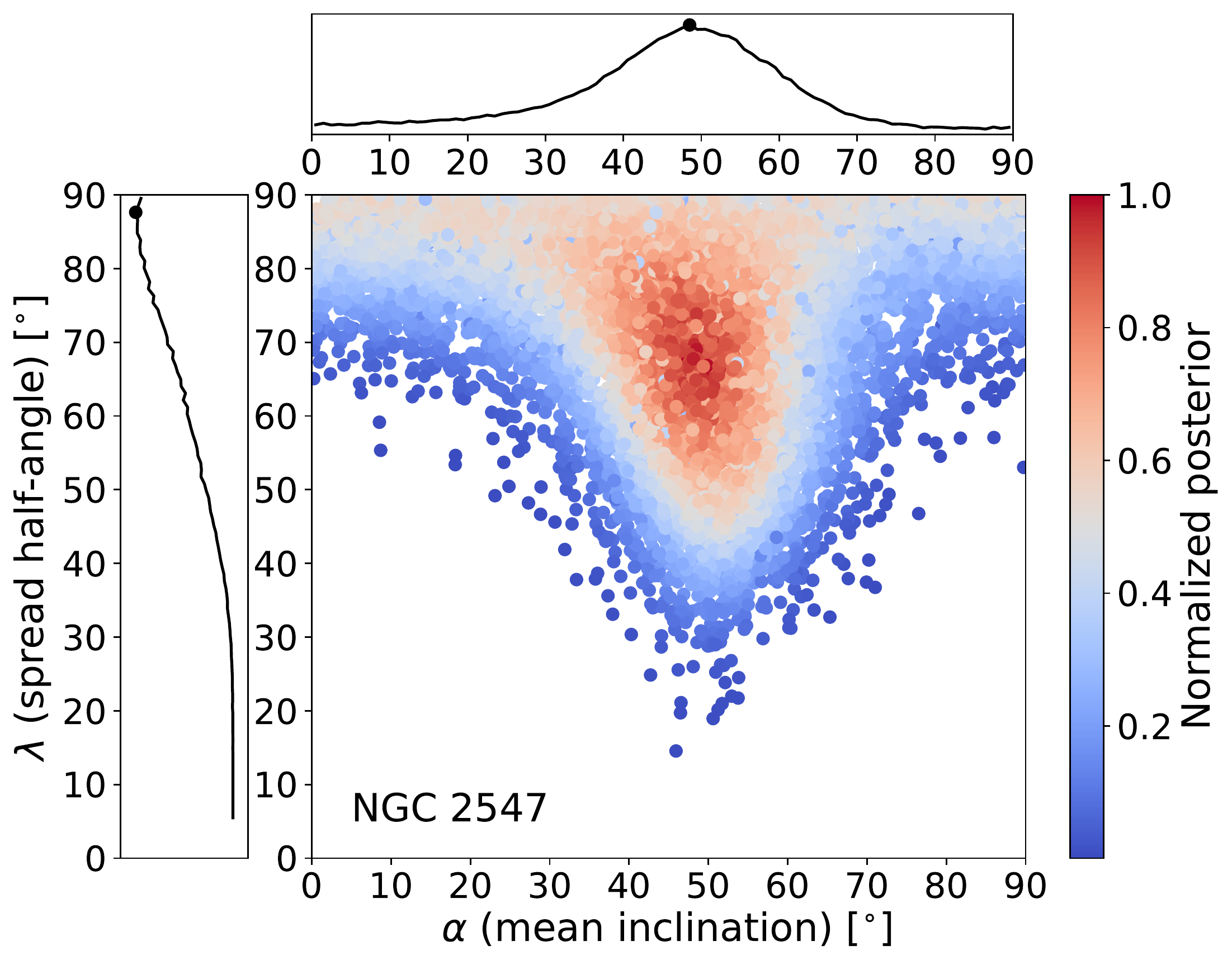}
    \includegraphics[scale=0.25]{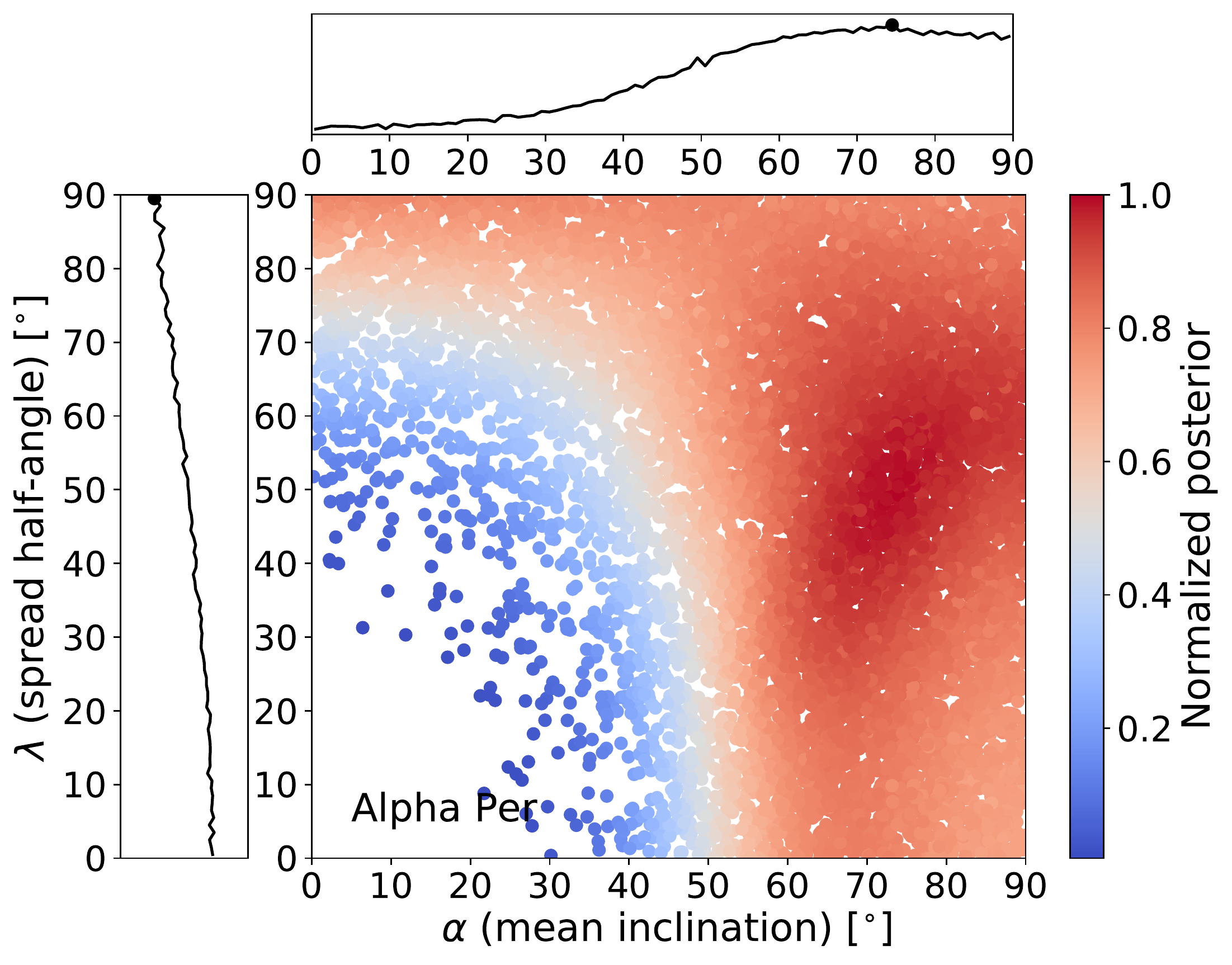}
    \includegraphics[scale=0.25]{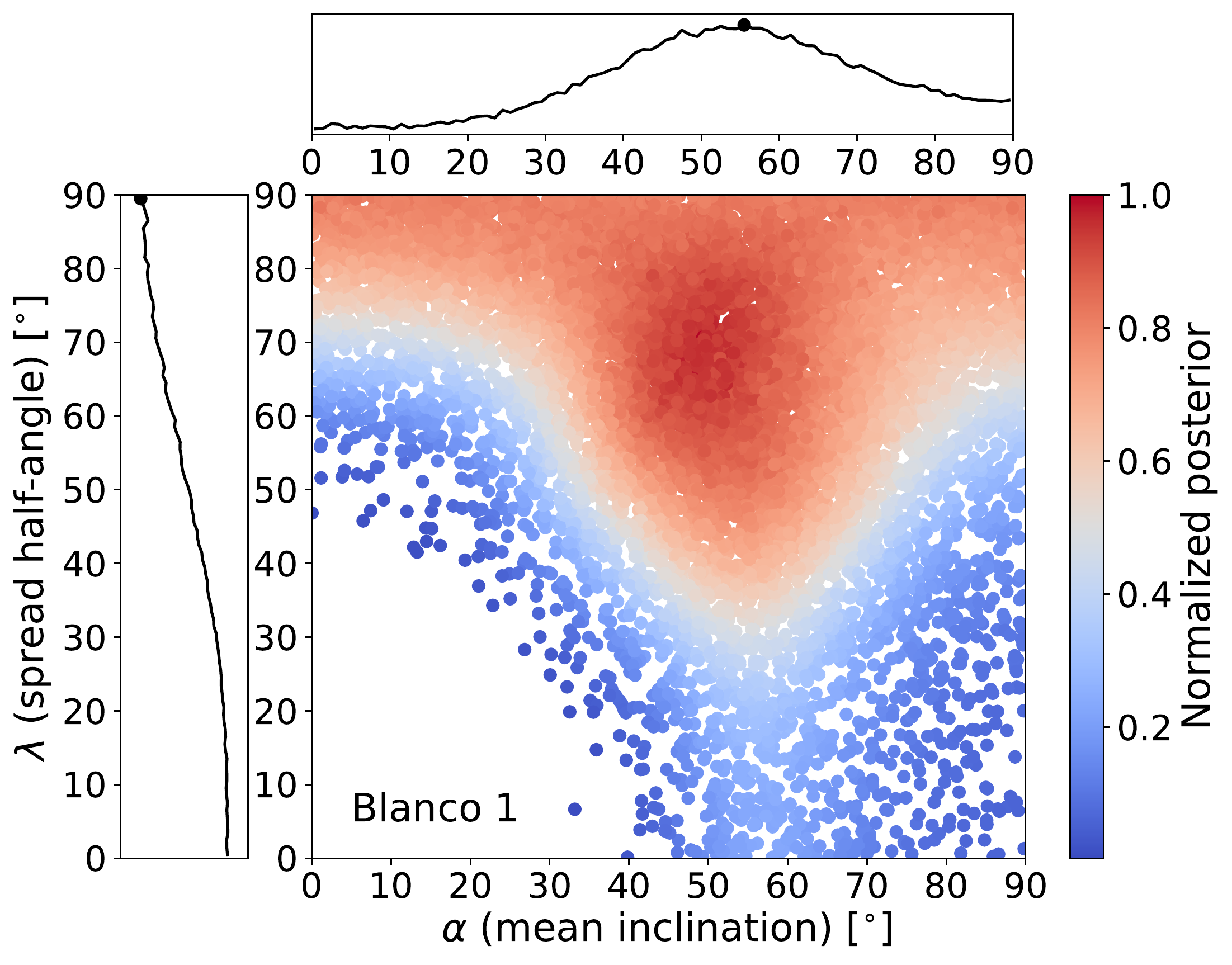}
    \includegraphics[scale=0.25]{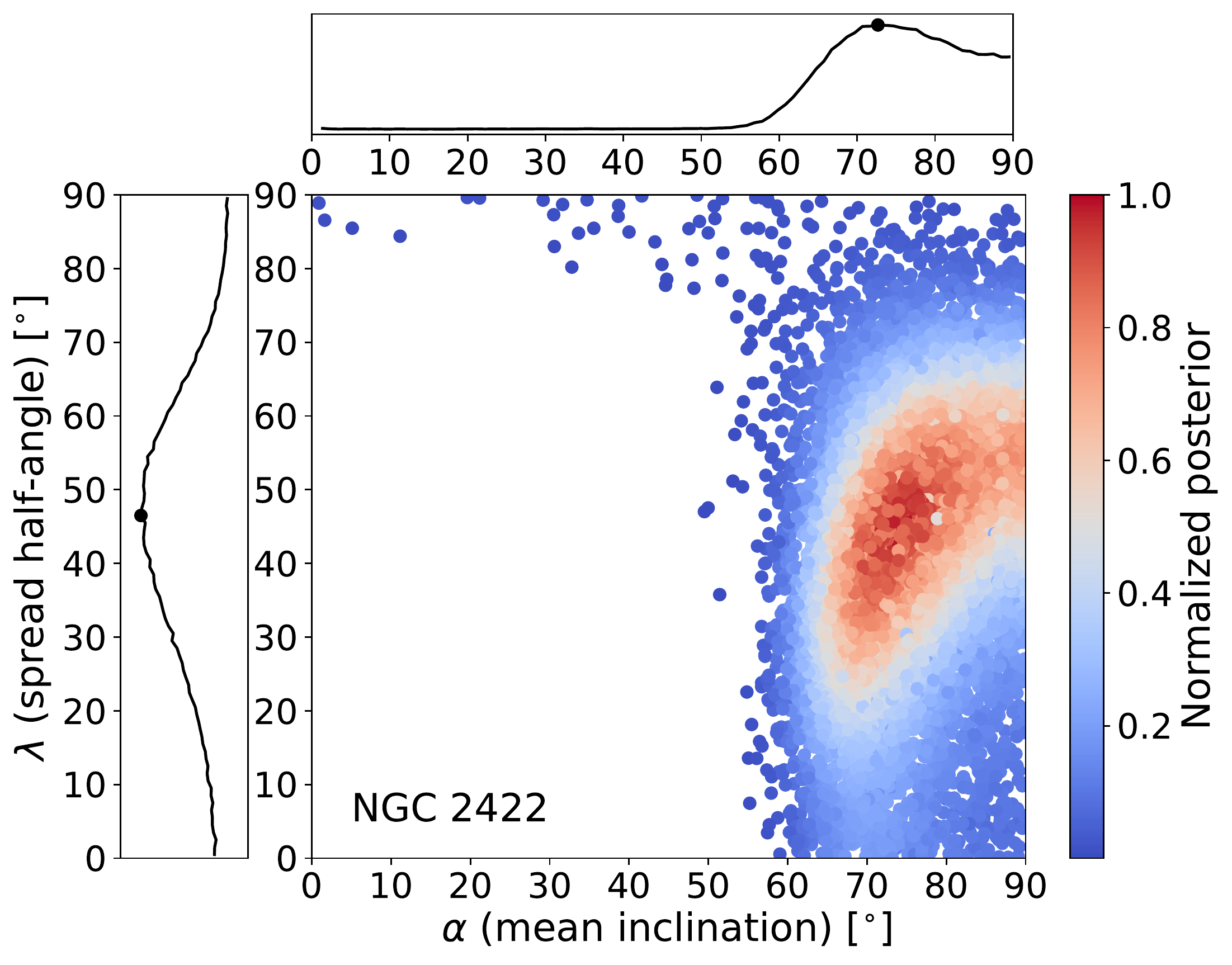}
    \includegraphics[scale=0.25]{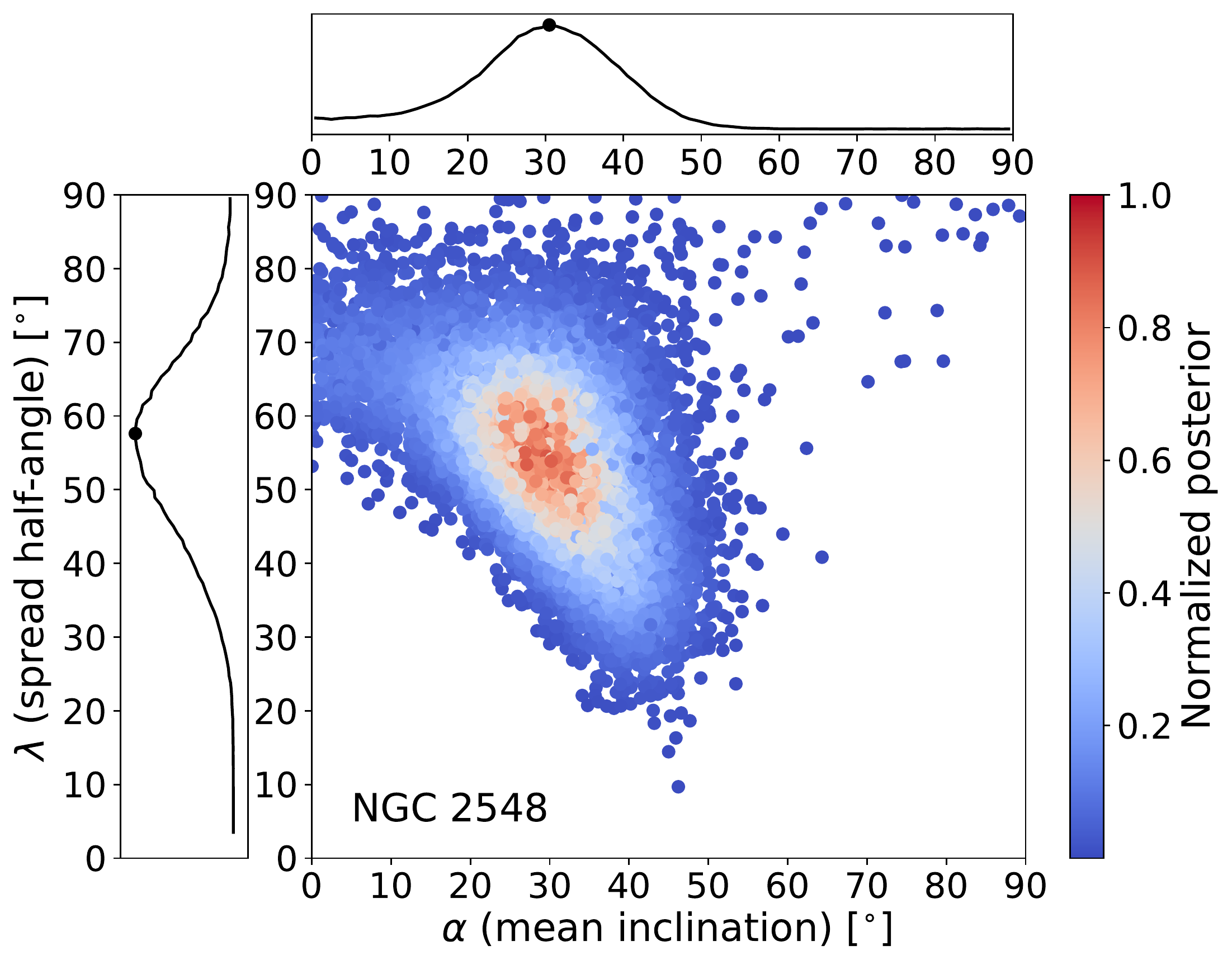}
    \caption{Two-parameter cone model log-posterior plots for all clusters, normalized by each cluster's maximum log-posterior value. Each panel also displays marginalized PPDs for $\alpha$ and $\lambda$. Each plot shows a randomly selected 2\% of the full MCMC chains.}
    \label{fig:likelihood_plots}
\end{figure*}

\begin{figure*}
    \centering
    \includegraphics[scale=0.375]{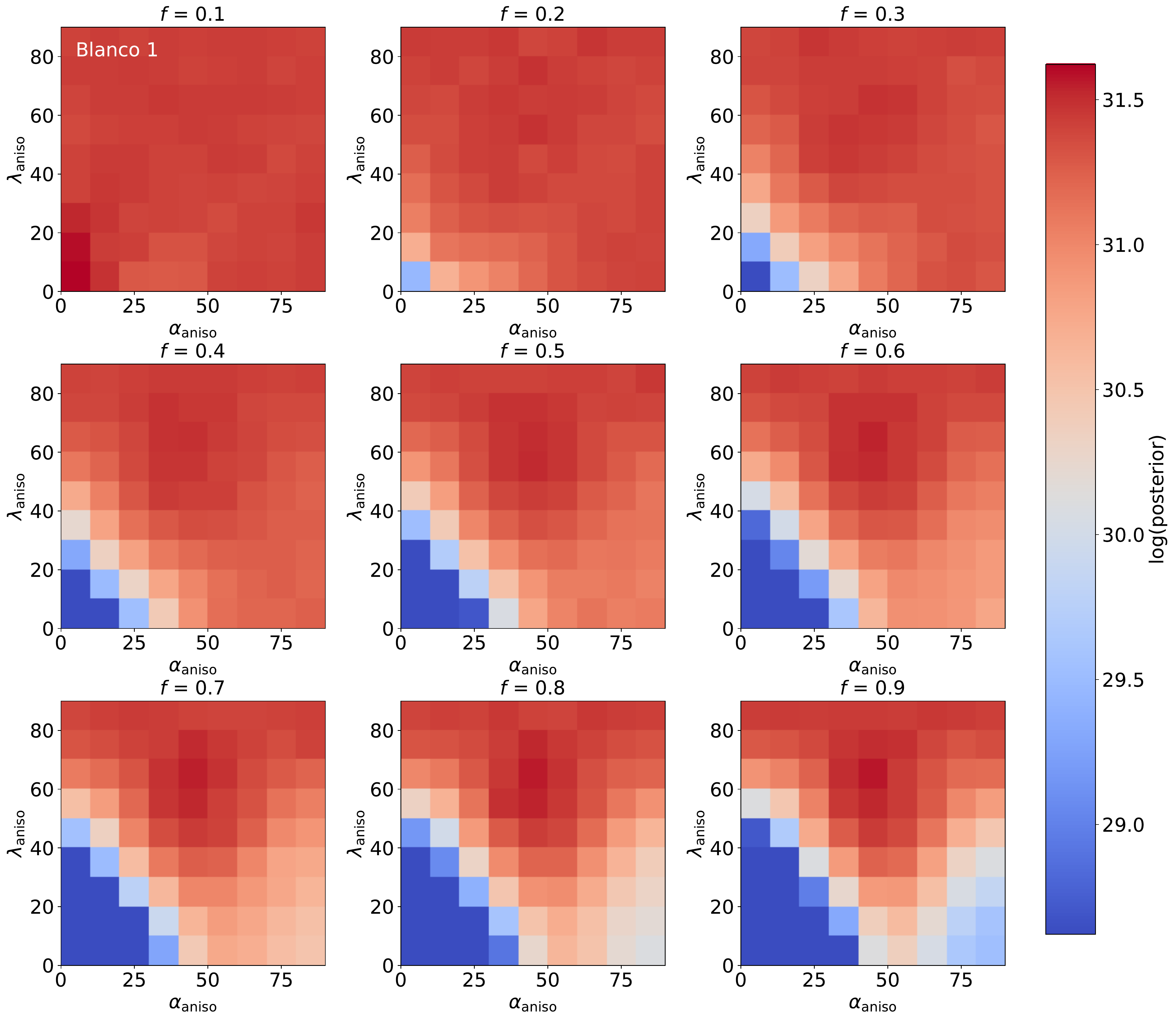}
    \caption{Blanco 1 log-posterior plots for a grid of three-parameter model values. Values of $\alpha_{\rm aniso}$ and $\lambda_{\rm aniso}$ range from 10$^{\circ}$ to 90$^{\circ}$ in 10$^{\circ}$ increments, and $f$ ranges from 0.1-0.9 in increments of 0.1. Each panel represents the results for specific value of $f$. For low values of $f$, the aligned subset's parameters are not constrained, and small amounts of systematic error can be overfit by the more complex model. As $f$ increases, the posterior plots approach their two-parameter counterpart, degenerate between isotropic and moderately aligned (see Figure \ref{fig:likelihood_plots}).}
    \label{fig:3pgridsearch_iso}
\end{figure*}

\begin{figure*}
    \centering
    \includegraphics[scale=0.375]{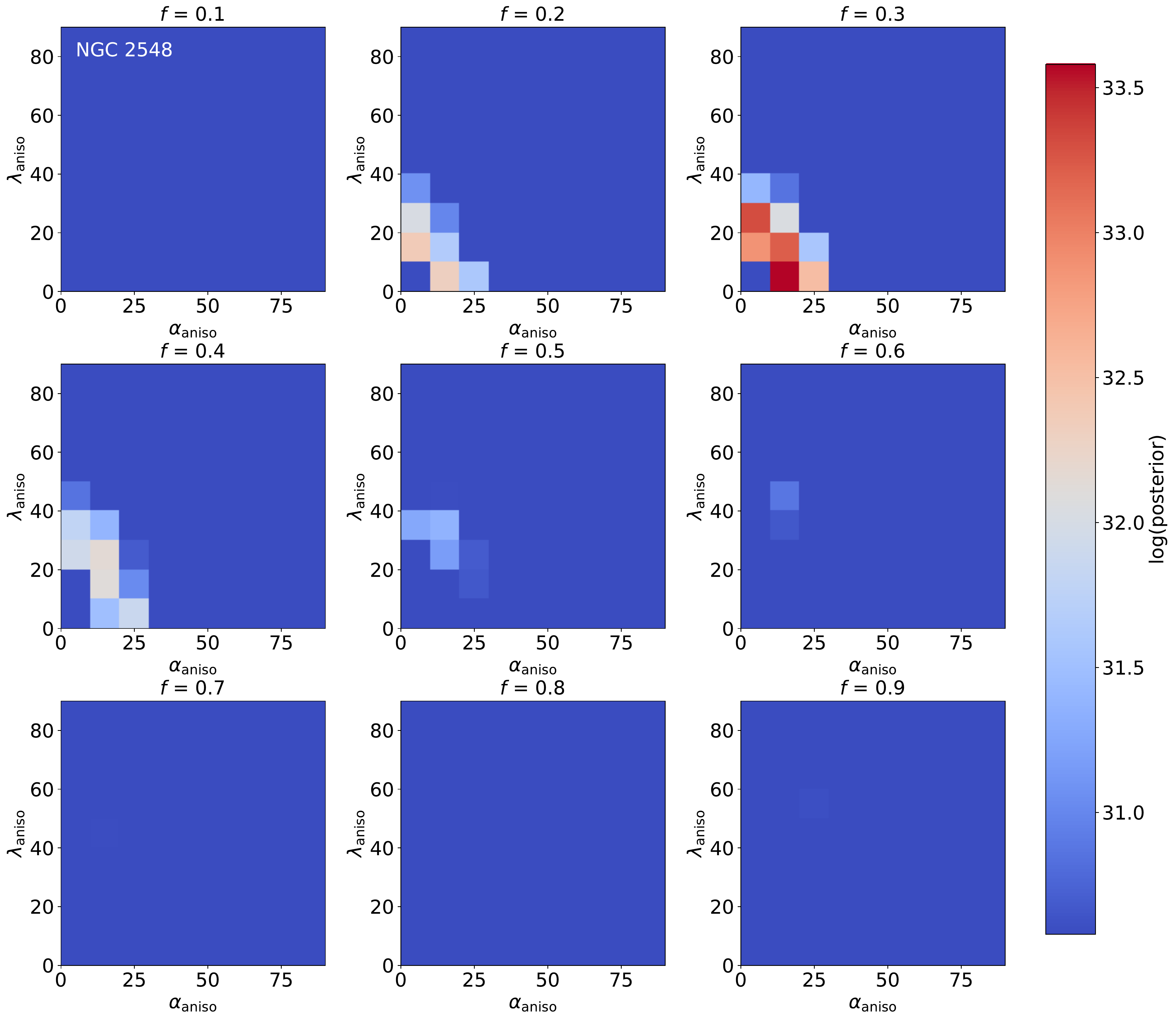}
    \caption{Same as Figure \ref{fig:3pgridsearch_iso}, but for a cluster not well-described by isotropy or moderate alignment (NGC 2548). There is a significant preference for a specific combination of three-parameter values. No panels are comparable to the two-parameter posterior plot for this cluster (see Figure \ref{fig:likelihood_plots}) because the corresponding posteriors are much smaller than the optimal three-parameter model fit.}
    \label{fig:3pgridsearch_aniso}
\end{figure*}

We plot two visualizations of this grid search in Figures \ref{fig:3pgridsearch_iso} and \ref{fig:3pgridsearch_aniso}. Each panel of these figures represents a chosen value of $f$. The posterior plots in each panel highlight the most probable values of $\alpha_{\rm aniso}$ and $\lambda_{aniso}$. The more similar each posterior plot looks to the two-parameter version shown in Figure \ref{fig:likelihood_plots}, the less distinctive that three-parameter fit is from the two-parameter fit. 

\begin{figure}
    \centering
    \includegraphics[scale=0.425]{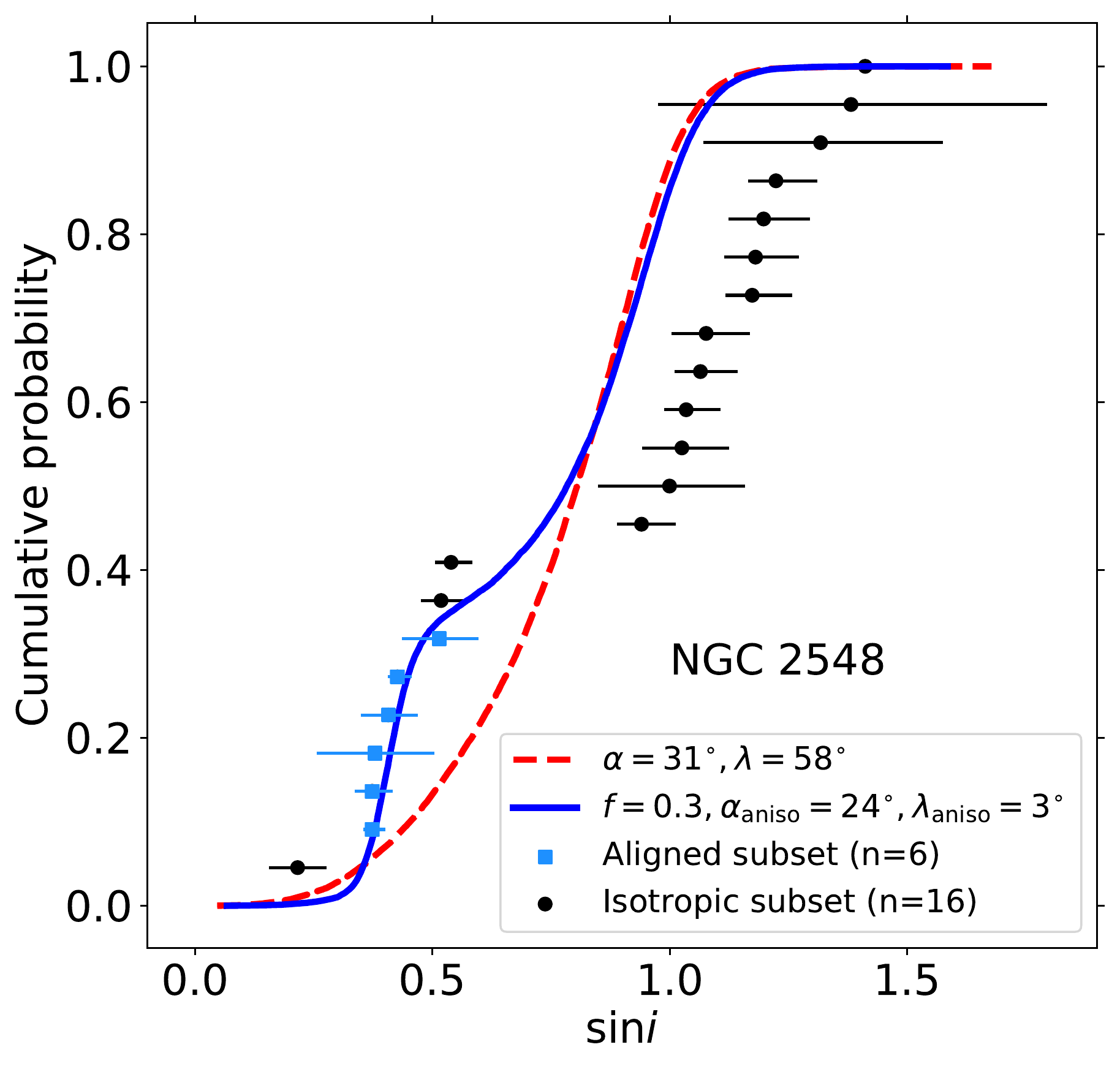}
    \caption{Three-parameter cone model fit to NGC 2548 inclinations (blue) compared to the optimal two-parameter model (red, dashed). Blue and black $\sin i$ points denote stars modeled as aligned and isotropic, respectively.}
    \label{fig:m48_3param}
\end{figure}

In one case, we noticed a significantly distinct three-parameter model solution and ran the full MCMC analysis to determine the optimal fit. We further examined the inclinations of stars in NGC 2548, since the grid search identified better three-parameter fits than the best-fitting two-parameter model, which yielded the greatest reduced $\chi^{2}$ value (6.6) among the clusters analyzed. Allowing for a subset of aligned stars, we determined a maximum-posterior three-parameter model solution of $\alpha_{\rm aniso} =$ 24$^{\circ}$, $\lambda_{\rm aniso} =$ 4$^{\circ}$, and $f =$ 0.3. The reduced $\chi^{2}$ value corresponding to this parameter combination is 3.4, and the log-posterior is 33.3. Figure \ref{fig:m48_3param} compares this cluster's two- and three-parameter model fits.

\begin{figure}
    \centering
    \includegraphics[scale=0.425]{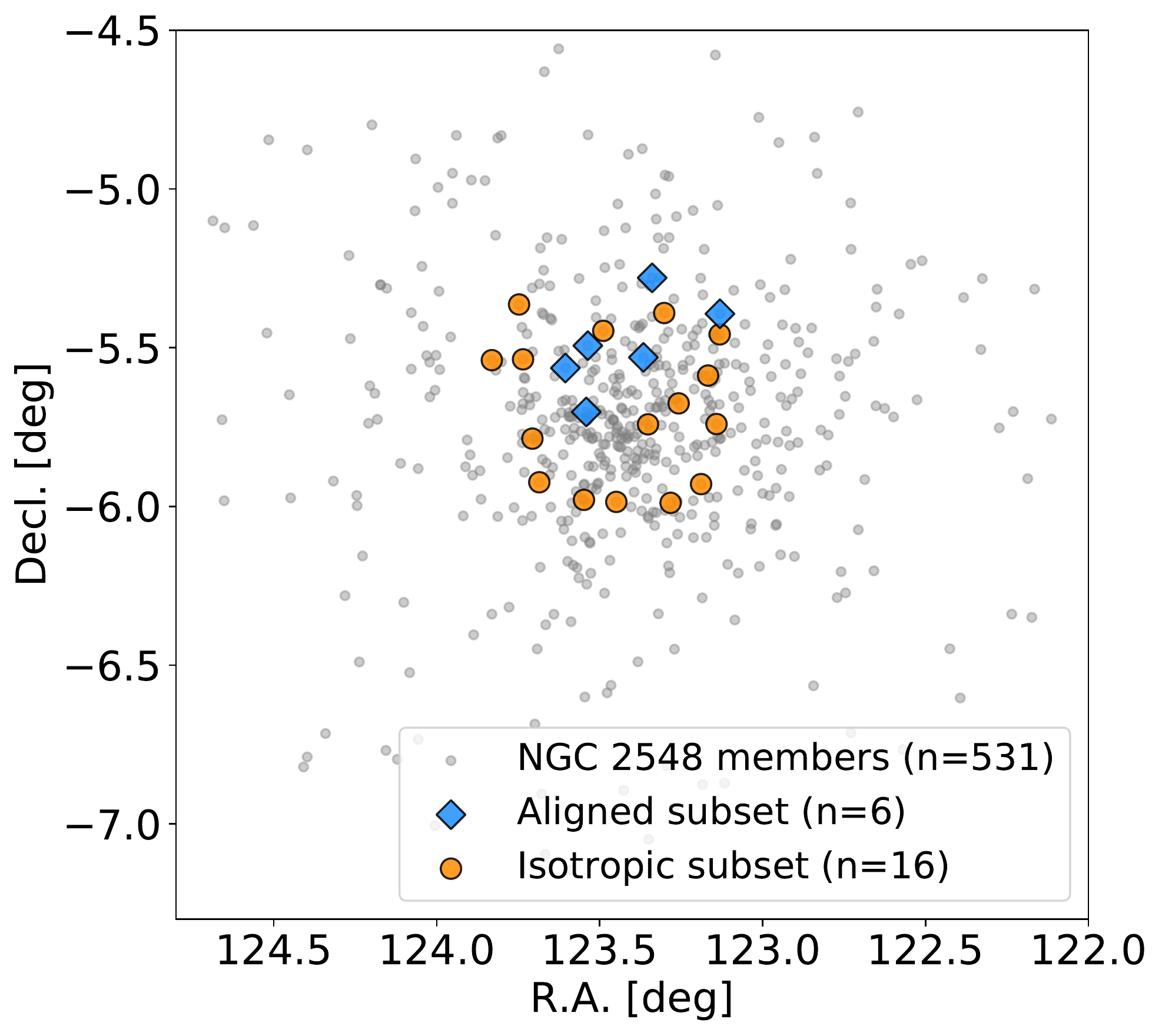}
    \caption{Cluster member positions for NGC 2548 stars (gray). Stars with $\sin i$ determinations are split into the same aligned (blue diamond) and isotropic (orange circle) subsets as shown in Figure \ref{fig:m48_3param}.}
    \label{fig:m48_radec_positions}
\end{figure}

\begin{figure}
    \centering
    \includegraphics[scale=0.425]{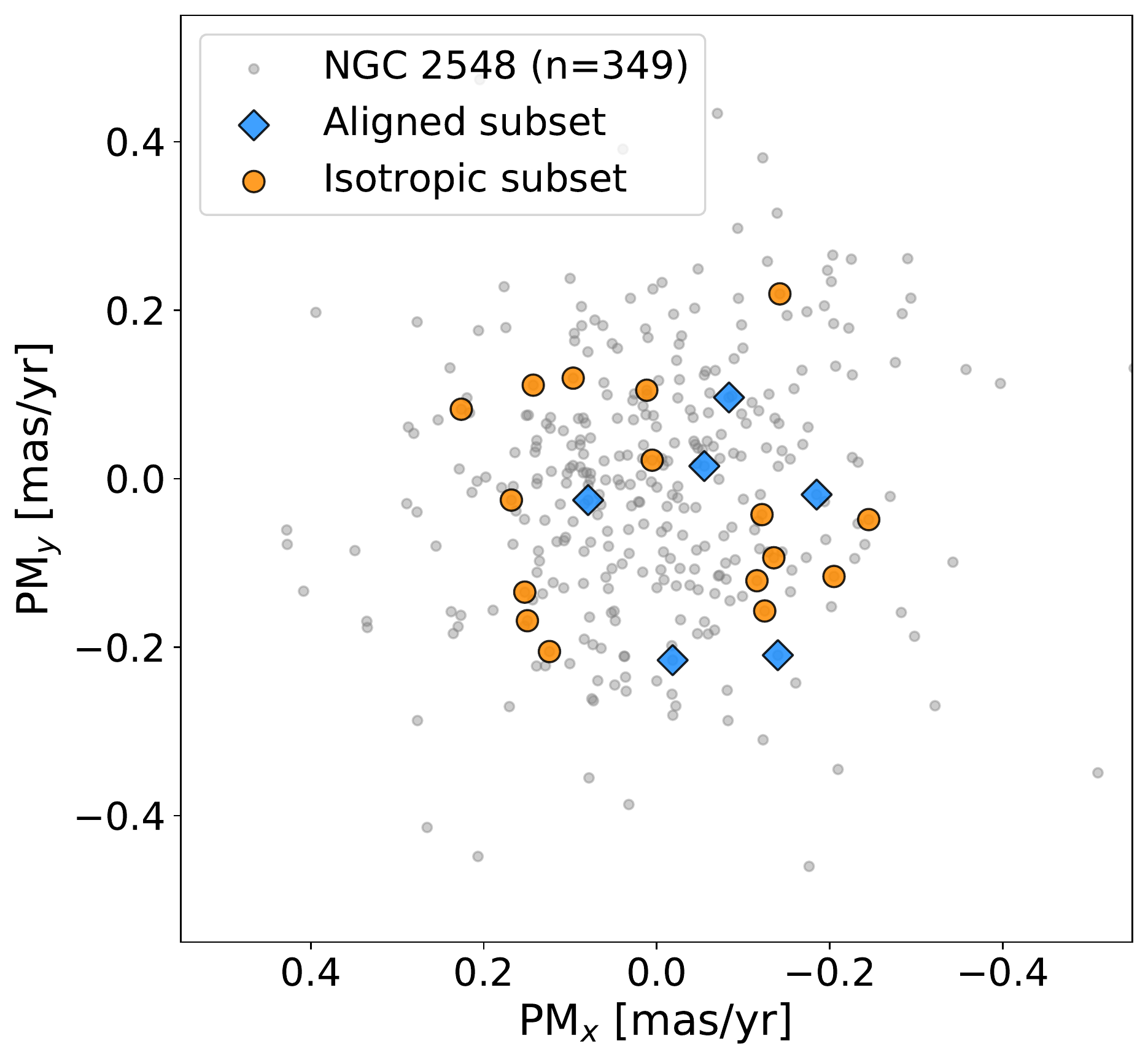}
    \caption{Internal proper motions for NGC 2548 stars on the single-star main sequence (gray). Stars with $\sin i$ determinations are plotted with the same split as described in Figure \ref{fig:m48_radec_positions}.}
    \label{fig:m48_pmxy}
\end{figure}

For NGC 2548, we selected six stars of the 22 corresponding to a mean inclination and fraction of stars consistent with our three-parameter model expectation. The inclinations of this group modeled as aligned are marked as blue in Figure \ref{fig:m48_3param}, while we label the remaining stars as an isotropic subset. As in \papertwo, we studied the positions and proper motions of the selected stars to search for groupings in their position angles and plane-of-sky positions (Figure \ref{fig:m48_radec_positions}) as well as internal proper motions (Figure \ref{fig:m48_pmxy}). 

A Kuiper test on the position angles of the selected group under the null hypothesis of random angles yielded a $p$-value of 0.08. A bootstrapped, two-sample, two-dimensional K-S test \citep[][]{ndtest1, ndtest2, numerical_recipes_3}\footnote{\texttt{ndtest} Python code written by Zhaozhou Li, available at \url{https://github.com/syrte/ndtest}}
comparing plane-of-sky positions for the selected group to those of the background group found a $p$-value of 0.09. Another two-sample 2D K-S test, this time comparing the internal proper motions of the selected and background groups, returned a $p$-value of 0.49.

\subsection{Internal Kinematics}

\begin{figure*}
    \centering
    \includegraphics[scale=0.3]{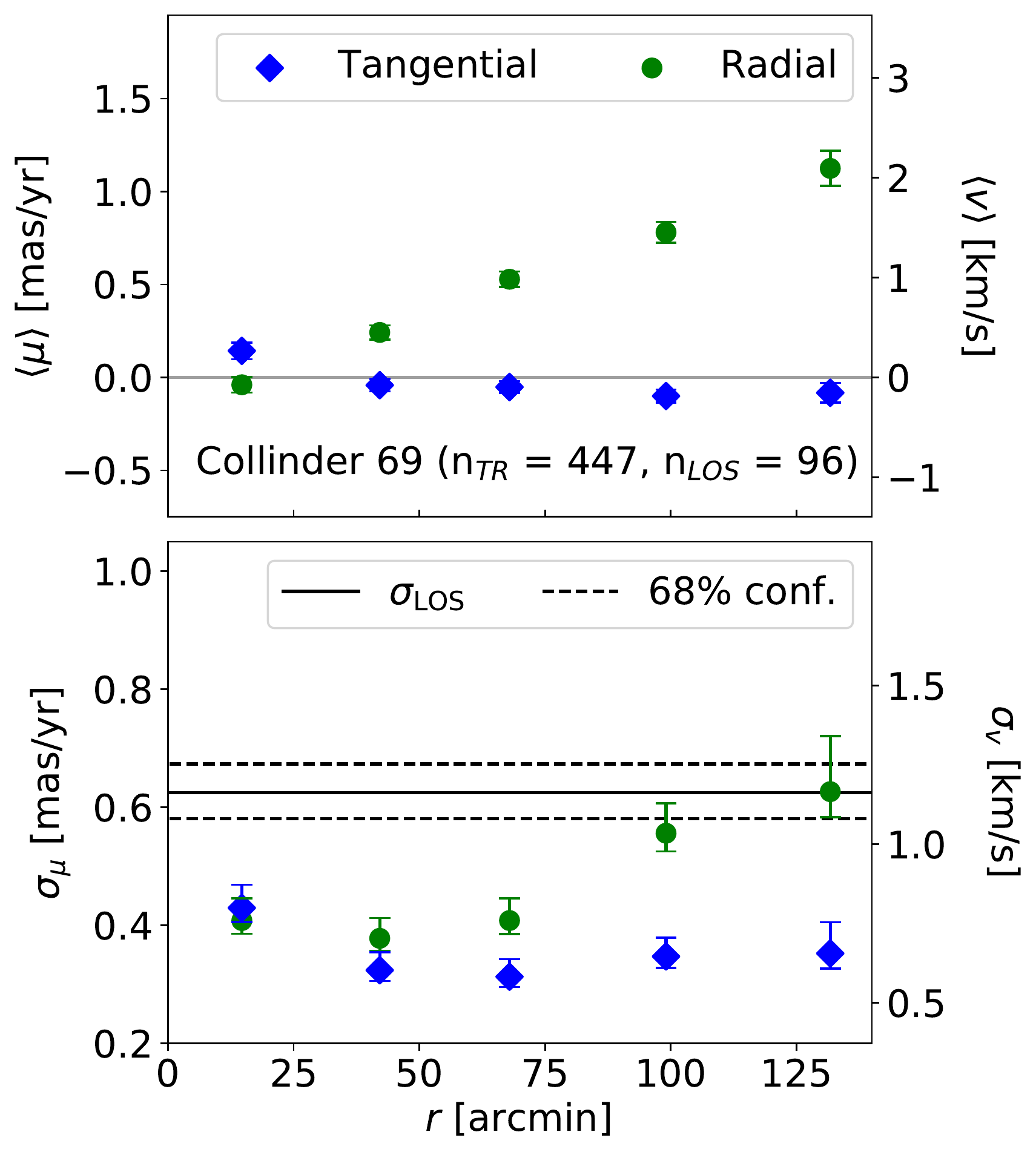}
    \includegraphics[scale=0.3]{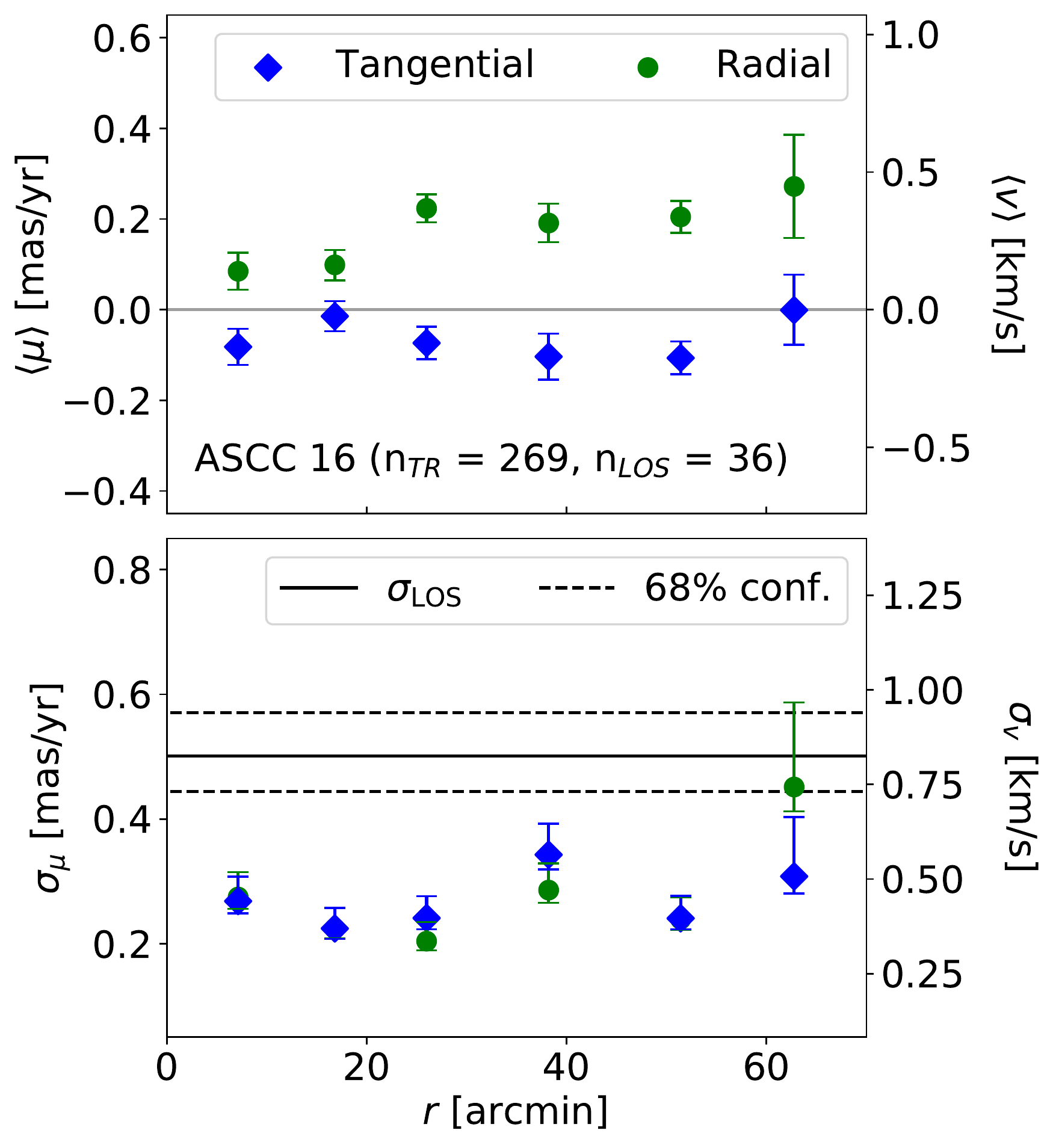}
    \includegraphics[scale=0.3]{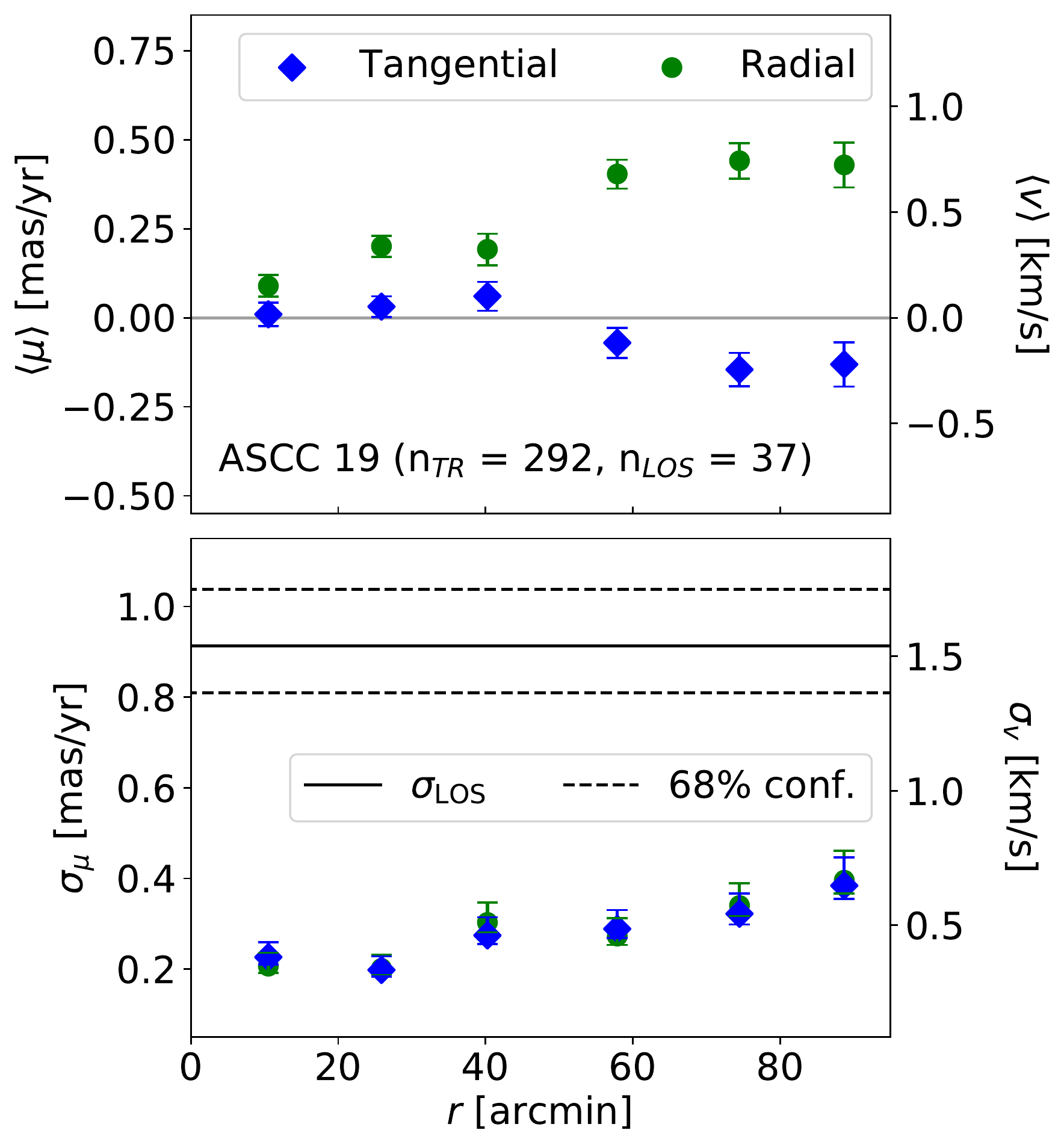}
    \includegraphics[scale=0.3]{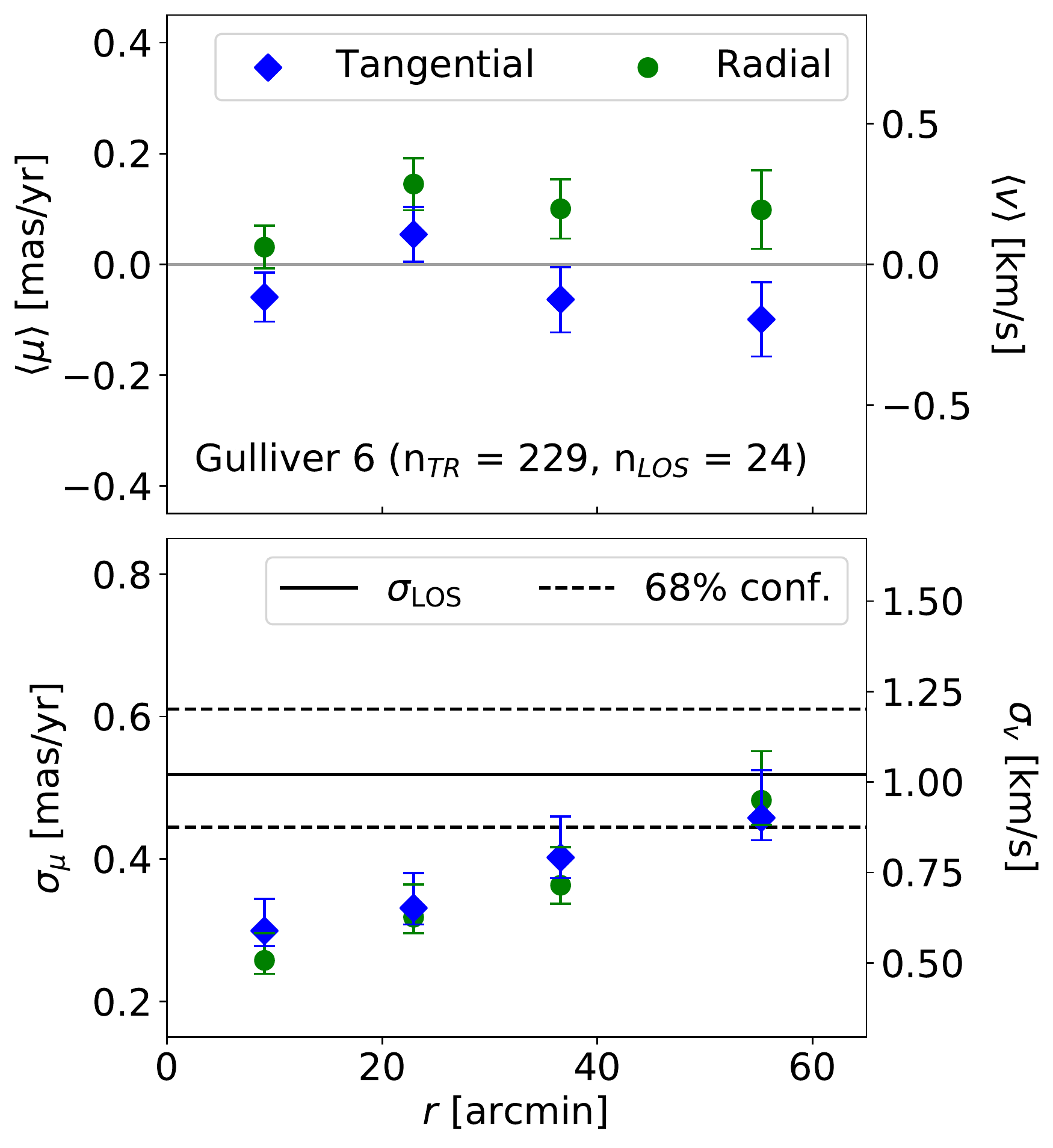}
    \includegraphics[scale=0.3]{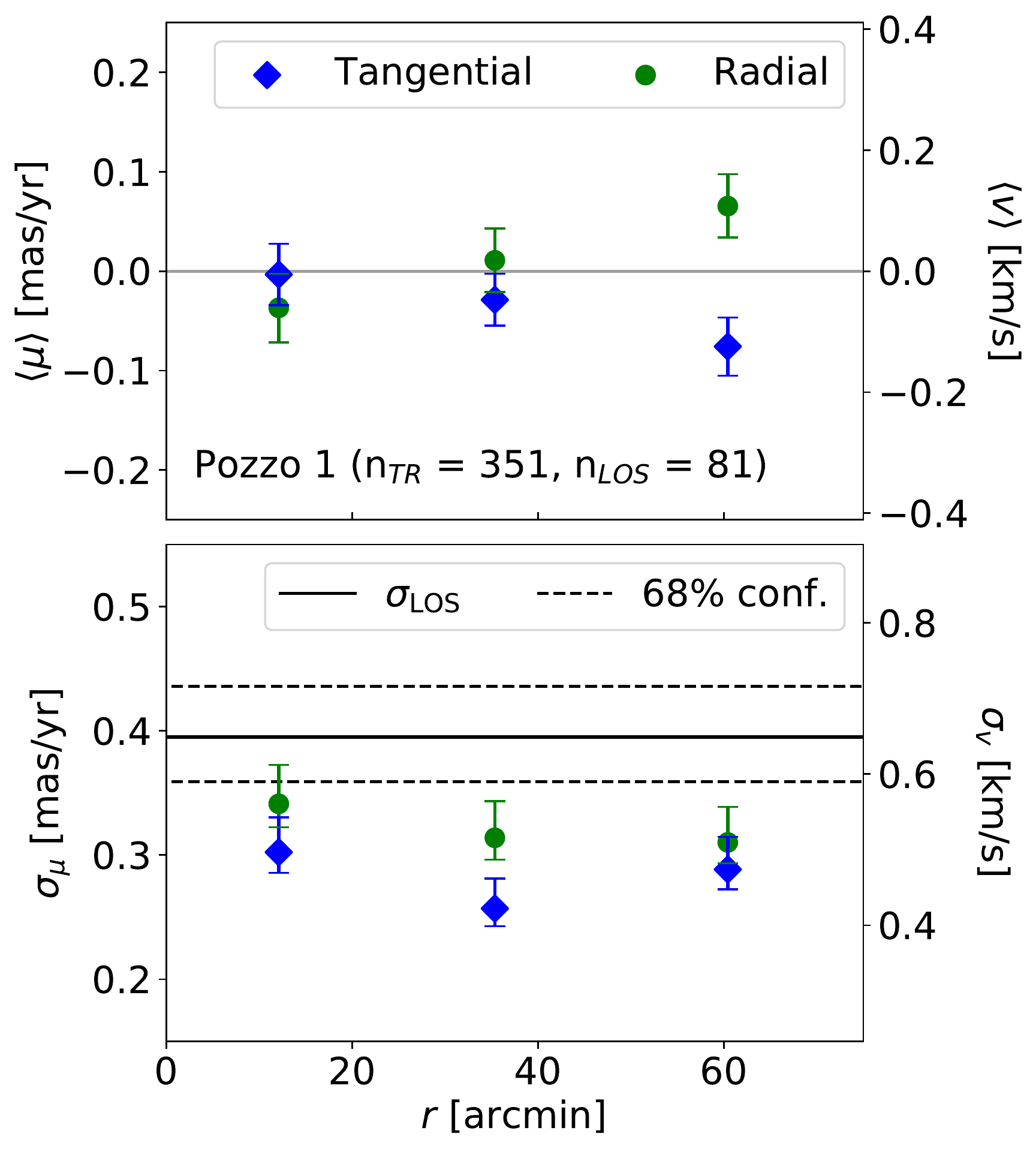}
    \includegraphics[scale=0.3]{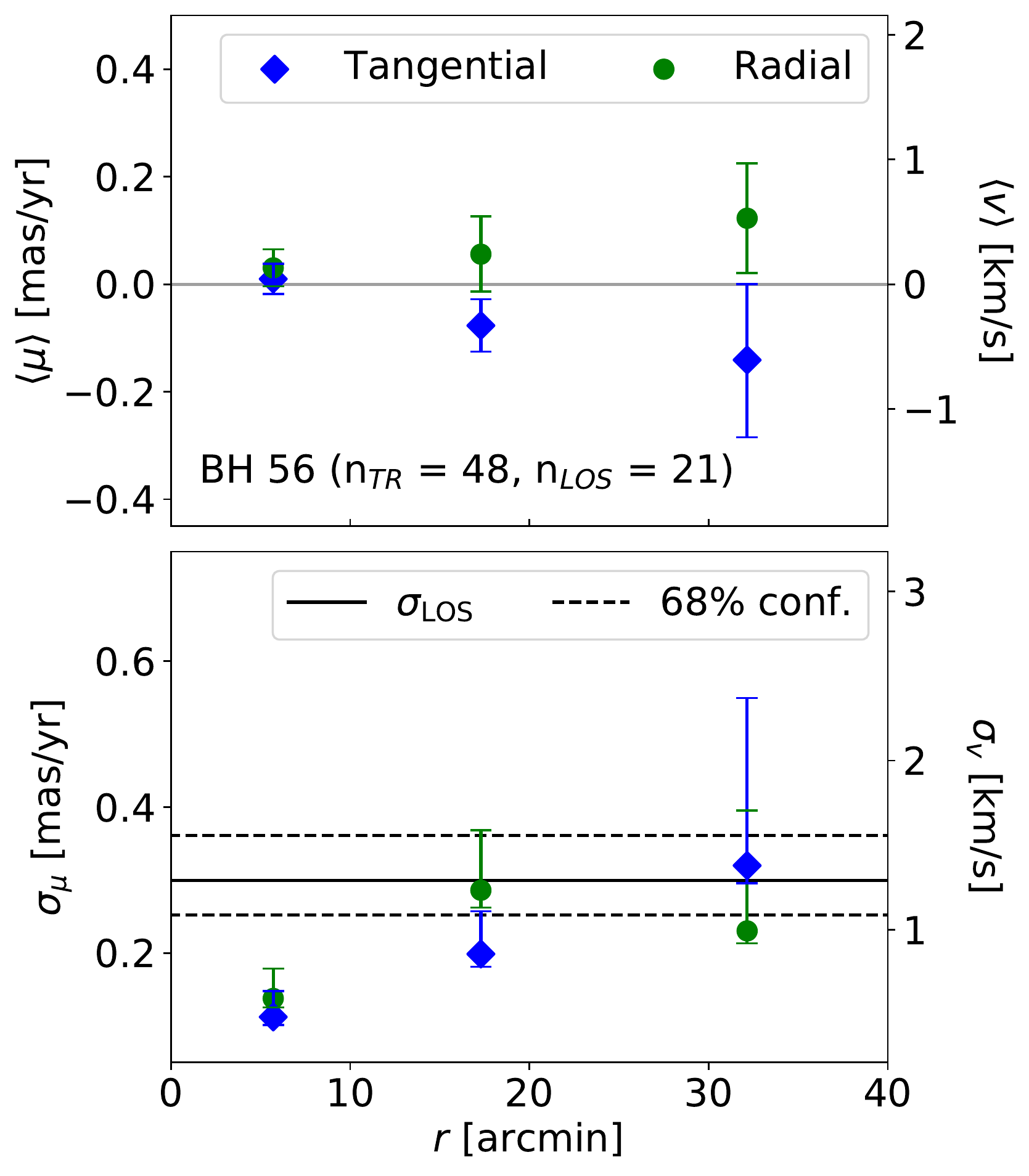}
    \includegraphics[scale=0.3]{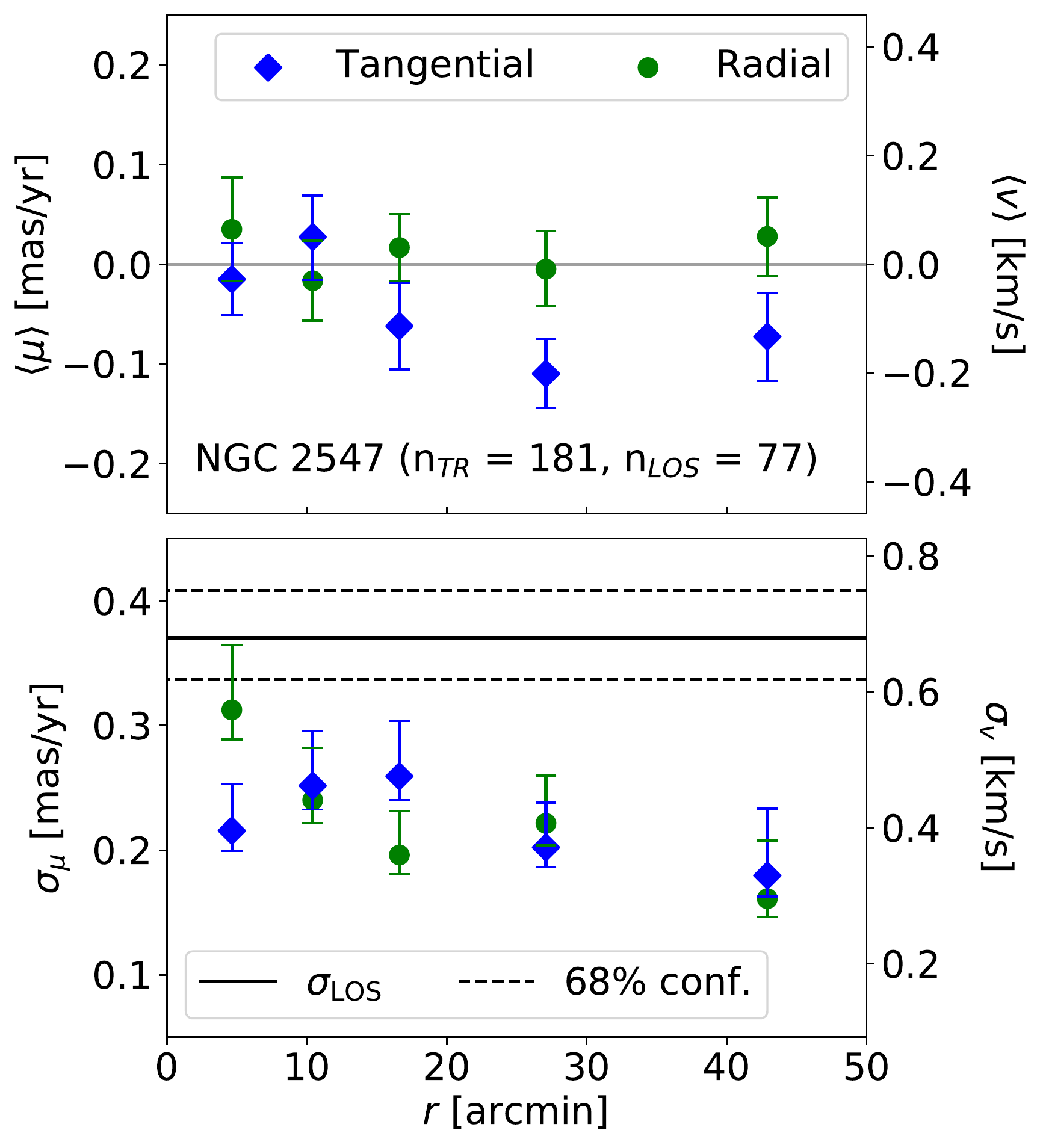}
    \includegraphics[scale=0.3]{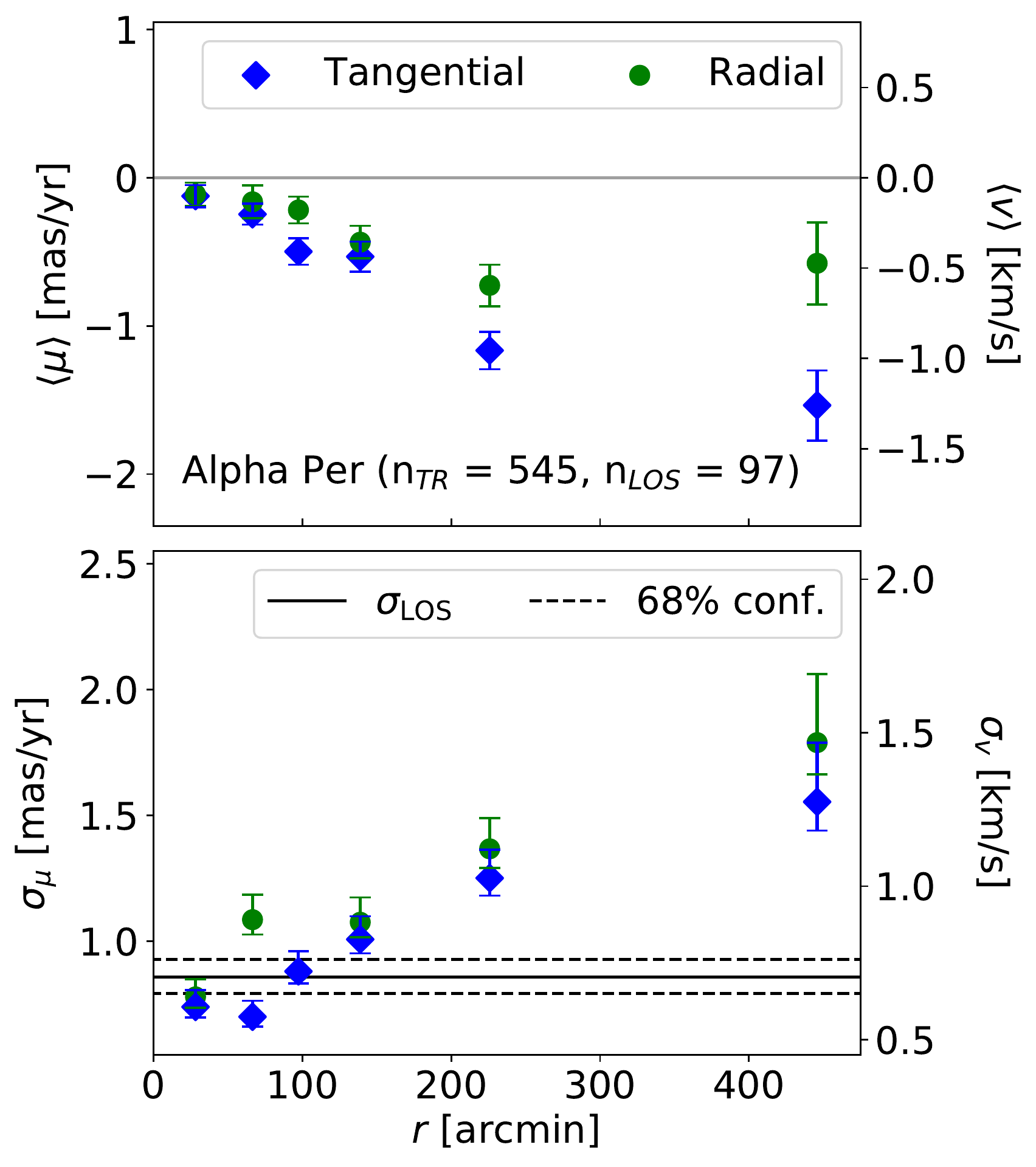}
    \includegraphics[scale=0.3]{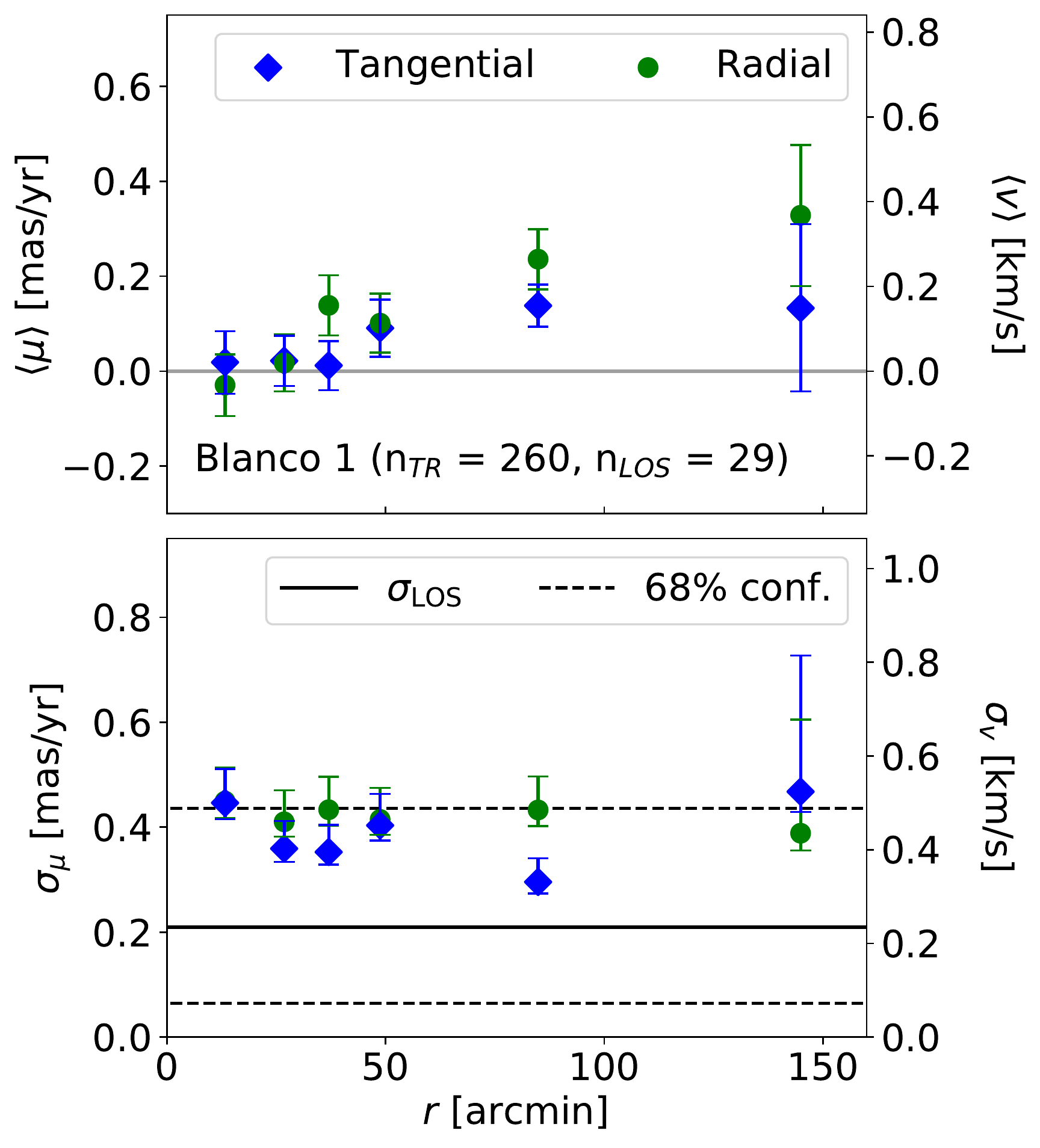}
    \includegraphics[scale=0.3]{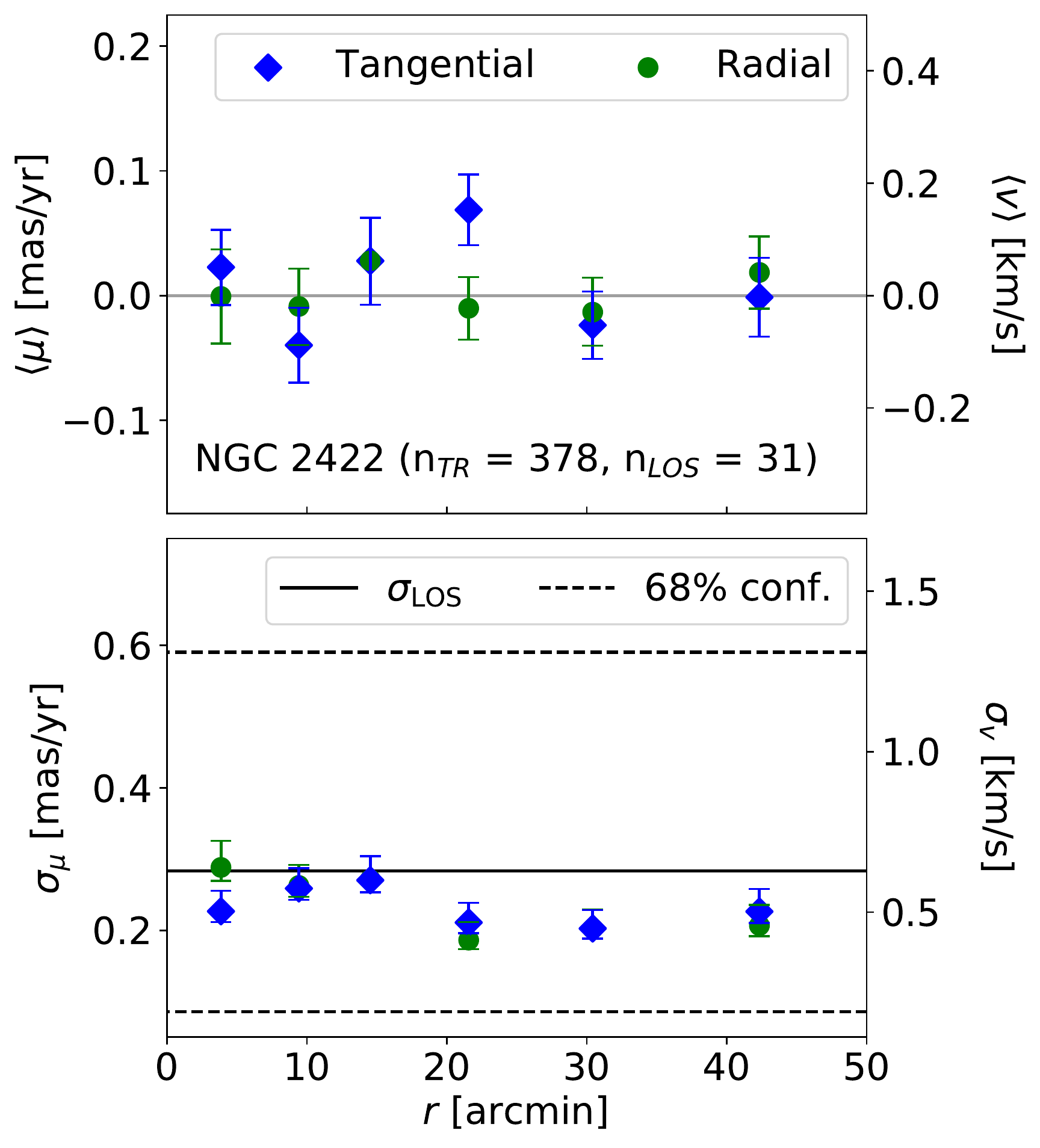}
    \includegraphics[scale=0.3]{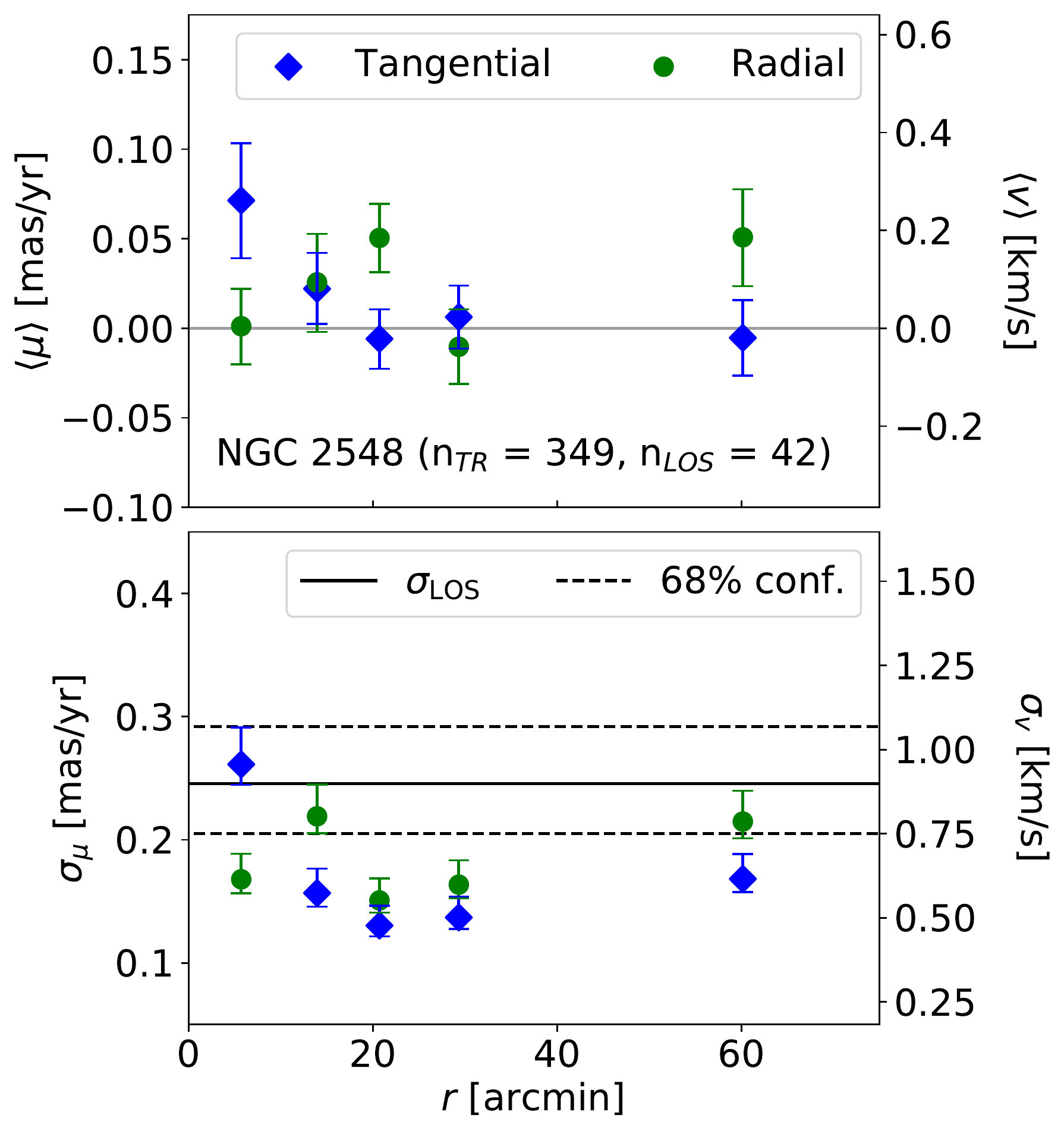}
    \caption{Tangential, radial and LOS internal kinematics for all clusters.}
    \label{fig:kinematics}
\end{figure*}

Figure \ref{fig:kinematics} plots our results for the binned internal kinematics analysis of each cluster, and Table \ref{tab:kinematics} lists our determinations of mean kinematic parameters in the tangential, radial and LOS directions. As in \papertwo, we have subtracted the predicted trend due to motion along the LOS from the radial values.

\begin{table*}
\centering
\caption{TESS-based rotation periods of stars in targeted clusters.}
\label{tab:periods}
\begin{sideways}
\begin{tabular}{cccccccc}
\hline
\hline
Gaia DR2 Source ID & TIC ID & RA [$^\circ$] & Dec [$^\circ$] & P$_{\rm mem}$ & $G_{\rm BP} - G_{\rm RP}$ & Period [d] & Clust. \\
(EDR3 Source ID) \\
\hline
3336144679286317824 & 200591845 & 85.358 & 9.198 & 0.3 & -0.1776 & 0.430 $\pm$\ 0.010 & Collinder 69 \\
(3336144679286317824) \\
3337914343251052416 & 436158760 & 83.993 & 9.532 & 1.0 & -0.062 & 0.7762 $\pm$\ 0.0083 & Collinder 69 \\
(3337914343251052416) \\
3334428715296579456 & 436153064 & 83.869 & 7.632 & 0.3 & 0.2957 & 1.94 $\pm$\ 0.45 & Collinder 69 \\
(3334428715296579456) \\
3339510731055206528 & 436098642 & 83.853 & 10.196 & 1.0 & 0.7991 & 1.53 $\pm$\ 0.11 & Collinder 69 \\
(3339510731055206528) \\
3339719260307854976 & 436013480 & 83.535 & 10.451 & 0.8 & 0.8421 & 0.862 $\pm$\ 0.010 & Collinder 69 \\
(3339719260307854976) \\
3336346882048786688 & 144667141 & 84.774 & 9.329 & 1.0 & 0.9401 & 1.397 $\pm$\ 0.019 & Collinder 69 \\
(3336346882048786688) \\
3334413906247430656 & 144610660 & 84.536 & 7.892 & 1.0 & 0.9939 & 0.6821 $\pm$\ 0.0083 & Collinder 69 \\
(3334413906247430656) \\
3337985841569936384 & 436098559 & 83.832 & 9.797 & 1.0 & 1.0265 & 1.833 $\pm$\ 0.015 & Collinder 69 \\
(3337985841569936384) \\
3338166986111279232 & 200525421 & 83.028 & 9.899 & 0.8 & 1.0687 & 2.069 $\pm$\ 0.017 & Collinder 69 \\
(3338166986111279232) \\
3334832476580230144 & 284200823 & 84.459 & 8.979 & 1.0 & 1.1148 & 2.406 $\pm$\ 0.038 & Collinder 69 \\
(3334832476580230144) \\
... & ... & ... & ... & ... & ... & ... & ... \\
\hline
\end{tabular}
\end{sideways}
\tablecomments{Table \ref{tab:periods} is published in its entirety in the machine-readable format. The electronic table includes data for NGC 2516 taken from \paperone\ for convenience.}
\end{table*}

\begin{table*}
\centering
\caption{Relevant quantities and determinations of the inclinations of open cluster members.}
\label{tab:results}
\begin{sideways}
\begin{tabular}{cccccccc}
\hline
\hline
Gaia DR2 Source ID & TIC ID & $T_{\rm eff}$ [K] & Radius [\rsun] & $v\sin i$ [km s$^{-1}$] & Period [d] & $\sin i$ & Clust. \\
(EDR3 Source ID) \\
\hline
3337914343251052416 & 436158760 & 15000 $\pm$\ 560 & 2.026 $\pm$\ 0.038 & 40.9 $\pm$\ 4.1 & 0.7762 $\pm$\ 0.0083 & 0.307$^{+0.037}_{-0.033}$ & C69 \\
(3337914343251052416) \\
3337985841569936384 & 436098559 & 5700 $\pm$\ 120 & 2.074 $\pm$\ 0.030 & 69.3 $\pm$\ 6.9 & 1.833 $\pm$\ 0.015 & 1.202$^{+0.14}_{-0.12}$ & C69 \\
(3337985841569936384) \\
3338281335319489024 & 284158151 & 5867 $\pm$\ 95 & 2.711 $\pm$\ 0.034 & 34.2 $\pm$\ 3.4 & 3.236 $\pm$\ 0.020 & 0.802$^{+0.094}_{-0.082}$ & C69 \\
(3338281335319489024) \\
3339437476093315584 & 436337191 & 5400 $\pm$\ 110 & 1.916 $\pm$\ 0.033 & 29.7 $\pm$\ 3.0 & 3.865 $\pm$\ 0.058 & 1.177$^{+0.14}_{-0.12}$ & C69 \\
(3339437476093315584) \\
3337979003982003200 & 436103228 & 5268 $\pm$\ 84 & 1.742 $\pm$\ 0.021 & 31.2 $\pm$\ 3.1 & 3.15 $\pm$\ 0.46 & 1.084$^{+0.21}_{-0.20}$ & C69 \\
(3337979003982003200) \\
3337936092965373568 & 436153907 & 4900 $\pm$\ 110 & 1.465 $\pm$\ 0.017 & 16.9 $\pm$\ 2.0 & 3.33 $\pm$\ 0.11 & 0.752$^{+0.10}_{-0.10}$ & C69 \\
(3337936092965373568) \\
3337958663016889600 & 436008528 & 5236 $\pm$\ 75 & 1.9 $\pm$\ 0.030 & 16.3 $\pm$\ 2.0 & 2.65 $\pm$\ 0.62 & 0.423$^{+0.13}_{-0.12}$ & C69 \\
(3337958663016889600) \\
3334861995892958720 & 436248202 & 4937 $\pm$\ 40 & 1.999 $\pm$\ 0.018 & 32.7 $\pm$\ 3.3 & 2.85 $\pm$\ 0.56 & 0.884$^{+0.21}_{-0.21}$ & C69 \\
(3334861995892958720) \\
3337903824874463872 & 436008466 & 4900 $\pm$\ 120 & 1.478 $\pm$\ 0.017 & 26.6 $\pm$\ 2.7 & 2.348 $\pm$\ 0.031 & 0.829$^{+0.097}_{-0.093}$ & C69 \\
(3337903824874463872) \\
3338281610197395712 & 284158141 & 4705 $\pm$\ 88 & 1.686 $\pm$\ 0.017 & 28.8 $\pm$\ 2.9 & 2.586 $\pm$\ 0.083 & 0.867$^{+0.11}_{-0.10}$ & C69 \\
(3338281610197395712) \\
... & ... & ... & ... & ... & ... & ... & ... \\
\hline
\end{tabular}
\end{sideways}
\tablecomments{Table \ref{tab:results} is published in its entirety in the machine-readable format. This table includes updated data for NGC 2516, Pleiades, Praesepe, and M35 data (Papers \citetalias{paper1_healy} and \citetalias{paper2_healy}) due to the revised procedures detailed in this paper.}
\end{table*}

\begin{table*}[]
\begin{center}
    \caption{Inclination distribution parameters based on the \citet[][]{jackson2010} two-parameter cone model. We list the number of rotation period and $\sin i$ values for each cluster, the values corresponding to PPD peaks for the spread half-angle $\lambda$ and mean inclination $\alpha$, and their 95\% confidence ranges. We also report the reduced $\chi^{2}$ and logarithm of the unnormalized posteriors corresponding to the PPD-peak parameters for each cluster. The final column lists the classification corresponding to the best-fitting model for each cluster's spin-axis distribution: isotropic/moderately aligned (I), unconstrained (U), or partially aligned (P).} 
    \label{tab:incl_distribs}
    \begin{tabular}{c|c|c|c|c|c|c|c}
    \hline
    \hline
    Cluster & N$_{\rm per}$ & N$_{\sin i}$ & $\lambda$ PPD peak (95\% conf. range) & $\alpha$ PPD peak (95\% conf. range) & $\chi_{\rm red}^{2}$ & $\ln({\rm post.})$ & Class. \\
    \hline
        
        Collinder 69 & 202 & 26 &
        86$^{\circ}$
        ($\lambda > 22^{\circ}$) & 54$^{\circ}$
        ($5^{\circ} < \alpha < 62^{\circ}$) & 1.1 & 24.2 & I \\
        
        ASCC 16 & 129 & 10 &
        86$^{\circ}$
        ($\lambda > 6^{\circ}$) & 88$^{\circ}$
        ($16^{\circ} < \alpha < 88^{\circ}$) & 1.0 & 9.1 & U \\
        
        ASCC 19 & 157 & 17 &
        89$^{\circ}$
        ($\lambda > 15^{\circ}$) & 53$^{\circ}$
        ($9^{\circ} < \alpha < 85^{\circ}$) & 0.6 & 18.0 & I \\
        
        Gulliver 6 & 121 & 10 & 
        59$^{\circ}$
        ($25^{\circ} < \lambda < 81^{\circ}$) &
        37$^{\circ}$
        ($4^{\circ} < \alpha < 52^{\circ}$) & 0.8 & 12.3 & U \\
        
        Pozzo 1 & 110 & 13 &
        90$^{\circ}$
        ($\lambda > 11^{\circ}$) & 67$^{\circ}$
        ($13^{\circ} < \alpha < 87^{\circ}$) & 0.6 & 11.4 & U \\
        
        BH 56 & 45 & 11 &
        89$^{\circ}$
        ($\lambda > 5^{\circ}$) & 85$^{\circ}$
        ($19^{\circ} < \alpha < 88^{\circ}$) & 0.9 & 9.8 & U \\
        
        NGC 2547 & 79 & 27 &
        89$^{\circ}$
        ($\lambda > 45^{\circ}$) & 51$^{\circ}$
        ($9^{\circ} < \alpha < 79^{\circ}$) & 1.7 & 45.1 & I \\
        
        Alpha Per & 294 & 30 &
        89$^{\circ}$
        ($\lambda > 6^{\circ}$) & 69$^{\circ}$
        ($12^{\circ} < \alpha < 87^{\circ}$) & 0.4 & 25.0 & I \\
        
        Blanco 1 & 187 & 29 &
        86$^{\circ}$
        ($\lambda > 21^{\circ}$) & 57$^{\circ}$
        ($9^{\circ} < \alpha < 85^{\circ}$) & 1.3 & 31.5 & I \\
        
        NGC 2422 & 159 & 21 &
        50$^{\circ}$
        ($11^{\circ} < \lambda < 71^{\circ}$) & 73$^{\circ}$
        ($62^{\circ} < \alpha < 88^{\circ}$) & 0.5 & 39.2 & U \\
        
        NGC 2548 & 135 & 22 &
        58$^{\circ}$
        ($35^{\circ} < \lambda < 74^{\circ}$) & 
        31$^{\circ}$
        ($10^{\circ} < \alpha < 45^{\circ}$) & 6.6 & 30.6 & P \\
        
    \hline
        NGC 2516 & 158 & 44 &
        90$^{\circ}$
        ($\lambda > 34^{\circ}$) & 53$^{\circ}$
        ($9^{\circ} < \alpha < 84^{\circ}$) & 0.2 & 72.0 & I \\
        
        Pleiades & 759 & 39 &
        90$^{\circ}$
        ($\lambda > 29^{\circ}$) & 53$^{\circ}$
        ($9^{\circ} < \alpha < 83^{\circ}$) & 0.4 & 55.6 & I \\
        
        Praesepe & 947 & 36 &
        87$^{\circ}$
        ($\lambda > 28^{\circ}$) & 52$^{\circ}$
        ($9^{\circ} < \alpha < 83^{\circ}$) & 0.8 & 41.8 & I \\
        
        M35 & 1146 & 69 &
        59$^{\circ}$
        ($37^{\circ} < \lambda < 81^{\circ}$) & 
        36$^{\circ}$
        ($6^{\circ} < \alpha < 50^{\circ}$) & 2.3 & 100.6 & P \\
        
    \hline
    \end{tabular}
\end{center}
\end{table*}

\begin{table*}
\begin{center}
\caption{Mean determinations of internal kinematics for each cluster.}
\label{tab:kinematics}
\begin{tabular}{cccccc}
\hline
\hline
 Cluster & Direction & $\theta_{\rm c}$ [$^{\circ}$] & $v_{0}$ [km s$^{-1}$] & $v_{\rm rot}$ [km s$^{-1}$] & $\sigma$ [km s$^{-1}$] \\
\hline

& Tangential & --- & $-0.037 \pm 0.033$ & --- & $0.685 \pm 0.024$ \\
Collinder 69 & Radial & --- & $0.842 \pm 0.046$ & --- & $ 0.953 \pm 0.033$ \\
& LOS & $110 \pm 10$ & $25.09 \pm 0.14$ & $0.90 \pm 0.23$ &
$1.162 \pm 0.086$ \\
\hline

& Tangential & --- & $-0.115 \pm 0.028$ & --- & $0.452 \pm 0.021$ \\
ASCC 16 & Radial & --- & $0.278 \pm 0.028$ & --- & $ 0.445 \pm 0.020$ \\
& LOS & $200 \pm 40$ & $20.33 \pm 0.24$ & $0.29^{+0.23}_{-0.19}$ &
$0.83^{+0.12}_{-0.09}$ \\
\hline

& Tangential & --- & $-0.065 \pm 0.030$ & --- & $0.502 \pm 0.022$ \\
ASCC 19 & Radial & --- & $0.485 \pm 0.029$ & --- & $ 0.500 \pm 0.021$ \\
& LOS & $340 \pm 40$ & $19.88 \pm 0.26$ & $0.67 \pm 0.35$ &
$1.54 \pm 0.19$ \\
\hline

& Tangential & --- & $-0.079 \pm 0.052$ & --- & $0.739 \pm 0.038$ \\
Gulliver 6 & Radial & --- & $0.268 \pm 0.051$ & --- & $ 0.762 \pm 0.037$ \\
& LOS & $260 \pm 90$ & $32.47 \pm 0.21$ & $0.25^{+0.27}_{-0.17}$ &
$1.02^{+0.18}_{-0.14}$ \\
\hline

& Tangential & --- & $-0.021 \pm 0.027$ & --- & $0.496 \pm 0.019$ \\
Pozzo 1 & Radial & --- & $0.042 \pm 0.028$ & --- & $ 0.526 \pm 0.020$ \\
& LOS & $110 \pm 90$ & $17.811 \pm 0.085$ & $0.090^{+0.096}_{-0.063}$ &
$0.651 \pm 0.065$ \\
\hline

& Tangential & --- & $-0.22 \pm 0.13$ & --- & $0.89 \pm 0.10$ \\
BH 56 & Radial & --- & $0.25 \pm 0.15$ & --- & $ 1.01 \pm 0.11$ \\
& LOS & $100 \pm 30$ & $23.49 \pm 0.31$ & $0.86 \pm 0.48$ &
$1.29 \pm 0.24$ \\
\hline

& Tangential & --- & $-0.080 \pm 0.033$ & --- & $0.431 \pm 0.024$ \\
NGC 2547 & Radial & --- & $0.017 \pm 0.034$ & --- & $ 0.446 \pm 0.025$ \\
& LOS & $310 \pm 50$ & $11.820 \pm 0.086$ & $0.16 \pm 0.12$ &
$0.679 \pm 0.065$ \\
\hline

& Tangential & --- & $-0.491 \pm 0.038$ & --- & $0.902 \pm 0.027$ \\
Alpha Per & Radial & --- & $-0.290 \pm 0.041$ & --- & $ 0.958 \pm 0.029$ \\
& LOS & $0 \pm 70$ & $-1.814 \pm 0.073$ & $0.090^{+0.093}_{-0.062}$ &
$0.704 \pm 0.056$ \\
\hline

& Tangential & --- & $0.064 \pm 0.030$ & --- & $0.472 \pm 0.021$ \\
Blanco 1 & Radial & --- & $0.112 \pm 0.030$ & --- & $ 0.485 \pm 0.022$ \\
& LOS & $180 \pm 110$ & $4.31 \pm 0.24$ & $0.24^{+0.26}_{-0.17}$ &
$0.24^{+0.25}_{-0.16}$ \\
\hline

& Tangential & --- & $0.016 \pm 0.028$ & --- & $0.531 \pm 0.021$ \\
NGC 2422 & Radial & --- & $0.001 \pm 0.029$ & --- & $ 0.553 \pm 0.021$ \\
& LOS & $10 \pm 70$ & $36.00 \pm 0.58$ & $0.81^{+0.73}_{-0.56}$ &
$0.63^{+0.68}_{-0.44}$ \\
\hline

& Tangential & --- & $0.066 \pm 0.036$ & --- & $0.662 \pm 0.026$ \\
NGC 2548 & Radial & --- & $0.078 \pm 0.038$ & --- & $ 0.687 \pm 0.027$ \\
& LOS & $0 \pm 90$ & $9.75 \pm 0.18$ & $0.21^{+0.22}_{-0.14}$ &
$0.90 \pm 0.16$ \\
\hline

\end{tabular}
\end{center}
\tablecomments{$\theta_{\rm c}$ is the position angle of each cluster's LOS rotation axis; $v_{0}$ is the velocity in the specified direction, $v_{\rm rot}$ is the LOS rotation velocity, and $\sigma$ is the velocity dispersion in the specified direction.}
\end{table*}

\begin{figure*}
    \centering
    \includegraphics[scale=0.3]{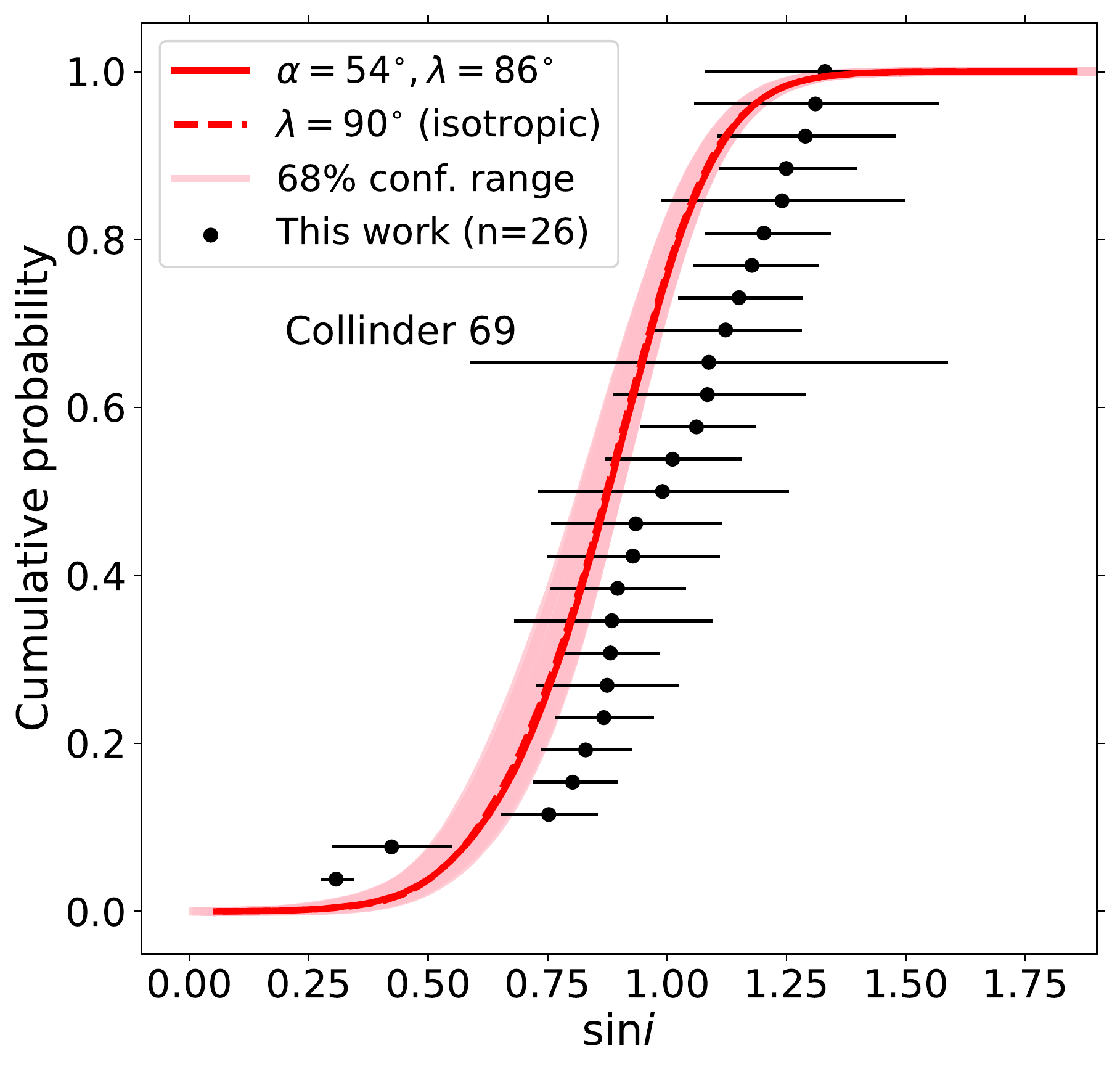}
    \includegraphics[scale=0.3]{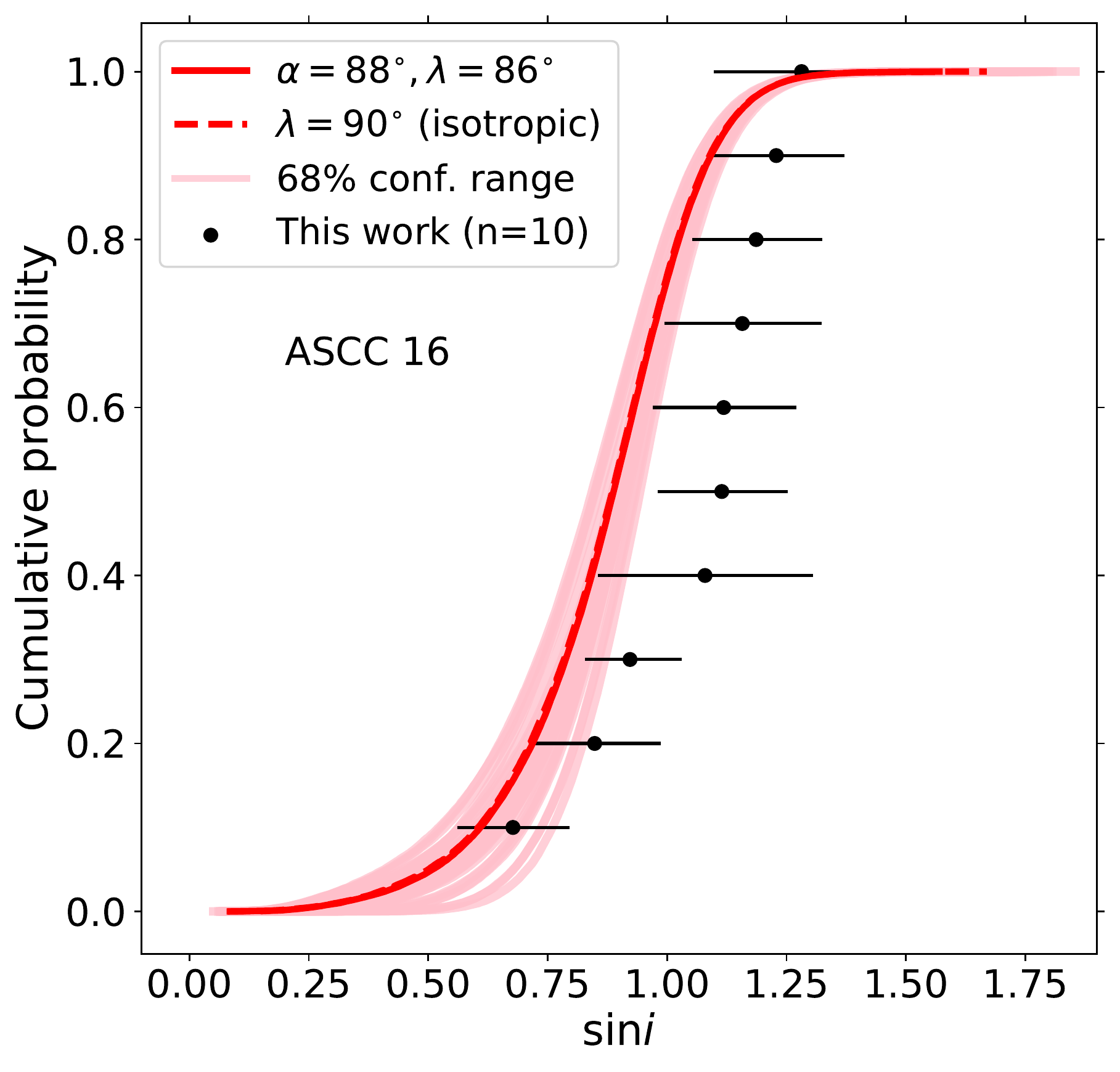}
    \includegraphics[scale=0.3]{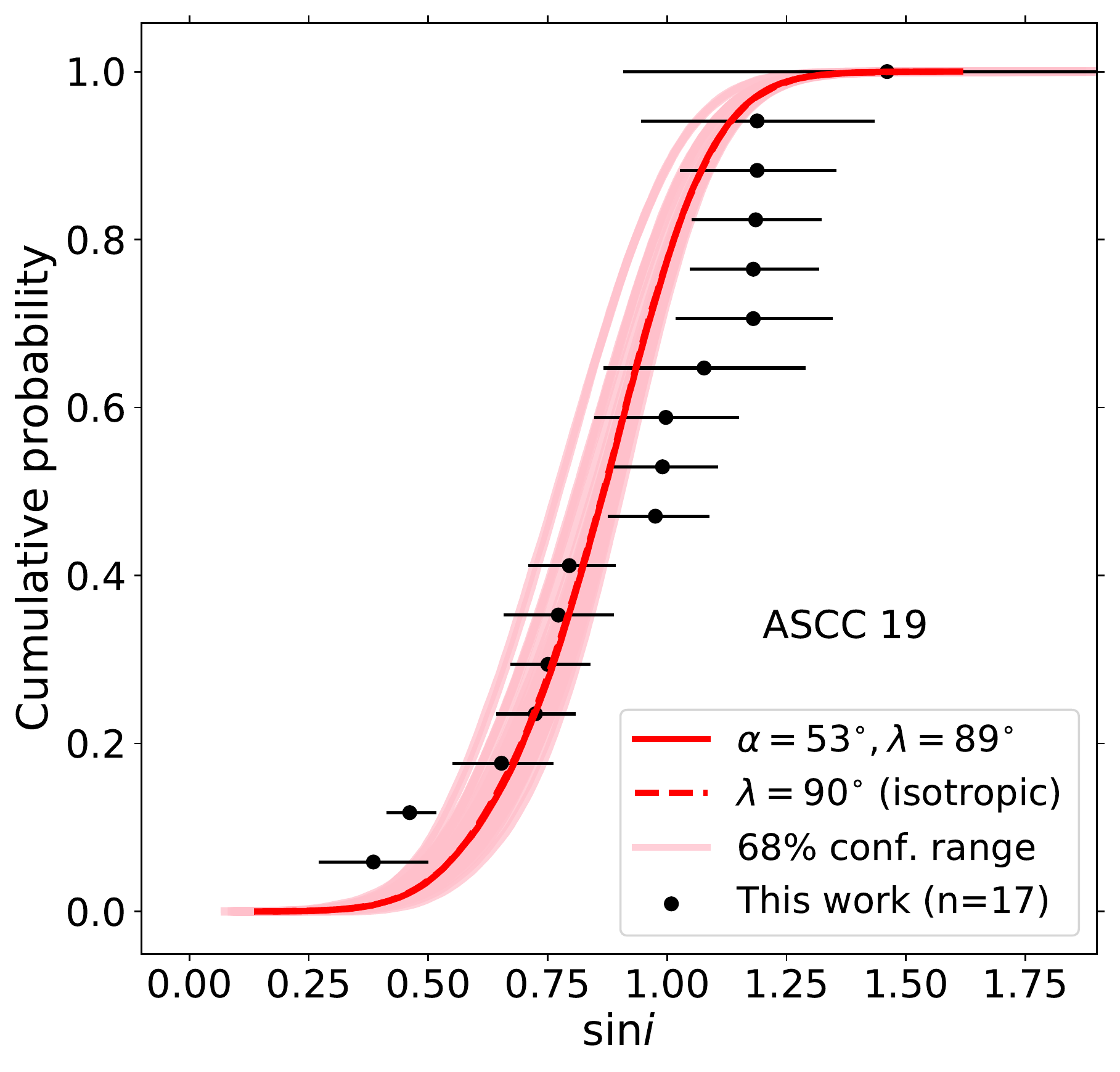}
    \includegraphics[scale=0.3]{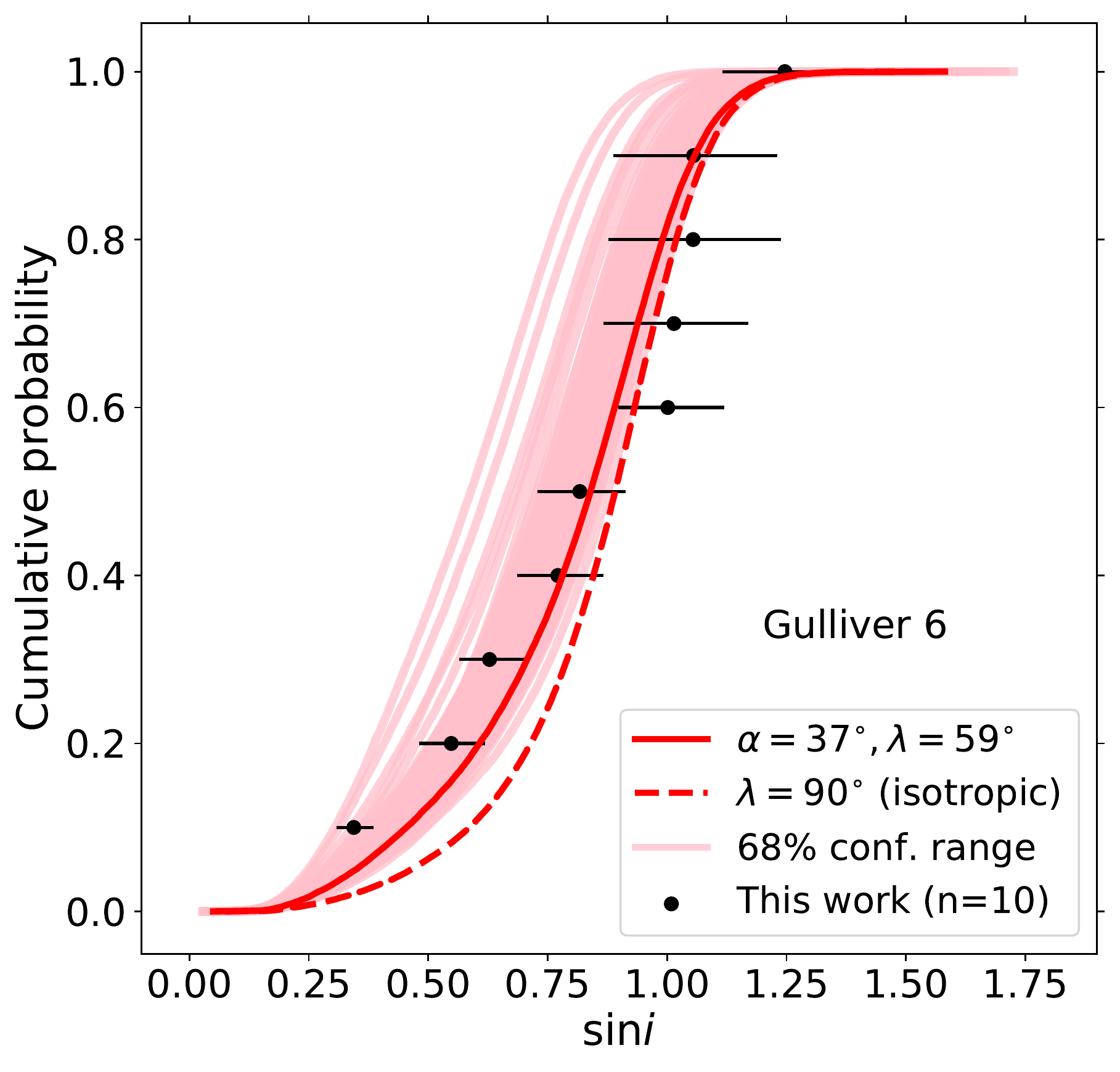}
    \includegraphics[scale=0.3]{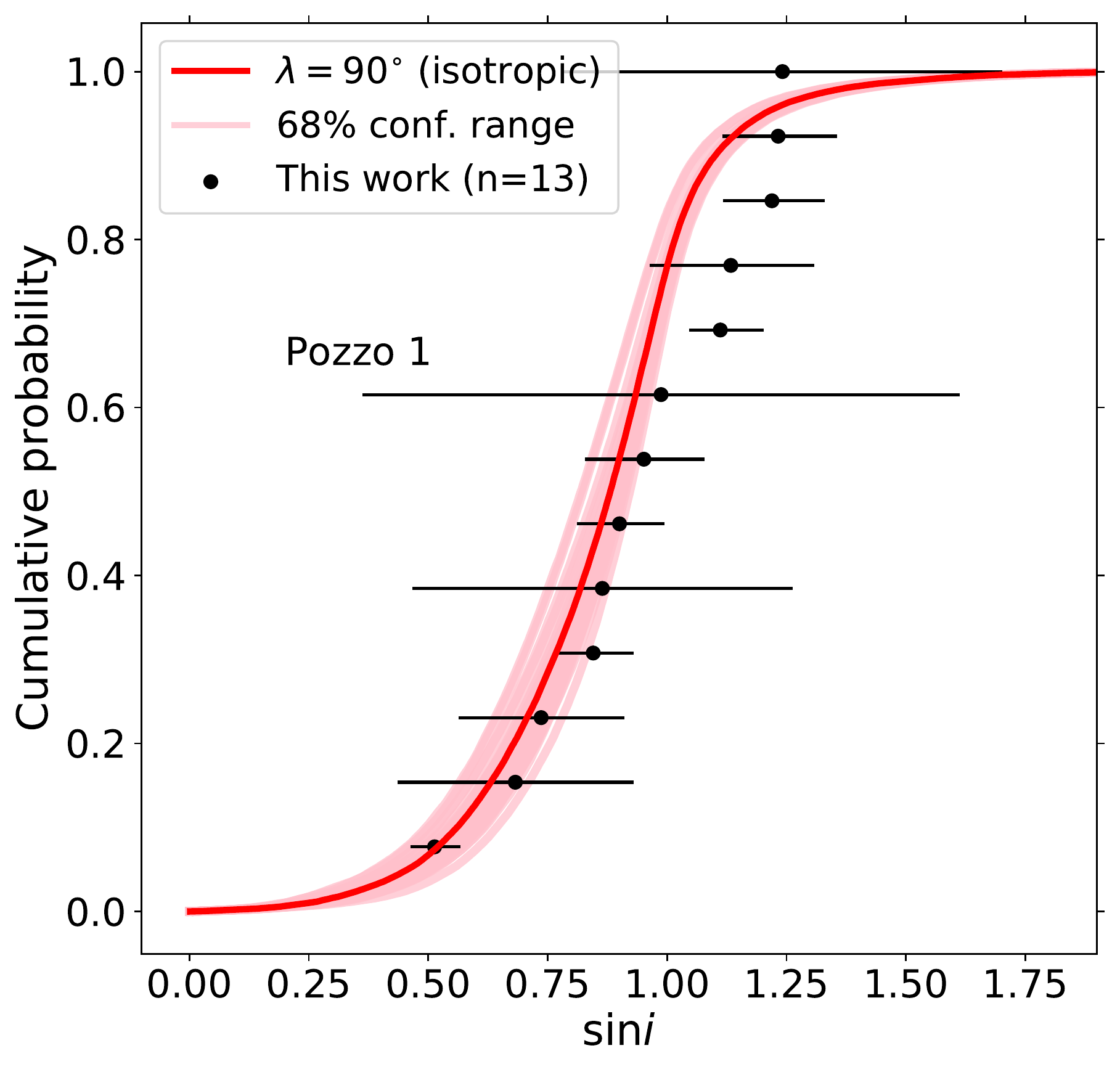}
    \includegraphics[scale=0.3]{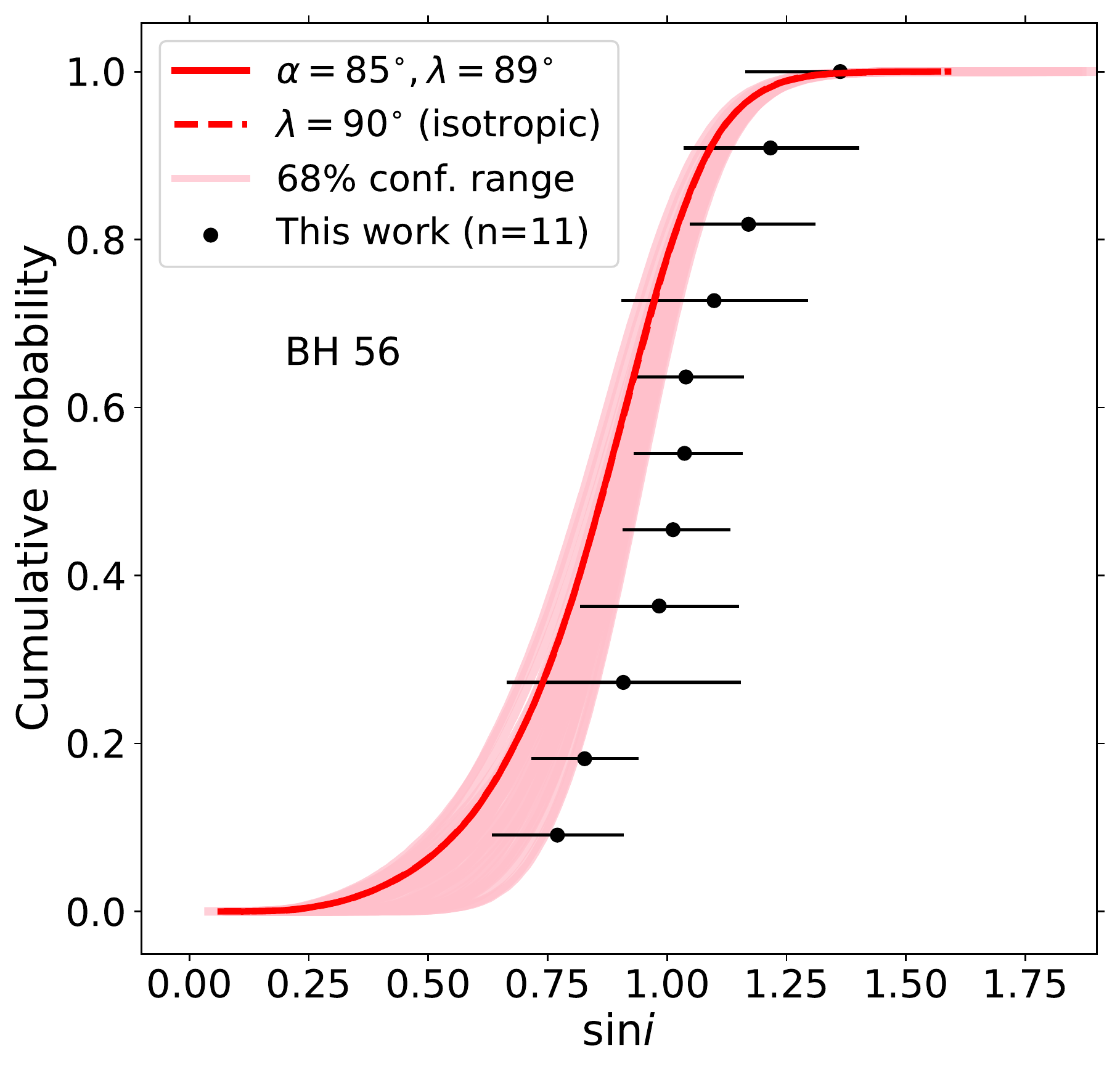}
    \includegraphics[scale=0.3]{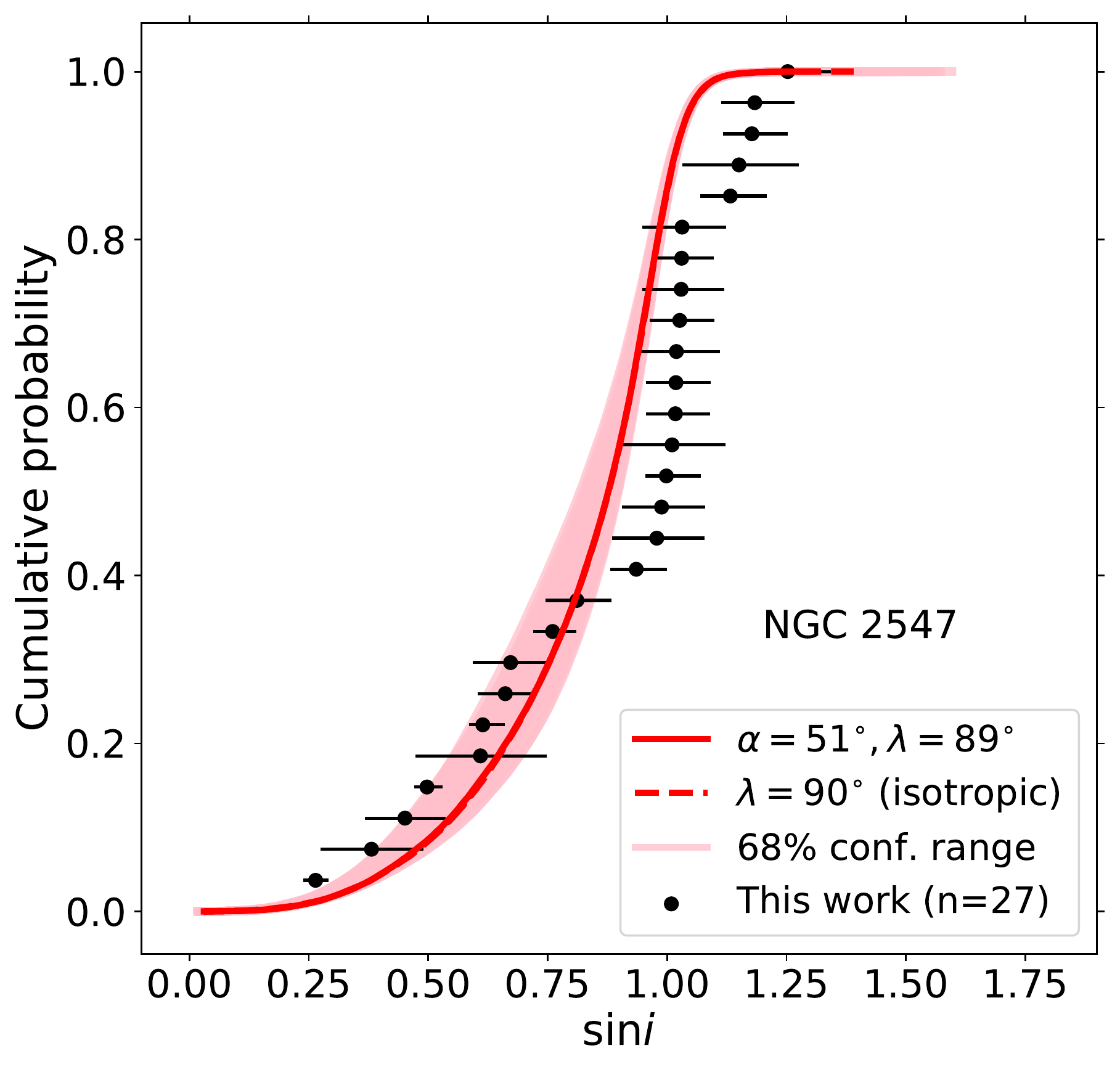}
    \includegraphics[scale=0.3]{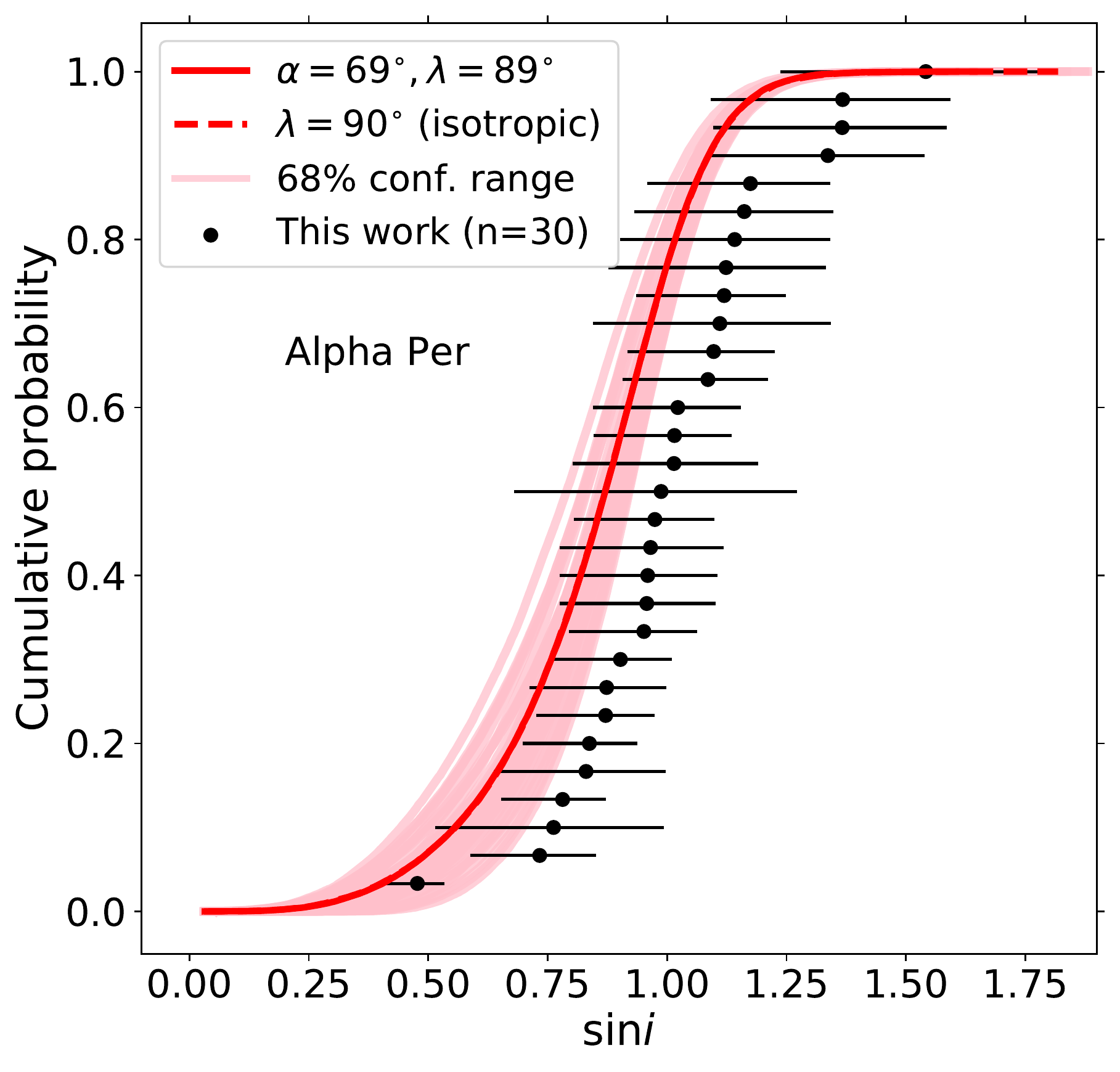}
    \includegraphics[scale=0.3]{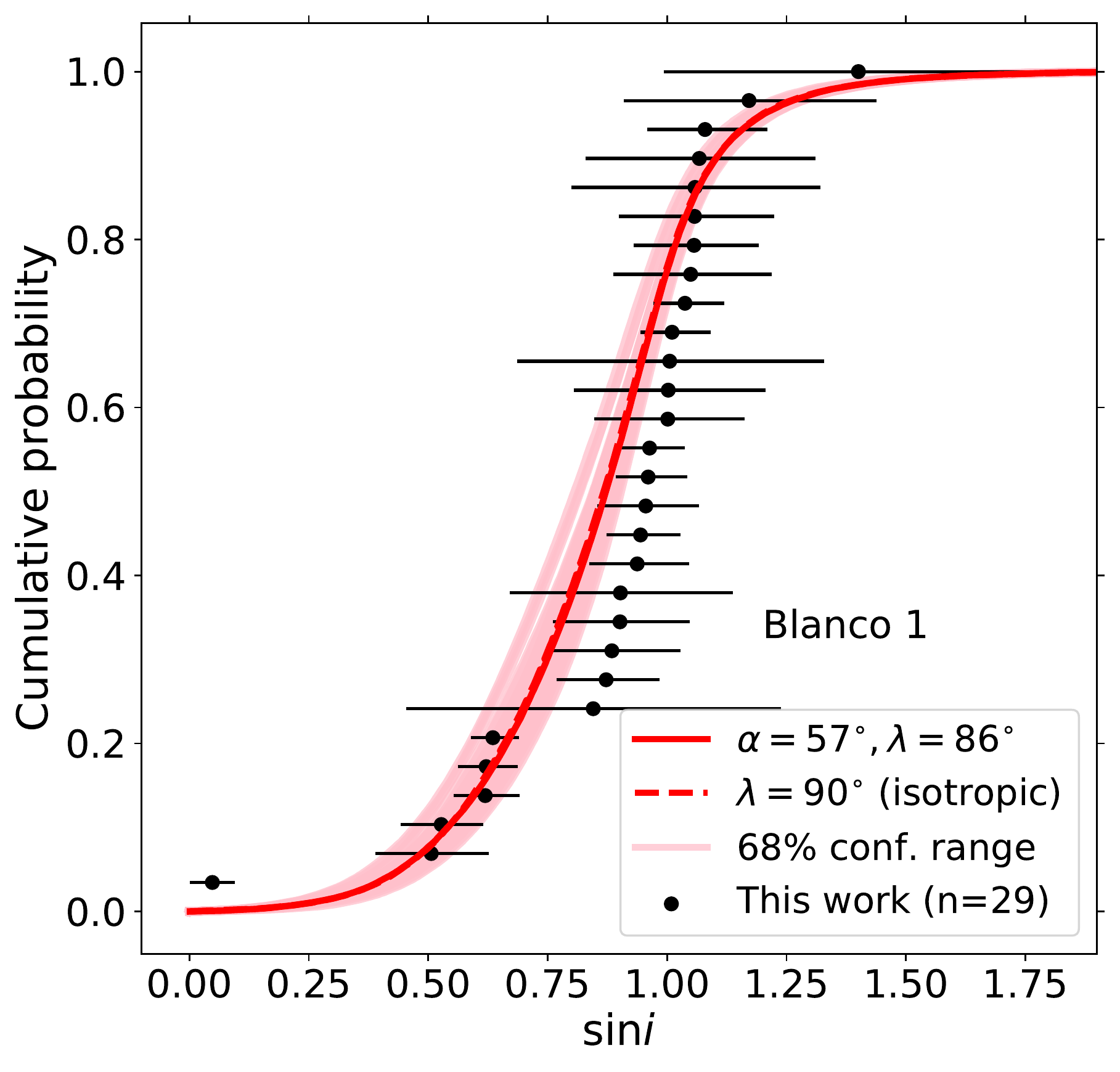}
    \includegraphics[scale=0.3]{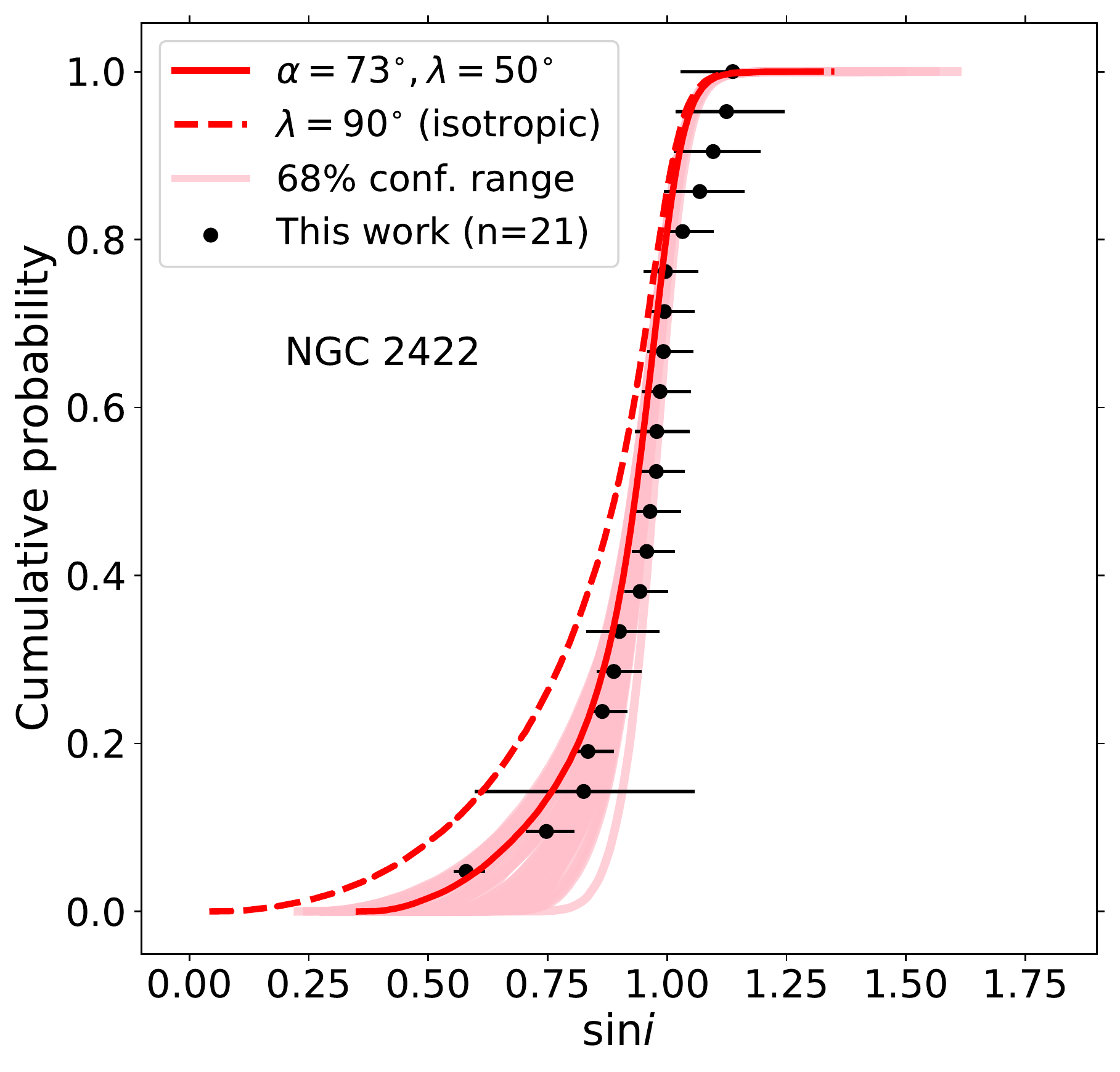}
    \includegraphics[scale=0.3]{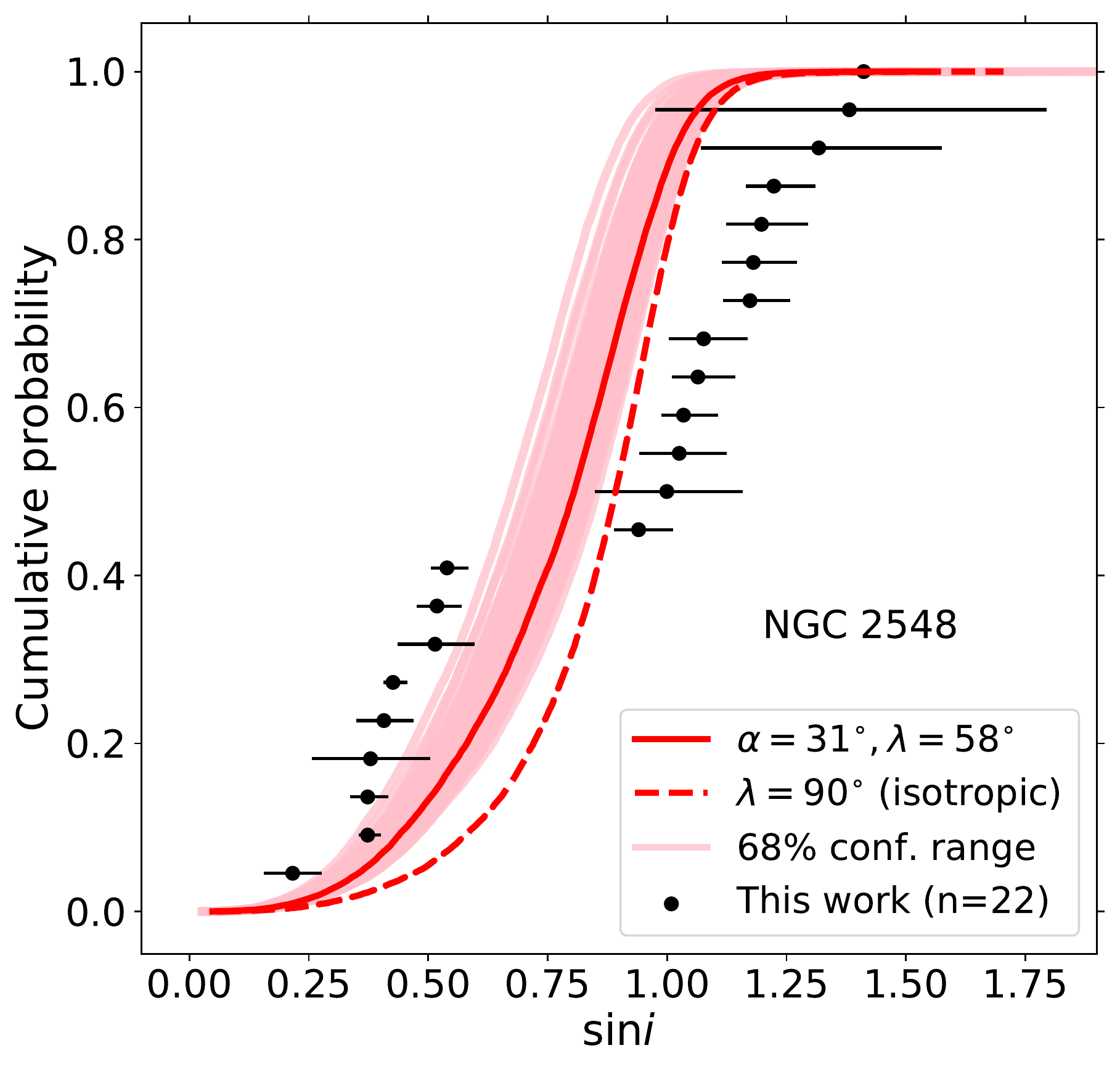}
    \caption{Inclination CDF plots for all clusters. Solid red curves correspond to PPD peaks for $\alpha$ and $\lambda$. Dashed curves show the isotropic distribution. Pink CDFs are generated with cone model parameters drawn from the 68\% confidence interval for each cluster.}
    \label{fig:ecdf_plots}
\end{figure*}

\section{Discussion} \label{sec:discussion}

In this section, we begin by discussing our interpretations of the results for the individual clusters in separate subsections. Since a small sample of inclinations taken from a random distribution can appear to be anisotropic (as quantified in simulations by \citet[][]{corsaro2017}), we only included clusters having 15 or more $\sin i$ values in our ensemble analysis. As a result, we do not discuss the inclination distributions of ASCC 16, Gulliver 6, Pozzo 1, and BH 56 in this section.

We continue with a discussion of the added inferences we make from the full constrained sample among the 15 clusters in this work and Papers \citetalias{paper1_healy} and \citetalias{paper2_healy}. We obtain ensemble insights using two different approaches: classifications of spin-axis distributions on a cluster-to-cluster scale, and allocations on a star-by-star scale based on model fitting results.

\subsection{Newly Analyzed Clusters}

\subsubsection{Collinder 69}
This cluster's CDF is shifted rightward (towards greater values of $\sin i$) compared with all possible model CDFs. Despite this shift, isotropic spins are preferred by the spread half-angle PPD. The Collinder 69 inclination distribution is likely influenced by magnetic radius inflation due to the young age of the cluster. The systematic shift may also be caused an overestimate of SDSS DR17 projected rotation velocities or TESS rotation periods.

Despite the visible offset between the model and data CDFs, we can still make statistical inferences about the alignment or isotropy of the cluster's spin-axis distribution, since we have made informed efforts to account for inflated radii and uncertainties in APOGEE $v\sin i$ values, as well as the influence of differential rotation on period measurements. By increasing (or establishing) the appropriate uncertainties on the data, we decrease the statistical significance of systematic errors at the expense of precision in the results. 

Nonetheless, we find that the spread half-angle for Collinder 69 has a meaningful constraint, with $\lambda > 22^{\circ}$ disfavoring tight alignment at 95\% confidence. The posterior plot for this cluster displays the familiar ``peninsula'' shape we first described in Section 4.2 of \papertwo: isotropic and near-isotropic regions of parameter space ($\lambda \gtrsim 65^{\circ}$) are favored, extending to moderate alignment at moderate mean inclinations ($30^{\circ} \lesssim \lambda \lesssim 65^{\circ}$; $35^{\circ} \lesssim \alpha \lesssim 75^{\circ}$). These regions of the posterior plot are surrounded by lower-probability parameter combinations. Similar to our results for NGC 2516 (\paperone) and the Pleiades and Praesepe (\papertwo), we cannot distinguish between isotropy and moderate alignment for Collinder 69. However, we make further inferences about this and other clusters showing similar peninsula-shaped posterior plots in Sections \ref{subsec:combined_insights_1} and \ref{subsec:combined_insights_2}.

\subsubsection{ASCC 19}
This cluster's posterior plot displays a peninsula shape that is comparable to that of Collinder 69. The log-posterior of the PPD-peak model is smaller for ASCC 19, reflecting the smaller number of $\sin i$ values for this cluster. The reduced $\chi^2$ value for the PPD-peak model is also smaller than Collinder 69, reflecting the better CDF fit for ASCC 19 and less systematic offset among inclination values.

The optimal model's value of $\chi^{2}_{\rm red} < 1$ indicates the possibility of overfitting or an overestimate of uncertainties. Because the 2D cone model has low complexity in using two parameters to model the CDF, overestimated uncertainties may be the more likely source of the reduced $\chi^{2}$ value. Since $v\sin i$ uncertainties are the dominant source of error in each $\sin i$ determination and we used an approximate estimate for the APOGEE $v\sin i$ uncertainties (10\% for all values above 10 km s$^{-1}$), it is possible that we did not accurately represent the uncertainty for every value. Absent further analysis of the APOGEE spectra, our uncertainty estimates still lead to informative constraints on spin alignment with a determination of $\lambda > 15^{\circ}$ at 95\% confidence. Overall, we classify ASCC 19 as either isotropic or moderately aligned.

\subsubsection{NGC 2547}
The PPD-peak result for NGC 2547 yields a greater log-posterior than for any of the previous clusters. The posterior plot forms a peninsula shape similar to those of Collinder 69, ASCC 19, and Pozzo 1, but in this cluster's case the extent of high-probability parameter space is smaller. The reduced $\chi^{2}$ value for NGC 2547 suggests a potential underfit or underestimated uncertainties, which may be the source of the tighter constraints. Since we still see the degenerate solution shared by the aforementioned clusters, we add NGC 2547 to the set of clusters classified as isotropic/moderately aligned.

\subsubsection{Alpha Per}
This cluster shows a rightward offset of nearly all of its $\sin i$ values compared with models. As a result, there is not a strong constraint on the cluster's spin-axis distribution. We have classified other such clusters in our sample as having an unconstrained spin-axis distribution when the model parameters are loosely constrained. However, a previous study of Alpha Per by \citet[][]{jackson2010} determined that the cluster has either isotropic or moderately aligned spins ($\lambda > 40^{\circ}$ at 90\% confidence and an optimal value of $\lambda = 90^{\circ}$). Since this scenario is consistent with our result, we adopt it as our qualitative interpretation of Alpha Per. Thus, we treat the cluster as having cone model parameters degenerate between isotropy and moderate alignment for the ensemble analysis.

\subsubsection{Blanco 1}
We classify this cluster as isotropic/moderately aligned, since its peninsula-shaped posterior plot matches the pattern we have previously identified. The three-parameter multi-panel plot in Figure \ref{fig:3pgridsearch_iso} is representative not only of Blanco 1 but also the other clusters classified as isotropic/moderately aligned. Studying this plot in more detail helps to explain differences between the two- and three-parameter models for the most commonly seen inclination distributions in our sample.

For low values of $f$, for which the model includes a small fraction of aligned stars, the $\alpha_{\rm aniso}$-$\lambda_{\rm aniso}$ posterior plot contains approximately uniform high posteriors for all parameter combinations. This means that no constraints can be placed on the orientation, spread half-angle, or even the existence of an aligned subset at low values of $f$. Instead, the three-parameter model uses its added complexity to achieve similar or slightly better fits than the two-parameter model. We do not interpret these results as indicative of an aligned subset of stars in these clusters, since the model will tend to overfit to slight systematic errors.

As $f$ increases, the $\alpha_{\rm aniso}$-$\lambda_{\rm aniso}$ posterior plots begin to resemble the peninsula shape of the two-parameter model MCMC. The background inclination distribution of the three-parameter model is always isotropic. Thus, if the ``aligned'' fraction of stars also appears isotropic, the resemblance suggests that the three-parameter model cannot find a better solution having an aligned subset of stars compared to a single, isotropic/moderately aligned inclination distribution for all cluster members.

\subsubsection{NGC 2422}
\label{subsec:ngc_2422}
This cluster displays a signature of anisotropic spins, with its PPD-peak model corresponding to moderate alignment at a nearly edge-on mean inclination. NGC 2422 $v\sin i$ values (reported by \citet{bailey2018_ngc2422_vsini}) have the smallest average fractional uncertainty among all the clusters in the sample at $\sim$ 3.3\%. These small uncertainties may either reinforce the preference of alignment over isotropy or overconstrain the model to an inaccurate set of parameters. The reduced $\chi^{2}$ value for this cluster is suggestive of possible overfitting. 

To further investigate this cluster's inclinations, we compared $v\sin i$ values reported by \citet[][B18]{bailey2018_ngc2422_vsini} and \citet[][J16]{jackson2016} for NGC 2516, a cluster we previously analyzed in \paperone. Among the 14 stars in common between the two studies, the B18 $v\sin i$ values are on average 12 $\pm$ 4\% (standard error) greater than those of J16. To see how the posteriors change for NGC 2422 if this discrepancy is caused by a systematic error in B18 $v\sin i$ values, we performed a new MCMC analysis after adding a 12\% fractional error in quadrature to the lower uncertainties of each $\sin i$ value. While we do not assume that the J16 values are ground-truth by default, we found that if the systematic error manifests as an underestimate in J16 $v\sin i$ values, then the inclinations for NGC 2516 are fit more poorly by all possible two-parameter combinations. If we instead treat the systematic error as an overestimate in B18 values, we achieve a better fit for the optimal NGC 2422 model (as quantified by the reduced $\chi^{2}$ value). We found that our new analysis of NGC 2422 resulted in an isotropic PPD-peak value of $\lambda = 90^{\circ}$, with the spread half-angle constrained to be $> 13^{\circ}$ at 95\% confidence.

We also experimented with subtracting 12\% from each $\sin i$ value while keeping the uncertainties unchanged. The resulting MCMC run yielded a similar result as the previous one, finding $\lambda = 89^{\circ}$ and $\lambda > 14^{\circ}$ at PPD-peak and 95\% confidence, respectively. These two additional analyses show that while the results we report for NGC 2422 favor spin alignment, the known discrepancy in $v\sin i$ values reported by B18 and J16 for NGC 2516 is enough to change the NGC 2422 result to one that favors isotropy. Thus, while we do not change the statistics we report for this cluster, we emphasize the potential for systematic $v\sin i$ error to be the cause of the anisotropic signature, and due to this uncertainty we do not include NGC 2422 in the ensemble analysis.

\subsubsection{NGC 2548}
\label{subsec:ngc_2548}
The three-parameter multi-panel plot for NGC 2548 (Figure \ref{fig:3pgridsearch_aniso}) represents the appearance of a three-parameter combination that achieves a significantly better solution than the two-parameter model (See Figure \ref{fig:m48_3param}). As opposed to Figure \ref{fig:3pgridsearch_iso}, the familiar peninsula shape does not appear for any value of $f$, nor does the two-parameter posterior plot become more visible as $f$ increases. Instead, the three-parameter model reports low log-posterior values for all regions of parameter space except for a relatively narrow range of values. We saw a similar plot for M35 (\papertwo), indicating the insufficiency of the two-parameter model to describe these clusters' inclination distributions.

The two-parameter model MCMC run for NGC 2548 disfavors isotropy, but the inclination distribution is the poorest-fit of our 15-cluster sample with a reduced $\chi^{2}$ value of 6.6. The three-parameter results offer an improvement in modeling values of $\sin i$ less than 0.5 but still fail to capture the distribution of larger values closer to and exceeding unity. Even when using the more complex model, the lowest achieved $\chi^{2}_{\rm red}$ value of 3.4 quantifies the high degree of underfitting compared to other clusters in the sample. The statistical tests on the position angles, locations and proper motions of potentially aligned stars did not have significant incompatibility with the null hypothesis of randomness.

It is worth noting that the $v\sin i$ values for both this cluster and M35, another cluster for which we found an aligned subset as a potential explanation, come from the same survey, the WIYN Open Cluster Study. M35 and NGC 2548 also saw the two largest fractions of $v\sin i$ measurements eliminated due to suspected systematic error (see Section \ref{subsec:systematic_errors} and Figure \ref{fig:per_teff_plots}). While this elimination process removed stars likely to yield nonphysical $\sin i$ values significantly greater than unity, it did not target $v\sin i$ values that result in low inclinations, since it is difficult to disentangle truly pole-on or slowly rotating stars from systematic underestimates that are still physically feasible. The common origin of the $v\sin i$ data for this cluster and M35 raises the possibility that a shared systematic error may be the source of both clusters' disagreement with the two-parameter models.

To unambiguously interpret the results of the three-parameter model, the sample size of any subset of cluster stars must be great enough to rule out apparent alignment, even if the full set of $\sin i$ values in the cluster is suitable for constraints on the inclination distribution using the two-parameter model. Since only 6 stars make up the subset of NGC 2548 inclinations modeled as aligned, the same limitations we faced with using the two-parameter model to describe ASCC 16, Gulliver 6, Pozzo 1, and BH 56 are applicable to NGC 2548 and the three-parameter model. While the visibly bimodal $\sin i$ distribution of this cluster is noteworthy and merits a different qualitative label than the aforementioned clusters in Table \ref{tab:incl_distribs}, we cannot interpret the results of the three-parameter model fit as indicative of spin alignment. Given our model's poor goodness-of-fit, the lack of a significant position angle, location, or proper motion grouping, and the qualities of the spectroscopic data described above, it is plausible that systematic error in $v\sin i$ caused the apparent alignment.

\subsection{Previously Analyzed Clusters}
NGC 2516 (\paperone), the Pleiades and Praesepe (\papertwo) all show minimal changes from their previously determined inclination distributions. 

For NGC 2516, there is a 33\% increase in the number of reported $\sin i$ values because we did not apply an 0.7 \msun\ threshold for this version of the analysis (see Section \ref{subsec:results_inclinations}). These three clusters all have peninsula-shaped posterior plots with PPD-peak mean inclination values within $1^{\circ}$ of each other. Optimal spread half-angle values are within $3^{\circ}$ of each other.

M35 saw a slightly larger change from its two-parameter model constraints in \papertwo. The 95\% confidence interval in the spread half-angle now extends to $\lambda = 81^{\circ}$ compared to its original value of $67^{\circ}$. This change is slightly more favorable of isotropy, and may be caused by the standardization of stellar magnitudes for SED fitting in this paper. We did not include $BVRI$ magnitudes from \citet[][]{m35_wocs_thompson2014} in this work, perhaps inducing a slight change in the resulting $\sin i$ values or their uncertainties. The maximum-posterior three-parameter model solution for M35 remains similar to the optimal parameters in \papertwo, favoring a small fraction of stars with tightly aligned spins along the LOS. The parameters we determined in this work's new analysis are $\alpha_{\rm aniso} = 6^{\circ}$, $\lambda_{\rm aniso} = 9^{\circ}$, and $f = 0.06$ (corresponding to $\sim$ 4 stars modeled as aligned, again too small a subset to conclude physical alignment).

\subsection{Combined Insights: Cluster-by-Cluster}
\label{subsec:combined_insights_1}
Our overall 15-cluster sample of spin-axis distributions, including the 11 studied in this work, three in \papertwo\ and one in \paperone, establishes patterns and permits us to draw ensemble-level insights not obtainable from any single cluster. 

\subsubsection{Unconstrained Clusters}
The small number of stars with $\sin i$ values for ASCC 16, Gulliver 6, Pozzo 1, and BH 56 precludes us from determining their spin-axis distributions. In addition, four of the five clusters whose spin-axis distributions we classify as unconstrained are among the youngest members of our sample. As we describe in Section \ref{subsec:systematic_errors}, these young clusters are especially likely to experience radius inflation due to magnetic activity. The probable presence of radius inflation in these clusters, manifesting as an unknown bias in $\sin i$, further complicates a constraint on spin alignment or isotropy for these clusters.

The fifth unconstrained cluster, NGC 2422, likely experiences less radius inflation than the other four, but the potential systematic $v\sin i$ error we identify in Section \ref{subsec:ngc_2422} creates enough uncertainty to similarly prevent constraints of the cluster's spin-axis distribution. The limitations illustrated by these five clusters emphasize that future work on this topic would benefit from observation campaigns dedicated to using a multi-object spectrograph to yield an increased number of $v\sin i$ values for as many Gaia-based cluster members as possible. Combined with the refinement and use of radius inflation predictions to mitigate systematic error in this parameter, future studies can provide greater constraints on younger and less populous open clusters.

\subsubsection{Clusters Modeled as Partially Aligned}
two clusters benefit from improved fits with the three-parameter model compared to insufficient fits with two parameter model (M35 and NGC 2548, see \papertwo\ and Section \ref{subsec:ngc_2548}). The maximum fraction of aligned stars favored by the three-parameter model is 0.3 (for NGC 2548). For the cluster-by-cluster insights, if we treat M35 and NGC 2548 as possibly partially aligned, this yields a fraction of 2/10 or 20\% consistent with partial alignment.

The small number of stars and potential for systematic error contributing to these classifications highlights a key difference of our star-by-star discussion in Section \ref{subsec:combined_insights_2}. With the star-by-star approach, these two clusters' stars are separated into isotropic and aligned subsets according to the $f$ value of the three-parameter model. As a result, the influence of the small number of potentially aligned stars is reduced, since instead of a single classification for each cluster, our results are weighted by the overall number of stars in the aligned and isotropic subsets.

\subsubsection{Isotropic or Moderately Aligned Clusters}
We have classified eight clusters having similar peninsula-shaped posterior plots suggesting either isotropic or moderately aligned spins. For any single cluster, we cannot demonstrate a preference for one of these scenarios or the other. However, the ensemble of similar constraints facilitates further analysis. The possible explanations for this eight-cluster subsample are that their spin-axis distributions are all moderately aligned, all isotropic, or a mix of both.

If all of the eight clusters in the peninsula-plot subsample have isotropic stellar spins, then at least 80\% of the 10 constrained clusters in our study have isotropic spins, while the remaining 20\% may be partially aligned (yet still with sizeable isotropic subsets, see Section \ref{subsec:combined_insights_2}). However, there remains the possibility that some or all of the isotropic/moderately aligned clusters are moderately aligned. We can put a limit on this number by comparing the ensemble of $\alpha$ values associated with each cluster's moderate alignment to an expected distribution of $\alpha$ values.

To estimate the ensemble distribution of $\alpha$ values resulting from the seven-cluster subsample, we used the PPDs from an MCMC run while limiting $\lambda < 65^{\circ}$ to exclude the region of parameter space corresponding to isotropy or near-isotropy (for which $\alpha$ is not well constrained, see Figure \ref{fig:likelihood_plots}). We approximated these PPDs as normal distributions and mapped values $< 0^{\circ}$ or $> 90^{\circ}$ into the 0-90$^{\circ}$ range (e.g. $-10{^\circ}$ became $10^{\circ}$, and $100^{\circ}$ became $80^{\circ}$). This ``compressed'' normal distribution well approximated the PPDs of $\alpha$ for each cluster, accurate to within a few degrees at the 5th, 16th, 32nd, 50th, 84th, and 95th key percentiles of the distribution.

We sampled $\alpha$ values from each approximated distribution, choosing a number of values proportional to each cluster's maximum log-posterior value. We thus weighted sampling by our confidence in each cluster's result, since the clusters with the largest $\ln(\rm post)$ values received the greatest representation. Combining all the results of this sampling across clusters to create a single posterior distribution for $\alpha$, we determined that the ensemble's $\alpha$ values in the aligned region of parameter space are $> 32^{\circ}$ at 95\% confidence and range from $\sim 37^{\circ}$ to $73^{\circ}$ at 80\% confidence.

Having estimated the range of observed $\alpha$ values, our ensemble analysis continued by establishing an expected $\alpha$ distribution to compare to the data. We considered two scenarios: random orientations of all clusters' mean inclinations, and uniform alignment. This latter scenario is simple but unlikely, because it requires a strong connection between isolated clusters amid the turbulent interstellar medium \citep[e.g.][]{ism_turbulence_2014}. It would also be inconsistent with the spread of LOS angles to each cluster in our sample (Galactic latitudes between $\sim -79^{\circ}$ and 33$^{\circ}$ (standard deviation of $29^{\circ}$) and Galactic longitudes between $\sim 15^{\circ}$ and $274^{\circ}$ (standard deviation $76^{\circ}$). 

If all clusters have their stellar spins oriented in the same three-dimensional direction, we would expect to observe a wide range of $\alpha$ values due to the varying LOS to each cluster. Instead, the comparable $\alpha$ values in our subsample contrast with the wide range of LOS directions, suggesting that not all the stellar spin-axis orientations in our subsample are intrinsically aligned with each other on a cluster-to-cluster scale. We therefore make the assumption of random $\alpha$ values among any aligned clusters, with the cumulative probability of a given $\alpha$ value described by
\begin{equation}
    P(\alpha) = 1-\cos \alpha
    \label{eq:aCDFprob}
\end{equation} \citep[see][]{jackson2010}.  

We continued by assuming that all clusters' $\alpha$ values were within the 80\% confidence range between $37^{\circ}$ and $73^{\circ}$. We then took the difference of cumulative probabilities from Equation \ref{eq:aCDFprob} at both of these values to find that the probability of a given aligned cluster having an $\alpha$ value within this range is $\sim 0.5$ (assuming random orientations between clusters). The independence between each cluster's spin-axis orientation further yields the probability of $n$ clusters having $\alpha$ values between $37^{\circ}$ and $73^{\circ}$ to be $0.5^{n}$ from a simplified binomial probability distribution.

For no fewer than 4 clusters to have $\alpha$ values within this range (with the remaining 4 isotropic, where $\alpha$ becomes meaningless), the probability is $p \sim 0.06$. Thus, the probability of no more than 4 clusters being aligned in this $\alpha$ range is $1 - p \sim 0.94$. Since the $\alpha$ range we used to determine this probability has a confidence level of 80\%, we place a ($0.8 * 0.94$) = 75\% confidence limit on the number of clusters having moderately aligned distributions at no more than four. It follows that no fewer than four of the eight clusters in this subsample have isotropic spins at 75\% confidence.

\subsubsection{Constraints on Spin Alignment}
As a result of the above discussion, we make the following constraints on spin alignment on a cluster-by-cluster scale at 75\% confidence. Since at least four our of eight isotropic/moderately aligned clusters have isotropic spins, combining them with the two potential partially aligned clusters places a lower limit of four out of ten clusters ($40\%$ having isotropic spin-axis orientations. We also place an upper limit on spin-aligned clusters at six out of ten (60\%). Of these six, up to four may have their entire spin-axis distribution moderately aligned, while up to two others are modeled as partially aligned and paired with an isotropic subset of stars. We can only place lower limits on the number of clusters with isotropic spin-axis orientations (and upper limits on the number of totally/partially aligned clusters) because it remains possible that the latter are isotropic and either masked by the degenerate parameters of their peninsula-shaped posterior plots or experiencing systematic error in their $\sin i$ values that is overfit by the three-parameter model. The number of isotropic clusters increases to 80\% if the entire eight-cluster subsample has randomly oriented spins.

For the eight-cluster subsample, we also performed an ensemble analysis of the spread half-angle posteriors in an analogous way to the mean inclinations. This time, we modeled the posteriors as normal distributions each having a mean $\lambda = 90^{\circ}$, the approximate PPD peak for all nine clusters. As for the ensemble mean inclination distribution, for values of each normal distribution $< 0^{\circ}$ or $> 90^{\circ}$, we mapped them such that they were within the range of 0-90$^{\circ}$. This approximation matched the key percentile values of each $\lambda$ PPD to within a few degrees. From the combination of posterior-weighted draws from each of these distributions, we determined that $\lambda > 25^{\circ}$ at 95\% confidence for eight-cluster subsample.

While these constraints rule out tight alignment among the eight isotropic/moderately aligned clusters, the best-fitting three-parameter models for M35 and NGC 2548 favor tight alignment. However, the parameters are poorly constrained compared to the eight-cluster subsample due to the small number of stars ($\sim 11$, see Section \ref{subsec:combined_insights_2}) modeled as aligned. In the next section, we discuss the interpretation of these results on a star-by-star basis, which can provide results less biased by potentially outlying $\sin i$ values on each cluster's classification.

\subsection{Combined Insights: Star-by-Star}
\label{subsec:combined_insights_2}
Overall in the 15-cluster sample, we determined 404 projected inclinations. The tally by cluster is in Table \ref{tab:incl_distribs}. The five unconstrained clusters contain 65 stars ($\sim 16\%$ of the sample). This leaves 339 stars which are part of clusters with constrained inclination distributions. The eight-cluster isotropic/moderately aligned subsample contains 248 stars out of those 339, or 73\%. At the lower limit of four out of eight isotropic spin-axis distributions in this subsample, we estimate that at least $\sim$ 124 stars are part of isotropic distributions. Since we cannot determine which clusters of the nine may have spin alignment, this value is not exact, but variable to a degree proportional to the number of $\sin i$ values for each cluster. The remaining $\sim$ 124 stars may be part of an aligned distribution, or may also have isotropic spins.

The most significant difference using the star-by-star approach compared to cluster-by-cluster is the treatment of clusters modeled as partially aligned. While these clusters have a single classification determined by their disagreement with the two-parameter model, the star-by-star approach can consider the fraction of stars aligned versus those modeled as isotropic. For M35 and NGC 2548, best-fitting aligned fractions of $f \sim$ 0.06 and 0.3 mean that a majority of these clusters' stars are modeled as isotropic. When we break down these stars into aligned and isotropic categories, we find that $\sim$ 11 stars out of the entire 339-star constrained sample are modeled as aligned. The remaining 80 stars in these two clusters are part of the isotropic background in the three-parameter model. 

This result means that overall, 97\% of stars with $\sin i$ determinations in our sample are consistent with being part of isotropic distributions. It is important to note that a particular degree of moderate alignment would present the same background $\sin i$ CDF as isotropy for the partially aligned clusters, a degeneracy analogous to the two-parameter model. The remaining 11 out of 339 stars ($\sim$ 3\%) are modeled as aligned, although the potential for systematic error discussed above for these clusters and the small sample size of these stars prevents interpreting that they are indeed physically aligned.

\subsection{Implications for Star Formation}
\label{subsec:combined_insights_physics}
The consistency of our results with a majority of isotropic spin-axis orientations suggests that among our sample of clusters with masses $\sim$ 200-4000 $M_{\odot}$ and ages $\sim$ 5-700 Myr, turbulence dominates ordered rotation in the clumps from which protostellar cores form. For isotropic distributions simulated by \citet[][]{corsaro2017}, there is a negligible amount of kinetic energy in rotation compared to turbulence. The remaining possibility of moderately aligned distributions leaves open the possibility that rotational kinetic energy for some clumps is $\gtrsim 10\%$ of their turbulent kinetic energy. Nonetheless, turbulence dominates rotation for at least some of the clusters in our sample. This is in agreement with recent measurements of a star-forming filament's specific angular momentum that found a value comparable with protostellar cores in other regions, supporting a star's inheritance of its spin-axis orientation from the surrounding turbulence \citep[][]{hsieh2021}.

\begin{figure}
    \centering
    \includegraphics[scale=0.5]{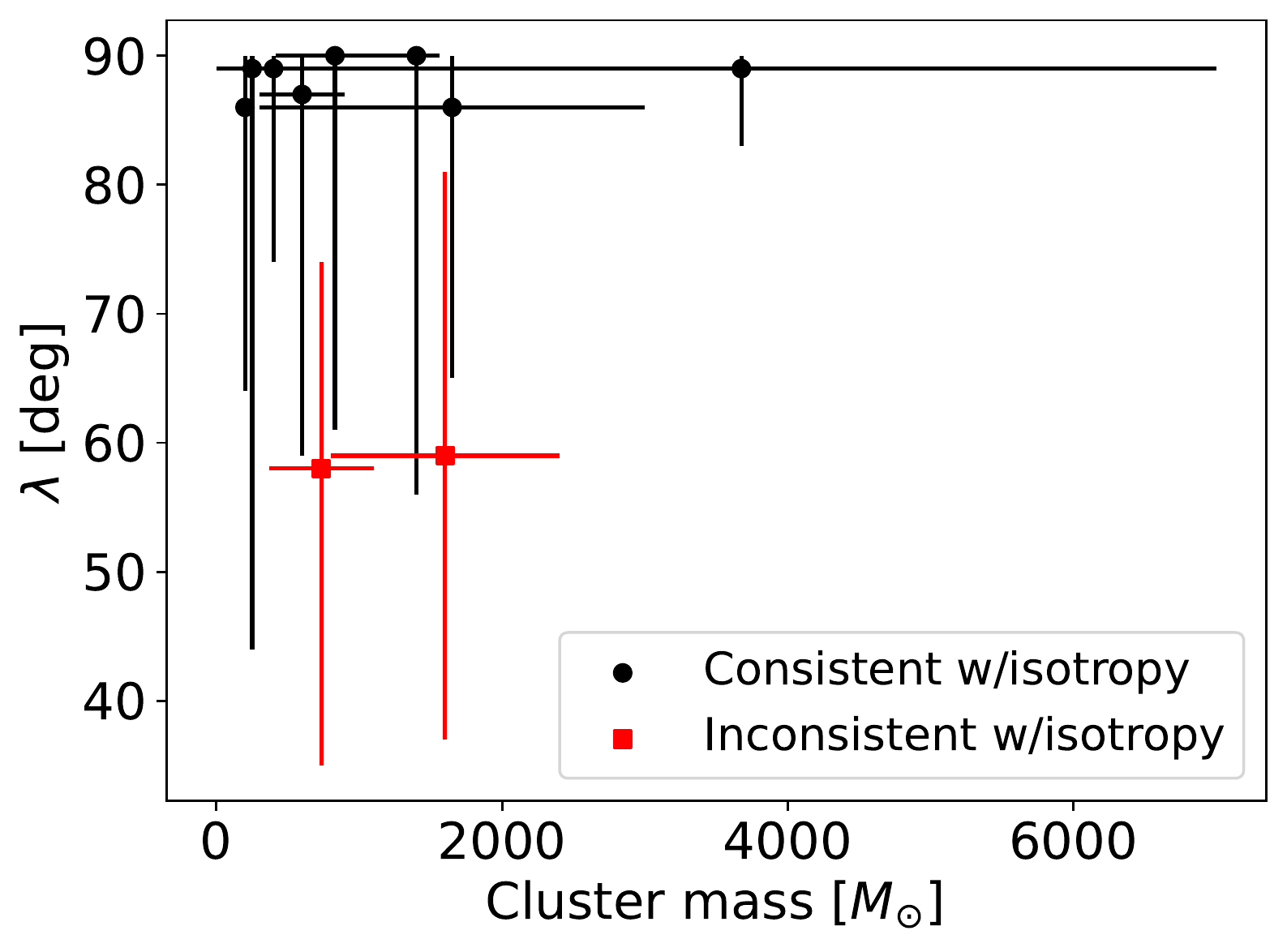}
    \caption{Spread half-angle posteriors versus cluster mass for the ten constrained clusters in our sample. Black circles represent the eight clusters consistent with isotropic spins, while red squares correspond to NGC 2548 and M35, whose spin-axis distributions are modeled as inconsistent with isotropy.}
    \label{fig:alignment_mass_plot}
\end{figure}

The mass of a cluster, if proportional to the mass of its progenitor clump, may have an important influence on the alignment of stellar spins. A clump with greater mass has a larger amount of angular momentum to impart onto individual protostars, potentially countering the randomization of turbulence with greater efficiency than in less massive clusters. To present our results in light of this possibility, we plot in Figure \ref{fig:alignment_mass_plot} the spread half-angle ($\lambda$) posteriors we determined for each of the ten constrained clusters versus the cluster mass and uncertainties from the literature. We plot M35 and NGC 2548 using red squares to highlight their inconsistency with isotropic spins, as opposed to the rest of the sample labeled by black circles.

The minimum mass among the two clusters inconsistent with isotropy is $\sim 800 M_{\odot}$, while the overall sample contains clusters with masses as small as $\sim 200 M_{\odot}$. A dependence of spin alignment on cluster mass could manifest as a trend in a diagram like Figure \ref{fig:alignment_mass_plot}.  However, owing to the uncertainties and sample size of clusters in our study, we cannot identify any significant mass dependence of spin alignment. Future work may add value to this plot increasing the number of points and reducing their uncertainties. The reported strong spin alignment in the massive open clusters NGC 6791 and NGC 6819 \citep[][]{corsaro2017}, with masses $\sim$ 5000 and 2600 $M_{\odot}$, respectively \citep{platais2011, kalirai2001}, further motivates the consideration of open cluster mass when studying the spins of their stars.

While \citet[][]{corsaro2017} did not model aligned subsets of stars in a cluster, observations and simulations suggest that star formation involves hierarchical substructure even within a single clump \citep[e.g.][]{juarez2019,lim2021,stefano2022}. The presence of such substructure during star formation suggests that the stars in a cluster may originate from many structures that were once distinguishable, and that spin alignment is more likely to manifest on these smaller scales compared to the overall cluster. In addressing clusters where a significantly populated subset of stars is discrepant with isotropic spins, the three-parameter model will continue to be useful in future studies.

That the majority of our results are consistent with isotropy also supports the turbulence-driven misalignment of protostellar rotation axes with their magnetic fields. Numerical simulations by \citet[][]{kuznetsova2020} found that magnetic fields tend to be aligned among protostars. If their rotation axes are isotropic, however, then these protostars will tend to be magnetically misaligned and will experience a reduced level of magnetic braking. This misalignment reduces the effect of the magnetic braking ``catastrophe,'' in which a level of braking strong enough to decrease a protostar's specific angular momentum to observed levels also prevents the formation of protoplanetary disks. The isotropy-induced prevention of excess angular momentum diffusion supports the formation of protoplanetary disks on a scale similar to their observed properties \citep[][]{gray2018, zhao2020_protostar_disk_review}.

\subsection{Implications for Radius Inflation}
\label{subsec:discussion_radius_inflation}

The systematic offsets observed in some fits to the data in Figure \ref{fig:ecdf_plots} suggest the possible presence of radius inflation to explain our model's underestimated $\sin i$ values (see Section \ref{subsec:systematic_errors}). By attributing this offset to radius inflation and assuming isotropic spins, we can use the SPOTS models to estimate the amount of radius inflation in each cluster.

We assumed isotropic spins in the 11 clusters for which isotropy is a possible solution (the eight with peninsula-shaped likelihood plots along with the unconstrained ASCC 16, Pozzo 1, and BH 56). We then shifted the isotropic model for each of these clusters towards larger $\sin i$ values (rightward on the plots) by the mean fractional radius inflation amount predicted by the cluster's SPOTS isochrone for all fractional spot coverage values. The minimum $\chi_{\rm red}^{2}$ yielded by adopting these values of $f_{\rm spot}$ revealed the amount of radius inflation necessary to mitigate the visible systematic offset between data and model for each of the 11 clusters.

\begin{figure}
    \centering
    \includegraphics[scale=0.4]{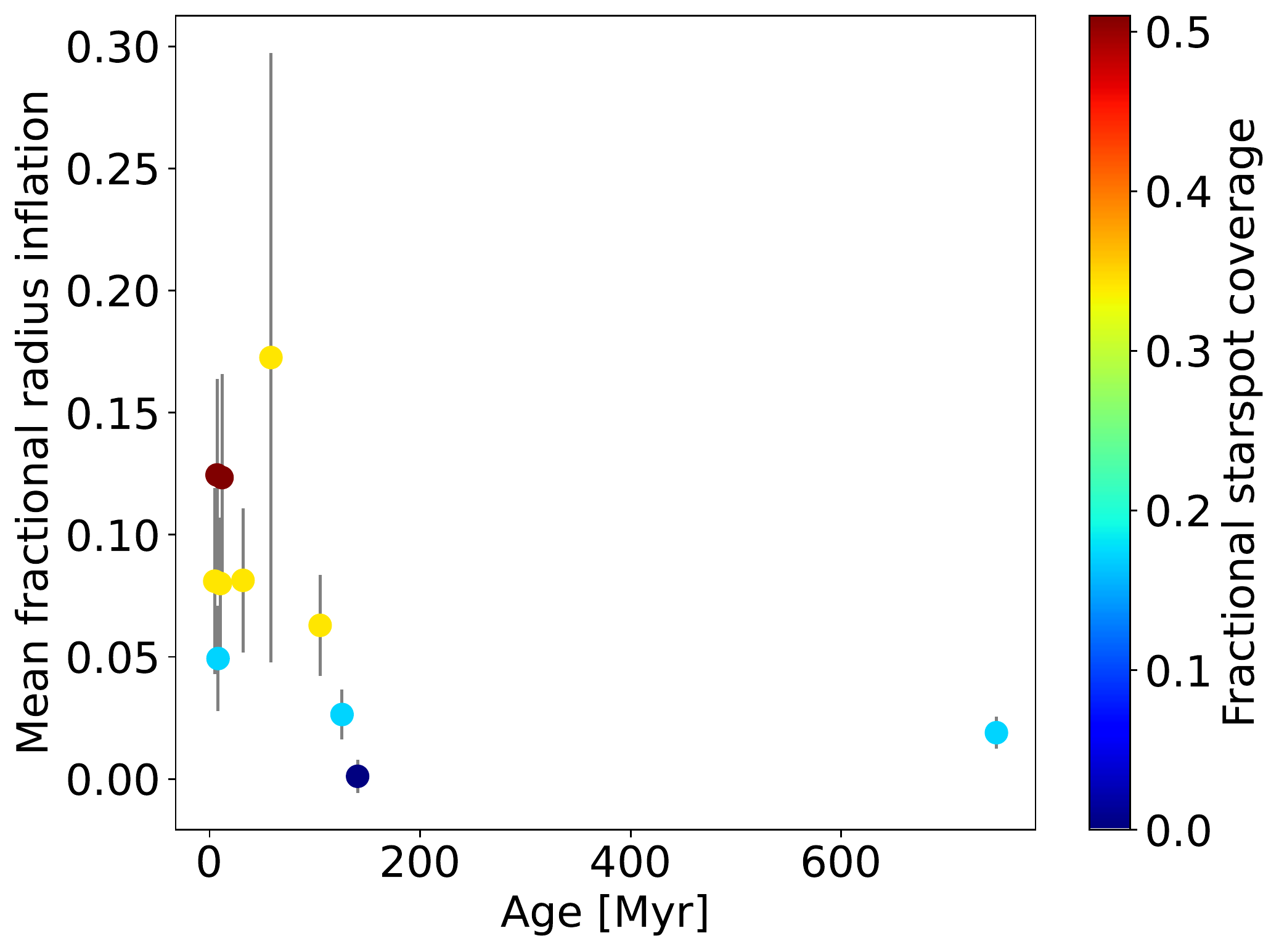}
    \caption{Mean (with standard deviation error bars) of the fractional radius difference between MIST isochrones and the SPOTS isochrone that achieves best fit to $\sin i$ data assuming an isotropic distribution. All isotropic/moderately aligned and two of three unconstrained clusters (ASCC 16 and BH 56) are plotted. Points are color-coded by the fractional starspot coverage associated with the best fit.}
    \label{fig:radius_inflation_age}
\end{figure}

Figure \ref{fig:radius_inflation_age} plots the mean fractional radius inflation amount corresponding to each cluster's optimal isotropic fit as a function of its age in Myr. The points are color-coded by the fractional spot coverage value of the optimal result, and the error bars come from the standard deviation of the SPOTS versus MIST radius predictions. We find that, under our assumption of isotropic spins, the typical radius inflation among our sample's mostly Sun-like stars on slow-rotator sequences is $\sim$ 10\% for clusters younger than $\sim 100$ Myr and drops to no more than a few percent for older clusters. The spot covering fractions associated with this amount of radius inflation range from 0 (one cluster best fit by the original MIST model) to 0.5 (two). 

Our radius inflation result is quantitatively similar to that of \citet[][]{somers2015}, who estimated up to $10\%$ radius inflation in pre-main sequence stars and up to $9\%$ for partially convective stars on the zero-age main sequence. The age threshold we determine for $\sim 10\%$ radius inflation is also in agreement with this work, since their analysis of the $\sim 100$ Myr-old Pleiades found that the cluster's stars on the slow-rotator sequence were consistent with unspotted models yielding no radius inflation. Finally, Section 2.2.2 of \citeauthor{somers2015} describes the observational evidence that pre-main sequence stars can have spot covering fractions up to $\sim 50\%$, supporting the maximum fraction that we estimate in our analysis.

\subsection{Simulations Informing Future Studies}
\label{subsec:simulations_3param}
To inform future studies searching for aligned subsets of stars in clusters, we performed Monte Carlo simulations of many partially aligned inclination distributions. Our goal was to estimate the ability of the three-parameter model to confidently identify different fractions of aligned stars in clusters. We generated $\sin i$ values using the three-parameter model, assuming random uncertainties corresponding to the mean values of the overall ensemble of measurements in this study ($\sim$ 11\%, 5\% and 1\% uncertainties respectively in $v\sin i$, $P$ and $R$). In our simulations, the possible values of $f$ ranged from 0.05-1.0 in increments of 0.05. For each value of $f$, we drew 5000 mean inclination values from a distribution of isotropic $\alpha_{\rm aniso}$ values, once again assuming no correlated spin alignment between clusters.

We fixed the spread half-angle $\lambda_{\rm aniso}$ in each simulation to a value informed by magnetohydrodynamic simulations of star formation. Considering that our work does not find evidence for the alignment of protostellar rotation axes and magnetic fields, we consulted these simulations to estimate the minimum level of misalignment that permits the formation of protoplanetary disks. Based on the results of \citet[][]{hennebelle2009}, a misalignment of $\sim 10^{\circ}$ can moderate magnetic braking at a protostellar mass-to-flux ratio consistent with observations to enable disk formation. We adopt twice this value, $\lambda_{\rm aniso} = 20^{\circ}$, to represent the cone within which stellar spin axes are uniformly distributed in our simulations. In doing so, we estimate a lower limit for the degree of spin alignment that a subset of stars can display.

We fit each set of $\sin i$ values with two CDFs: one for an isotropic distribution, and one for the three-parameter values used to generate each inclination distribution. We compared the ratio of posterior values describing each CDF's fit to the simulated data. A posterior value 19 times greater for the three-parameter model CDF than the isotropic CDF corresponds to a 95\% confidence determination of partial alignment over isotropy. We performed this process for simulated sample sizes of $N_{\sin i}$ = 10, 30, and 100.

\begin{figure*}
    \centering
    \includegraphics[scale=0.55]{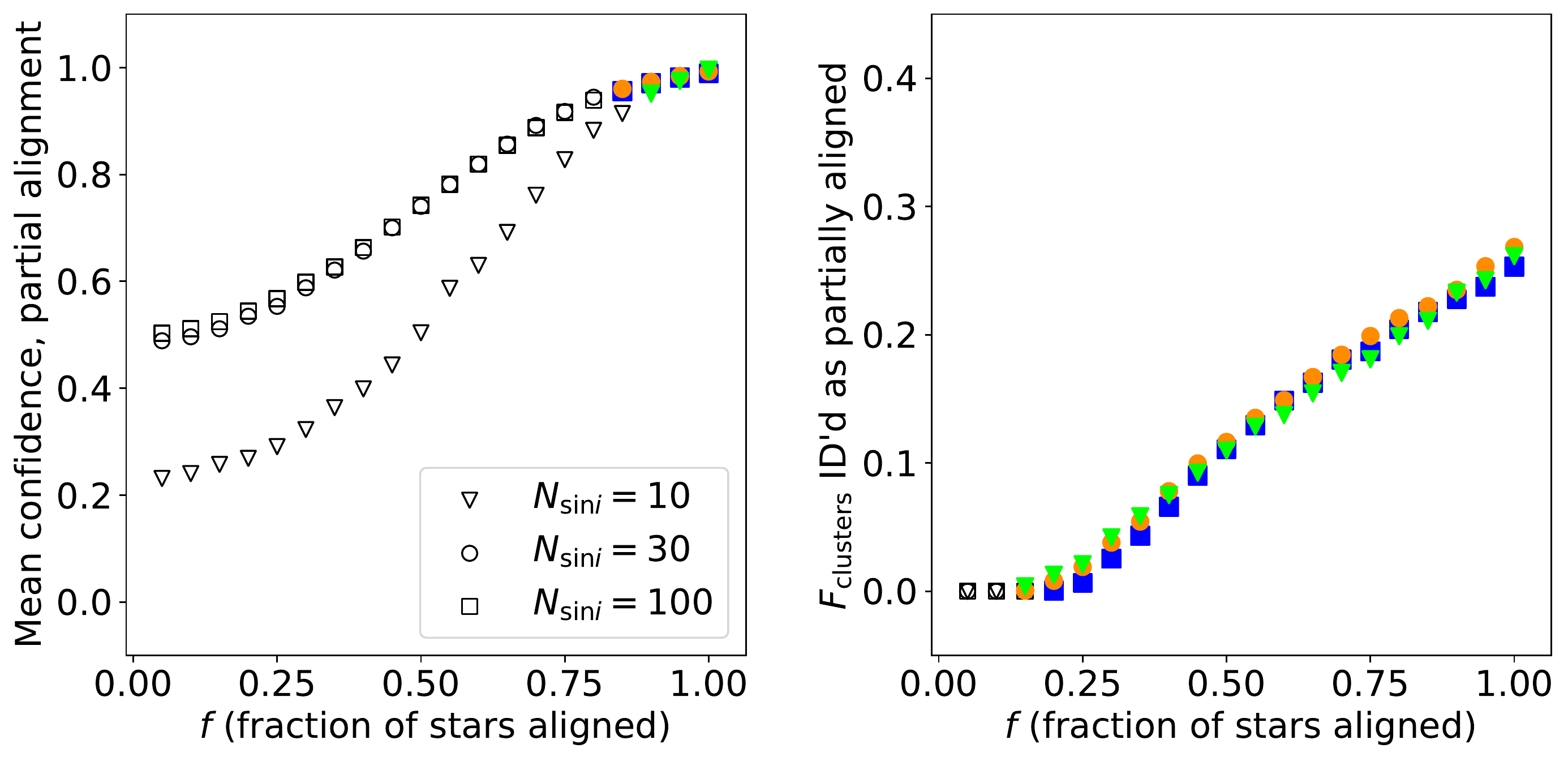}
    \caption{Partial alignment simulation results for sample sizes $N_{\sin i}$ = 10, 30, and 100. Left panel: mean confidence in partial alignment over isotropy versus fraction of aligned stars $f$. Colored symbols indicate mean confidence $\geq$ 95\%. Right panel: fraction of clusters $F_{\rm clusters}$ identified as partially aligned (at 95\% confidence over isotropic) versus $f$. Colored symbols indicate a fraction $> 0$.}
    \label{fig:3param_simulations}
\end{figure*}

Figure \ref{fig:3param_simulations} plots the results of these simulations. The left panel plots the mean confidence level favoring partial alignment over isotropy as a function of the fraction of aligned stars $f$. The right panel plots the fraction of clusters $F_{\rm clusters}$ for which partial alignment was favored over isotropy at $\geq$ 95\% confidence. Since every cluster in these simulations was generated from the three-parameter model with $f > 0$, the isotropic model is never the correct solution. The degree to which it is comparable with the appropriate three-parameter model solution offers insight into the outcomes of future studies of stellar spins in open clusters.

We found that the minimum value of $f$ required to produce a 95\% confidence determination of partial alignment, averaged over $\alpha_{\rm aniso}$ values, was 0.9 for a distribution of 10 $\sin i$ values and 0.85 for both 30 and 100 values (see colored symbols in the left panel of Figure \ref{fig:3param_simulations}). This suggests that a large fraction of stars needs to be aligned in a randomly selected cluster to yield a strong signature of partial alignment. We also determined that the fraction of clusters with 95\% confidence identifications of partial alignment is zero until $f = 0.15$, after which it rises to $\sim 0.1$ at $f = 0.5$ and increases until approaching $\sim 0.25$ as $f$ approaches 1 (see colored symbols in the right panel of Figure \ref{fig:3param_simulations}).

Approximately 98\% of the partially aligned spin-axis distributions detected at $\geq$ 95\% confidence had $\alpha_{\rm aniso} < 45^{\circ}$. This shows that aligned spins with orientations closer to edge-on will be very difficult to confidently detect given the assumed alignment model, measurement uncertainties and rotation velocity distribution. For a collection of ten clusters each having half of their stars' spins partially aligned (with the other half isotropic), our simulations suggest that one will be identified as partially aligned by the three-parameter model at 95\% confidence versus isotropy. Improving on the current $\sim 10\%$ precision for $v\sin i$ values, which was the dominant source of uncertainty in our study compared to $P$ and $R$, would permit a greater number of high-confidence identifications of partial alignment in a given cluster sample.

We found little dependence of the mean confidence level and $F_{\rm clusters}$ on the sample size ($10 \leq N_{\sin i} \leq 100$). There is a 5\% decrease in the minimum $f$ to achieve a mean confidence of 95\% between 10 and 30 $\sin i$ values. In practice, the possibility for apparent alignment due to a small sample of stars and systematic error effects on the spectro-photometric parameters informing $\sin i$ determinations supports the analysis of as many homogeneously obtained data points as possible. However, returns diminish between $\sim$ 30 and 100 $\sin i$ values, and upon reaching this level observing time becomes better spent studying the stars of another cluster.

\subsection{Kinematics}
The mean tangential velocities for ASCC 16, Alpha Per, and NGC 2547 are discrepant with zero at greater than 95\% confidence. In each case, the direction of the motion is counterclockwise on the sky. Since the Galactic latitude of each of these clusters is negative, their counterclockwise rotation opposes the galactic disk angular momentum. This is consistent with models presented by \citet{kroupa2022}. Since the inclination distributions of these clusters are either unconstrained or potentially isotropic, we cannot definitively compare the mean inclination of their spin-axis distributions to the orientation of their bulk rotation.

Six clusters display significant radial kinematics. Of these, five (Collinder 69, ASCC 16, ASCC 19, Gulliver 6, and Blanco 1) are directed outward from cluster center. Collinder 69, the youngest cluster in our sample ($\sim 5$ Myr) also shows a significant signature of rotation about an axis perpendicular to the LOS, and it has a significant discrepancy between velocity dispersions in the radial and tangential directions. The expansion of young clusters is expected as a result of the expulsion of their gas \citep[e.g.][]{kroupa2001}, and \citet[][]{kuhn2019} found signatures of expansion in 75\% of a sample of 28 clusters and associations between 1-5 Myr in age.

One cluster (Alpha Per) has its radial motion directed inward. \citet[][]{makarov2006_alphaper_contraction} also found a signature of contraction for the stars in this cluster by comparing their LOS and astrometric velocity dispersions. As we detailed in Section 5.8 of \papertwo, the mean radial motions we report should be interpreted with caution due to the potential bias introduced by the $G < 18$ magnitude cutoff for cluster members in our sample.

\section{Conclusions} \label{sec:conclusion}
We have analyzed stellar rotation periods, radii, and projected rotation velocities to determine the orientation of stellar spin axes in 11 open clusters. Using two- and three-parameter models imagining stellar spin axes as uniformly distributed in a cone of varying mean inclinations and spread half-angles, we place constraints on the inclination distributions of six of these clusters. We combine these results with those of four additional clusters from Papers \citetalias{paper1_healy} and \citetalias{paper2_healy}, resulting in a combined sample of 15 clusters. We perform an ensemble analysis on ten of these clusters with constrained spin-axis distributions (and their associated 339 projected inclination values) to provide insight into the initial conditions of star formation.

We find that eight clusters (80\%) are consistent with isotropic spins. These clusters have a mean mass $\sim$ 1100 $M_{\odot}$ (range $\sim$ 200-4000 $M_{\odot}$) and mean age $\sim$ 100 Myr (range $\sim$ 5-700 Myr). Among these clusters, this result suggests the dominance of turbulence over ordered rotation in their progenitor clumps of molecular gas. Isotropic rotation axes also support a resolution to the protostellar magnetic braking ``catastrophe'' via reduced angular momentum diffusion by a misaligned magnetic field.

An isotropic model degeneracy with parameters describing moderate alignment prevents us from definitively reporting these clusters to have isotropic spins. Instead, we motivate and use the assumption of uncorrelated spin-axis orientations between clusters to strongly rule out that all eight clusters are aligned and to establish an upper limit of four potentially aligned clusters at 75\% confidence. This limit corresponds to no fewer than four isotropic clusters (50\%) in the overall ten-cluster sample. Using the two-parameter cone model, we constrain the mean inclination of potential spin alignment in Collinder 69, ASCC 19, NGC 2547, Alpha Per, Blanco 1, NGC 2516, the Pleiades, and Praesepe to be between 37$^{\circ}$ and 73$^{\circ}$ at 80\% confidence and $> 32^{\circ}$ at 95\% confidence. We also constrain the spread half-angle to be $> 25^{\circ}$ at 95\% confidence. For these eight clusters, we thus rule out tight alignment and mean inclinations oriented along the LOS.

We find that two clusters' inclination distributions are better fit by a three-parameter model that superposes aligned and isotropic subsets of stars. However, the combination of the small number of stars modeled as aligned and the potential for systematic error in their $\sin i$ values prevents an interpretation of spin alignment. The majority of these stars' fellow cluster members are modeled as isotropic, resulting in a 97\% fraction of stars consistent with isotropic spins across our ten-cluster sample.

Future studies of stellar inclinations in open clusters could increase the number of clusters studied and further reduce systematic error in order to improve constraints on spin-axis distributions. Having a greater number of clusters can strengthen the lower limit on isotropic versus moderately aligned clusters. A greater number of stars per cluster can be beneficial as well, although with diminishing returns on spin-axis constraints beyond sample sizes larger than $\sim$ 30 stars \citep[e.g.][]{jackson2010}. Therefore, increasing the number of clusters is more useful to improve the lower limit of isotropic clusters. A larger cluster sample size will also facilitate inquiry into the mass dependence of spin alignment.

We recommend multi-object spectrograph studies targeting Gaia-based open cluster members across the sky. The projected rotation velocity remains the least available and most inhomogeneous quantity of the three parameters of Eq. \ref{eq:spectrophot}. The above proposed studies would relieve that shortage and provide more homogeneous $v\sin i$ results. In addition, further understanding magnetic radius inflation (see Section \ref{subsec:discussion_radius_inflation}) would reduce a key systematic error that we encountered, especially for the youngest clusters in our sample. Increasing the number of clusters shares priority with reducing systematic error in all parameters of Eq. \ref{eq:spectrophot} for future work.

As illustrated by the 11 stars modeled as aligned in M35 and NGC 2548, even a small number of stars can lead to a statistically significant departure from isotropy, which increases with the fraction of aligned stars (see Section \ref{subsec:simulations_3param}). Increasing the number of $\sin i$ values for clusters that are not well fit by the two-parameter model will reduce the possibility of illusory alignment and yield tighter constraints on physical alignment (again facing diminishing returns beyond $\sim$ 30 stars). The orientation of partially aligned subsets has a large influence on their detectability, with pole-on orientations much easier to differentiate from an isotropic subset. A key improvement for future studies will come from increasing data quality and minimizing systematic errors, helping to confidently determine whether signatures of alignment are physical or instrumental in nature.

While TESS rotation periods were plentiful and essential to this work, the upcoming PLATO mission's smaller angular resolution will reduce the ambiguity due to blending with which we contended \citep[][]{plato2014}. PLATO will also enable asteroseismic studies of stellar inclination that, while subject to their own unique biases, provide additional direct measurements to compare with the spectro-photometric inferences of $\sin i$.

\section*{Acknowledgments}
We are grateful to the referee for providing helpful comments which strengthened this paper. We thank David Sing and Eric Feigelson for helpful thoughts and insights. G.K. acknowledges grant K-129249 from the National Research, Development and Innovation Office. Some of the data presented in this paper were obtained from the Mikulski Archive for Space Telescopes (MAST) at the Space Telescope Science Institute. The specific observations analyzed can be accessed via \dataset[10.17909/t9-yk4w-zc73]{https://doi.org/10.17909/t9-yk4w-zc73}.
This paper includes data collected with the TESS mission, obtained from the MAST data archive at the Space Telescope Science Institute (STScI). Funding for the TESS mission is provided by the NASA Explorer Program. STScI is operated by the Association of Universities for Research in Astronomy, Inc., under NASA contract NAS 5–26555. We acknowledge support from the STScI Director's Research Funds under grant number D0101.90264.

This work has made use of data from the European Space Agency (ESA) mission
{Gaia} (\url{https://www.cosmos.esa.int/gaia}), processed by the {Gaia}
Data Processing and Analysis Consortium (DPAC,
\url{https://www.cosmos.esa.int/web/gaia/dpac/consortium}). Funding for the DPAC
has been provided by national institutions, in particular the institutions
participating in the {Gaia} Multilateral Agreement.

This publication makes use of data products from the Two Micron All Sky Survey, which is a joint project of the University of Massachusetts and the Infrared Processing and Analysis Center/California Institute of Technology, funded by the National Aeronautics and Space Administration and the National Science Foundation. This publication makes use of data products from the Wide-field Infrared Survey Explorer, which is a joint project of the University of California, Los Angeles, and the Jet Propulsion Laboratory/California Institute of Technology, and NEOWISE, which is a project of the Jet Propulsion Laboratory/California Institute of Technology. WISE and NEOWISE are funded by NASA.

This work made use of the Third Data Release of the GALAH Survey \citep[][]{galah_dr3}. The GALAH Survey is based on data acquired through the Australian Astronomical Observatory, under programs: A/2013B/13 (The GALAH pilot survey); A/2014A/25, A/2015A/19, A2017A/18 (The GALAH survey phase 1); A2018A/18 (Open clusters with HERMES); A2019A/1 (Hierarchical star formation in Ori OB1); A2019A/15 (The GALAH survey phase 2); A/2015B/19, A/2016A/22, A/2016B/10, A/2017B/16, A/2018B/15 (The HERMES-TESS program); and A/2015A/3, A/2015B/1, A/2015B/19, A/2016A/22, A/2016B/12, A/2017A/14 (The HERMES K2-follow-up program). We acknowledge the traditional owners of the land on which the AAT stands, the Gamilaraay people, and pay our respects to elders past and present. This paper includes data that has been provided by AAO Data Central (\url{datacentral.org.au}).

The Pan-STARRS1 Surveys (PS1) and the PS1 public science archive have been made possible through contributions by the Institute for Astronomy, the University of Hawaii, the Pan-STARRS Project Office, the Max-Planck Society and its participating institutes, the Max Planck Institute for Astronomy, Heidelberg and the Max Planck Institute for Extraterrestrial Physics, Garching, The Johns Hopkins University, Durham University, the University of Edinburgh, the Queen's University Belfast, the Harvard-Smithsonian Center for Astrophysics, the Las Cumbres Observatory Global Telescope Network Incorporated, the National Central University of Taiwan, the Space Telescope Science Institute, the National Aeronautics and Space Administration under Grant No. NNX08AR22G issued through the Planetary Science Division of the NASA Science Mission Directorate, the National Science Foundation Grant No. AST-1238877, the University of Maryland, Eotvos Lorand University (ELTE), the Los Alamos National Laboratory, and the Gordon and Betty Moore Foundation.

The national facility capability for SkyMapper has been funded through ARC LIEF grant LE130100104 from the Australian Research Council, awarded to the University of Sydney, the Australian National University, Swinburne University of Technology, the University of Queensland, the University of Western Australia, the University of Melbourne, Curtin University of Technology, Monash University and the Australian Astronomical Observatory. SkyMapper is owned and operated by The Australian National University's Research School of Astronomy and Astrophysics. The survey data were processed and provided by the SkyMapper Team at ANU. The SkyMapper node of the All-Sky Virtual Observatory (ASVO) is hosted at the National Computational Infrastructure (NCI). Development and support of the SkyMapper node of the ASVO has been funded in part by Astronomy Australia Limited (AAL) and the Australian Government through the Commonwealth's Education Investment Fund (EIF) and National Collaborative Research Infrastructure Strategy (NCRIS), particularly the National eResearch Collaboration Tools and Resources (NeCTAR) and the Australian National Data Service Projects (ANDS).

This research has made use of the SIMBAD database, operated at CDS, Strasbourg, France. This research has made use of the VizieR catalogue access tool, CDS, Strasbourg, France. The original description of the VizieR service was published in A\&AS 143, 23. This work made use of NASA’s Astrophysics Data System Bibliographic Services. This research has made use of the SVO Filter Profile Service (\url{http://svo2.cab.inta-csic.es/theory/fps/}) supported from the Spanish MINECO through grant AYA2017-84089.

\vspace{5mm}

\facilities{TESS \citep{ricker2015}, Gaia \citep{gaia2016,gaia2018,gaia_edr3_2020}, SDSS:APOGEE-2 \citep[][]{majewski2017, sdss_dr17}, AAT:GALAH \citep[][]{desilva2015, galah_dr3}, CTIO:2MASS (\citealp{skrutskie2006} ,\dataset[10.26131/IRSA2]{https://doi.org/10.26131/IRSA2}), WISE (\citealp{wise_wright2010, catwise_2020}, \dataset[10.26131/IRSA1]{https://doi.org/10.26131/IRSA1}, \dataset[10.26131/IRSA126]{https://doi.org/10.26131/IRSA126}), {Hipparcos}:Tycho-2 \citep{perryman1997,hoeg2000}, GALEX \citep[][]{galex}, LAMOST \citep[][]{cui2012}, Pan-STARRS \citep[][]{panstarrs-1, panstarrs-1_database}, SkyMapper (\citealp[][]{skymapper_dr2}, \dataset[10.25914/5ce60d31ce759]{https://doi.org/10.25914/5ce60d31ce759})}

\software{\texttt{astropy} \citep{astropy2018},
\texttt{astroquery} \citep{ginsburg2019},
\texttt{corner} \citep{corner},
\texttt{eleanor} \citep{feinstein2019},
        \texttt{emcee} \citep{foremanmackey2013},
        \texttt{iPython} \citep{ipython},
        \texttt{isochrones} \citep{morton2015},
        \texttt{lightkurve} \citep{lightkurve2018},
        \texttt{matplotlib} \citep{matplotlib},
        \texttt{multinest} \citep{feroz2009_multinest},
        \texttt{ndtest},
        \texttt{numpy} \citep{numpy1,numpy2},
        \texttt{pandas} \citep{pandas},
        \texttt{pymultinest} \citep{buchner2014_pymultinest},
        \texttt{scikit-learn} \citep{scikit-learn},
        \texttt{scipy} \citep{scipy},
        \texttt{statsmodels} \citep{statsmodels},
        \texttt{TESScut} \citep{tesscut},
        \texttt{TOPCAT} \citep{taylor2005_topcat},
        \texttt{uncertainties} \citep{uncertainties},
        \texttt{zeropoint} \citep{lindegren2020_gaia_edr3_parallax_bias}
          }

\bibliography{sample631}{}
\bibliographystyle{aasjournal}

\listofchanges

\end{document}